\documentclass{article}

\usepackage{arxiv}

\usepackage[utf8]{inputenc} % allow utf-8 input
\usepackage[T1]{fontenc}    % use 8-bit T1 fonts
\usepackage{hyperref}       % hyperlinks
\usepackage{url}            % simple URL typesetting
\usepackage{booktabs}       % professional-quality tables
\usepackage{amsfonts}       % blackboard math symbols
\usepackage{nicefrac}       % compact symbols for 1/2, etc.
\usepackage{microtype}      % microtypography
\usepackage{lipsum}
\usepackage{graphicx}
\graphicspath{ {./images/} }

\usepackage{natbib}
\usepackage{placeins}
\usepackage{subfig}
\usepackage{url}
\usepackage{amsmath}
\usepackage{xltabular}
\usepackage{booktabs}
\usepackage{graphicx}
\usepackage[font = footnotesize, labelfont = bf]{caption} % to adjust caption text size
\usepackage{float} % for H float
\usepackage{appendix}
\title{Incorporating Missingness in a Framework for Generating Realistic Synthetic Randomized Controlled Trial Data}

\author{
 Niki Z. Petrakos \\
  Department of Epidemiology, Biostatistics and Occupational Health \\
  McGill University \\
  Montreal, Canada \\
  \texttt{niki.petrakos@mail.mcgill.ca} \\
   \And
 Erica E. M. Moodie \\
  Department of Epidemiology, Biostatistics and Occupational Health\\
  McGill University\\
  Montreal, Canada \\
  \texttt{erica.moodie@mcgill.ca} \\
  \And
 Nicolas Savy \\
  Institut de Mathématiques de Toulouse; UMR5219 \\
  Université de Toulouse; CNRS. UT2J \\
  Toulouse, France \\
  \texttt{nicolas.savy@math.univ-toulouse.fr} \\
}

\begin{document}
\maketitle
\begin{abstract}
The current literature regarding generation of complex, realistic synthetic tabular data, particularly for randomized controlled trials (RCTs), often ignores missing data. However, missing data are common in RCT data and often are not Missing Completely At Random. We bridge the gap of determining how best to generate realistic synthetic data while also accounting for the missingness mechanism. We demonstrate how to generate synthetic missing values while ensuring that synthetic data mimic the targeted real data distribution. We propose and empirically compare several data generation frameworks utilizing various strategies for handling missing data (complete case, inverse probability weighting, and multiple imputation) by quantifying generation performance through a range of metrics. Focusing on the Missing At Random setting, we find that incorporating additional models to account for the missingness always outperformed a complete case approach.
\end{abstract}

% keywords can be removed

\section{Introduction}\label{introduction_sec1}

Generating completely synthetic data that adequately capture real-life complexities, particularly within the context of small tabular randomized controlled trial (RCT) data, has garnered much attention in recent years due to the utility of synthetic data in determining optimal trial design \citep{Chen2021,SarramiForoushani2021,Friedrich2024,Zwep2024}. However, it still remains unclear how best to generate synthetic data in an RCT context, where data are in a tabular form, follow a specific temporal ordering, and often have a small sample size. For tabular data, each column is a random variable such that all columns collectively follow some joint distribution, which is almost always unknown. Each row in a tabular data set represents an observation from this (unknown) joint distribution. Additionally, variables of different types (continuous, categorical, multi-modal, etc.) can all exist in the same table, thus leading to a complex joint distribution that is hard to learn and mimic. The longitudinal aspect of RCTs is also important to consider for the data generation task. Generally, RCTs begin with collecting baseline data before randomizing participants to treatment arms, then potentially following participants at multiple time points after treatment randomization, and finally measuring the trial outcome. Though RCT data are longitudinal, they differ from the more-commonly studied context of time series data generation, as RCTs generally involve only a handful of time points over a relatively short period of time (e.g., a span of several weeks or months) and do not always follow regular time intervals. Moreover, while much of the data generation literature has shown success in generating synthetic data of high fidelity in other contexts (e.g., image data) via deep learning algorithms \citep{Denton2015,Radford2016}, the number of participants in a given trial can range from a few dozen to a few thousand, thus rendering the application of many deep learning generative models to the RCT data generation context challenging due to small sample sizes. 

Another limiting aspect of the current data generation literature (which largely falls within the machine learning, or ML, research domain) is that most studies completely ignore missing data. Even among studies focused on the health data context, the example data sets used to demonstrate the effectiveness of a generative model contain no missing data despite this feature being a common obstacle in health data. On the occasion that the real data sets do contain missing values, missingness is often handled as a data pre-processing step, where all participants with missing values are removed and the data generation procedure follows using only data from participants who were completely observed, i.e., the complete cases \citep{Neves2022,Mendikowski2023,Farhadyar2024}. Following this pre-processing, generative models are fit to the real data complete cases such that the generated synthetic data follow a distribution that is close to that of the real complete cases. This implies an assumption on the part of the researchers that the distribution of the complete synthetic data (since generally, no synthetic missingness is generated either) is close to the distribution of only the \textit{observed} real data. However, it is well established that missingness in trials is unlikely to be Missing Completely At Random (MCAR), where missingness is independent of all other information \citep{Little2012}. Rather, missing values in trials likely follow a Missing At Random (MAR) or Missing Not At Random (MNAR) mechanism, where the former means that missingness depends on observed data and the latter means that missingness depends on the missing values themselves. Hence, it is likely that the distribution of the complete real RCT data is different from the distribution of the observed real RCT data due to participant drop-out depending on certain characteristics, thus rendering the use of only complete cases to be biased. In other words, the generated synthetic data distribution likely does not mimic the (complete) real data distribution. 

\begin{figure}[h!]
    \centering
    \includegraphics[width=\linewidth]{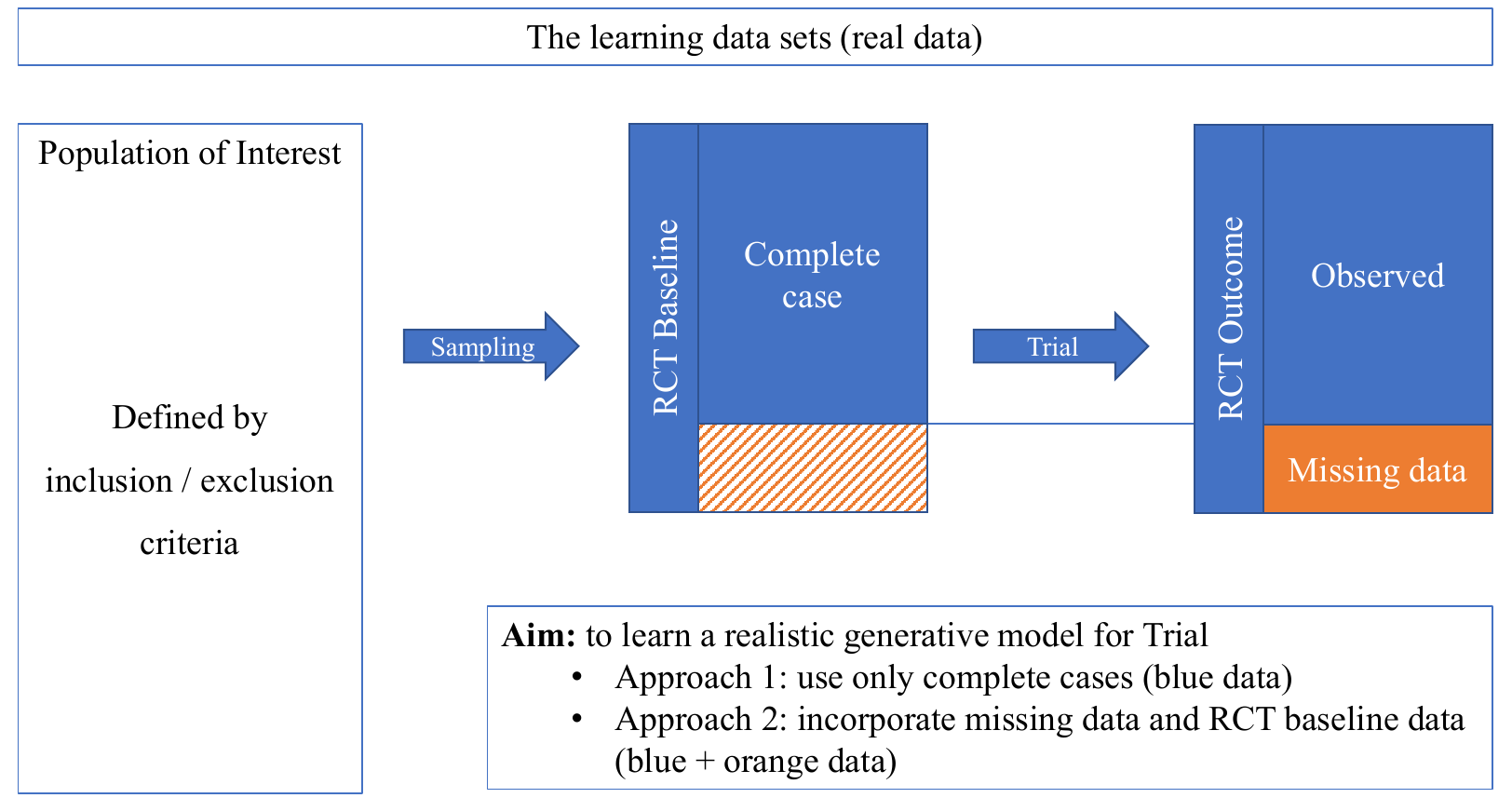}
    \caption{Schema demonstrating the flow of available data for fitting generative models in the context of an RCT. The orange box at the RCT outcome stage represents the portion of missing data, and the shaded orange box at the RCT baseline stage represents the observations at baseline that would be omitted under a complete case approach. Sampling is assumed to be representative and thus the RCT baseline data set has the same distribution as the population of interest. Under MCAR, the distribution of the blue data is the same as the distribution of the orange (missing) data. Note that Approach 1 is what currently dominates the literature, whereas we propose that Approach 2 is needed, and examine different implementations of this.}
    \label{fig:missingdataschema}
\end{figure}

In \autoref{fig:missingdataschema}, we illustrate the phenomenon described above, whereby synthetic generation of RCT data often omits the key feature of missingness. The overarching goal in an RCT is to learn about the treatment effect within a specific population of interest, defined by certain inclusion/exclusion criteria. From this population, a baseline cohort is sampled and serves as the RCT study sample. While missing data can occur at baseline, it is more common for baseline data to be complete in an RCT and missingness to arise subsequently; we will focus on this setting in the present work. Then, the trial commences with participants randomized to various treatment arms and after a pre-specified time, the RCT outcome is measured; interim measurements between treatment receipt and the final outcome may also occur. It is often the case that some participants have missing data due to attrition. In a task focusing on generating realistic data on the basis of such a trial, these (partially observed) real RCT data are then used to fit generative models, where omitting observations with missingness as a data pre-processing step (i.e., removing rows with missing data in the outcome) necessarily means removing data at baseline as well (shown by the shaded rectangle in \autoref{fig:missingdataschema}). Ideally, the distribution of the real data complete cases would be the same as the distribution of the entire sampled cohort (and hence also the population of interest) -- this is the case under MCAR. Then, generative models would learn the distribution of the population of interest and thus generate synthetic data with the desired distribution. However, RCT data are rarely MCAR and thus the distribution of the complete cases is \textit{systematically different} from that of the population of interest. This leads to generative models learning a distribution that does not represent well the population of interest and thus hinders the goal of learning about the treatment effect within this population. Thus, simply ignoring and omitting the missing values as a data pre-processing step before generating synthetic data in the context of RCTs is inherently problematic. Moreover, since the goal of synthetic data generation is to generate data that capture the complexities in the real data, and since missing values are one such important complex aspect of real data, it is pertinent to account for missingness in the data generation procedure. 

\FloatBarrier

Of course, not all data generation studies restrict to using complete cases. In \cite{Wang2023} and \cite{Li2023}, a single imputation of missing values was performed before proceeding with the fitting of generative models. Similarly, in \cite{Walia2020}, authors performed mode imputation as a data pre-processing step. However, a single imputation does not take into account the uncertainty of the imputation procedure, and mode imputation is unlikely to adequately reflect the complexity in the data including dependence between variables, thus hindering the overall goal of generating realistic synthetic data. Additionally, filling in missing values in a data pre-processing step does not naturally allow for the generation of synthetic missing values, thus further detracting from the ability to generate synthetic data that are realistic. Even when data generation methods address missingness (in a manner other than restricting to complete cases) and allow for the generation of missing values in the synthetic data, there is very limited investigation of whether the true missingness mechanism (i.e., the reason for values being missing) was well-captured in the synthetic data \citep{Zhao2021}. Indeed, the current state-of-the-art method for generating tabular data is the Conditional Tabular Generative Adversarial Network (CTGAN), implemented in the popular python library Synthetic Data Vault \citep{Patki2016}; when fitting a CTGAN using this library, missing values are taken into account ``under the hood'' and synthetic missingness is also generated. However, there is no documentation describing how missingness is handled (nor is there mention of missing data in the seminal paper on CTGAN \citep{Xu2019}), and the generated data with missingness can result in distributions that are notably different from the original data (see \cite{Petrakos2025} for an example of a use-case). Unfortunately, this lack of documentation handling missing values in the data generation procedure is common \citep{Choi2017,Hyun2020,Koloi2023}. 

Missing data is a well-studied topic in the ML literature, focusing mainly on deep learning methods for imputing missing values, rather than with the broader goal of generating completely synthetic data \citep{Stekhoven2011,Yoon2018,Gondara2018,Nazabal2020,Kazijevs2023,Liu2023,Lim2025}. A significant drawback of many recent deep learning studies dedicated to missing data is that it remains unclear what missingness mechanism is assumed for the experiments presented \citep{Jarrett2022}. In instances when the missingness mechanism is defined, it is typically MCAR \citep{Liu2023}. Many studies argue that assuming MCAR and performing experiments in this setting is satisfactory because it is a common assumption in the ML literature \citep{Yoon2018,Ma2020,Spinelli2020,Neves2022,Pingi2024}.  However, MCAR is not realistic for missingness in trials and indeed, is not realistic in most real world examples \citep{Little2012}. This raises the question of whether the plethora of deep learning methods for imputing missing values applies to other missingness settings that are more likely to occur in real data collection settings, particularly health or trials contexts.

Among the few studies that assumed a missingness mechanism other than MCAR and that focused on missing data imputation for tabular health data, some involved temporal data and a select few others involved multi-modal data (i.e., variables of different types). No studies considered both multi-modal and temporally ordered data \citep{Liu2023}. Furthermore, in the deep learning missing data literature, there is limited discussion of missingness patterns. In longitudinal data where variables follow a temporal ordering, \textit{monotone missingness} occurs when a missing value for a given row in the data set at some point in time implies that there will be a missing value for that row at all future time points. The more general case when this missingness pattern does not hold is often referred to as \textit{non-monotone missingness}. Monotone missingness is particularly pertinent in the RCT context due to participant attrition. Hence, at the time when the patient drops out, and for all future time points, there are missing values in the data collected for such a patient. Non-monotone missingness is also applicable to RCTs, as participants can miss certain follow-up visits but return for later visits. It is therefore useful to study both the non-monotone and monotone scenarios when accounting for missing observations to generate synthetic RCT data.

The question motivating this work is whether there are any existing methods for synthetic data generation (particularly in the RCT context or at least in tabular health data) that also adequately account for missingness. To our knowledge, few such methods exist. Much research has been dedicated to either generating complete synthetic data using (often completely-observed) real data, or developing methods for imputing missing values, but not within the broader context of generating an entirely new, synthetic data set. \cite{Gomez2025} proposed a novel data augmentation method for blood glucose forecasting and tested their method on several real data sets, including one RCT. MAR and MNAR missingness mechanisms were assumed, but generative performance was based on forecasting and prediction results rather than investigating the closeness of the (multivariate) synthetic data distribution to the real data distribution. Another study demonstrated how their generative method involving a recurrent autoencoder was able to learn the complex real data distribution with missing values as well as impute the missing values, and to do so, an MAR mechanism was assumed \citep{Bianchi2019}. However, in the experimentation involving learning the real data distribution, the true missingness mechanism imposed by the authors on the real data was in fact MCAR. The few other studies that combined the tabular health data generation task (often involving longitudinal aspects) with the ability to handle missing values did not define the missingness mechanism \citep{Mosquera2023,Weng2024,Eckardt2024} or did not allow for the generation of synthetic missing values \citep{Neves2022,Weng2024}; none mentioned missingness patterns. \cite{Mosquera2023} noted that their approach in quantifying generative performance involved a single generated data set and did not incorporate additional variance that stemmed from the data generation procedure. \cite{Neves2022} combined multiple imputation with data generation, but did so under MCAR and did not consider multi-modal data. \cite{Eckardt2024} demonstrated how synthetic missing values could be generated by categorizing missing values to be their own stratum, though it remains unclear how this method would apply to continuous variables. In this work, we bridge this gap of accounting for missing data in synthetic data generation procedures, specifically within the context of small tabular RCT data involving longitudinal and multi-modal data. In particular, we investigate how best to account for various missingness scenarios likely to occur in an RCT setting, including considering both non-monotone and monotone missingness patterns, such that the complete and observed synthetic data distributions mimic the complete and observed real data distributions, respectively, while also generating synthetic missingness that follows the pattern of missingness in the original data source.

To address how best to generate complex, realistic synthetic data in the context of RCTs, we previously compared several synthetic data generation frameworks and determined that sequential data generation that separated the data generation task into steps that followed the temporal ordering of the real data outperformed a simultaneous data generation approach \citep{Petrakos2025}. More specifically, the modelling choices that led to the best performance (of those tested) were to use an R-vine copula at baseline, sample from a pre-defined probability distribution to generate synthetic randomized treatment assignment, and then fit separate regression models for each post-randomization variable (i.e., variables collected at follow-up visits including the trial outcome) where the predictors in the regression models were all variables collected at previous time points in the trial. Hence, the final regression model was used to generate the primary trial outcome, where all other variables in the trial were included as predictors. Additionally, an extra step to induce randomness when generating post-randomization variables via regression was implemented, where the final generated value was sampled from the set of admissible values, calculated as a function of the predicted value and a residual from the fitted model. Here, we extend our previously-proposed framework for generating synthetic tabular RCT data by incorporating additional models within the framework to both account for the missingness mechanism in the real data and generate synthetic missing values, all while still producing a completely synthetic data set that is close to, and captures the complex characteristics of, the real data.

We turn to classical approaches to handling missingness that can seamlessly be incorporated into the regression approaches used in the previously-proposed framework to generate post-treatment randomization variables, both within the context of non-monotone missingness and monotone missingness. More details can be found in \autoref{sec:missingdatabackground}, which introduces notation and pertinent information from the missing data literature. \autoref{sec:sequentialdatageneration} presents the proposed data generation frameworks accounting for missing data, \autoref{sec:experimentalmethods} describes the real data example, the missingness scenarios considered in our empirical comparisons, and the metrics used for quantifying generative performance, \autoref{sec:results} presents empirical results, and \autoref{sec:discussion} discusses strengths and limitations of the presented work as well as potential avenues for future research.

\section{Missing Data Background and Notation}
\label{sec:missingdatabackground}

In the missing data literature, missingness mechanisms are often defined as either MCAR, MAR, or MNAR \citep{Rubin1976}. Let $R$ represent the indicator of being observed, where $R=1$ denotes observed and $R=0$ denotes missing. As this work focuses on the RCT context, let $\mathbf{X}$ denote baseline data, $A$ denote randomized treatment assignment, and $Y$ denote the primary trial outcome. As an example, say $Y$ had missingness and all other variables were completely observed. Then under MCAR, missingness is independent of all other variables; in other words, $$\text{P}(R=1|\textbf{X},A,Y)=\text{P}(R=1).$$ In contrast, under MAR, missingness is dependent on the observed data: $$\text{P}(R=1|\textbf{X},A,Y)=\text{P}(R=1|\textbf{X},A).$$ Under MNAR, the missingness is dependent on the missing values (and potentially the observed values as well) and hence the previous equality no longer holds. In the following work, we assume baseline data and treatment assignment are completely observed, whereas post-randomization variables (those measured at follow-up visits) and the final outcome may be missing. We denote post-randomization variables as $Z_k, \ k=1,...,K$ where $K$ is the number of follow-up visits at which measurements were taken not including the final trial visit. Hence, let $R_{Z_k}$ be the indictor of being observed for $Z_k$ and $R_Y$ be the indicator of being observed for $Y$. The simulations presented in this paper include MCAR but mainly focus on MAR. MNAR is not explored because strong, often untestable assumptions are needed in order to model the missingness mechanism when it depends on the unobserved values (which by definition are unknown), a difficult task even with expert domain knowledge \citep{Dong2013}. 

The easiest approach to missing values is to simply omit them from the data before performing any data analysis; this is known as a complete case (CC) analysis. If data are MCAR, then the observed data should not systematically differ from the unobserved data, and hence CC will lead to unbiased estimators in subsequent statistical analyses. However, since a non-negligible proportion of observations may be omitted depending on the proportion of missingness, CC is inefficient due to reduced sample size. If data are not MCAR, then the observed data distribution will differ from the unobserved data distribution, which means CC would typically lead to biased estimators. There are some instances where CC is appropriate (though, it remains inefficient). If the indicator of being a complete case is independent of $Y$ given other covariates, then a CC analysis estimating the parameters of the conditional distribution of $Y$ given those covariates would result in unbiased estimators \citep{White2010}. Additionally, with only one covariate (say, a single $X$), CC is also unbiased if data are MCAR or MNAR with the mechanism that the probability of being a complete case is only a function of $X$ and not of $Y$ \citep{White2010}. However, both of these cases are generally not applicable to an RCT setting, as the former is only equivalent to an MAR setting if missingness is restricted to $Y$ and the latter is too simplistic since RCTs generally involve not only a vector $\mathbf{X}$ but also $A$ and $Z_1,...,Z_K$.

Two other well-established methods for handling missingness are inverse probability weighting (IPW) and multiple imputation (MI). Although IPW relies on the complete cases for the analysis model, all observations are used in the analysis to construct the weights which are then used to reweight observed cases so as to ensure that the weighted sample is representative of the population with both fully observed and missing data. These weights are equal to the inverse of the probability of being observed for each individual, which is estimated by fitting what is often termed a \textit{missingness model}. This is usually a logistic regression model with $R$ as the outcome and any variables deemed relevant to the missingness mechanism as the predictors. (Note that $R$ is always completely observed.) IPW is a common method for accounting for non-response in survey data \citep{Little2024}, which can be thought of as a similar procedure as participant drop-out in a trial \citep{Seaman2013}. Further, IPW has been shown to be effective in reducing bias due to missingness under the MAR setting \citep{Little2024} and can be applied to variables of different types. Hence, IPW lends itself well to our setting. 

While IPW involves modelling the probability of being observed, MI instead models the conditional distribution of the missing values given the observed data -- this model is often referred to as an \textit{imputation model} \citep{Rubin1987}. MI proceeds by first imputing $m$ complete data sets using the fitted imputation models, and then the statistical analysis of interest is performed on each of the $m$ imputed data sets which results in $m$ parameter estimates. The final step is to pool parameter estimates to yield one final parameter estimate, and variability of the MI estimator combines variance within data sets and across data sets in order to reflect uncertainty in the imputation procedure in addition to sampling variability; this is often done following Rubin's rules \citep{Rubin1987}. In this paper, MI via Chained Equations (MICE) is performed \citep{Raghunathan2001,vanBuuren2007} using predictive mean matching or PMM \citep{White2011}. MICE assumes MAR and can accommodate variables of multiple types, and thus is suitable in the context of RCT data. PMM is often preferred over other approaches to imputation as it involves sampling from an empirical conditional distribution rather than a parametric distribution, which is often the case in traditional MI \citep{Austin2021}.
Ideally, the observed data should explain well the missing values and hence imputed values for a given variable across the $m$ complete data sets should not differ drastically (though, variation is expected across the imputed data sets).

The following section describes how IPW or MI can be integrated into our previously-proposed framework for synthetic data generation to both account for missing values in the real data and generate synthetic missingness such that the generated synthetic data are close to the real data. CC is used as a baseline comparator to determine whether ignoring missingness entirely in a data pre-processing step by removing individuals with any missing values (as is often done in the data generation literature) is sufficient. We subsequently refer to this procedure as ``all stage'' CC. We also consider a ``by stage'' CC approach, where generative models are sequentially fit to different subsets of the real data at each step in the trial, based on the individuals with observed data at each time point. In other words, at baseline and treatment allocation, all individuals are included in the by stage CC approach; this approach is also known as available case analysis \citep{LittleRubinBook2002}. Then at each time point $k$ and for the final trial outcome, all available information is used to fit the generative models. Note that under monotone missingness, the observed information corresponds to subsets of individuals that are nested such that the final subset is contained within the subset at time $K$, which is contained within the subset at $K-1$, etc. Since the overarching goal is to generate complex and realistic synthetic data, the following notation is introduced: real data variables are denoted by a superscript $r$ and synthetic data variables are denoted by a superscript $s$; i.e., the real RCT data are comprised of $\{ \textbf{X}^r,A^r,Z_1^r,...,Z_K^r,Y^r \}$ and synthetic RCT data are comprised of $\{ \textbf{X}^s,A^s,Z_1^s,...,Z_K^s,Y^s \}$. We focus on generating synthetic data that match the observed data on all characteristics including the extent of missingness; however, the final step of creating missingness in the synthetic data can be omitted if complete synthetic data are preferred.

\section{Sequential Data Generation Under Varying Missingness Scenarios}
\label{sec:sequentialdatageneration}

Before demonstrating how missingness or imputation models can be incorporated in a sequential data generation procedure, we first describe the sequential procedure that we previously proposed in \cite{Petrakos2025}, which utilizes data from an RCT involving four possible treatment arms and a binary outcome. Additionally, the two post-randomization variables are count variables and hence synthetic values are deemed admissible (i.e., are kept) if they are non-negative. For the remainder of the paper, we set $K=2$ such that there are two variables collected at follow-up, denoted $Z_1$ and $Z_2$. First, an R-vine copula model is fit to the real, completely observed baseline data $\textbf{X}^r$, and $\textbf{X}^s$ is generated by sampling from the fitted copula model with the number of synthetic samples set to be equal to the number of real samples. Next, $A^s$ is generated by sampling from a pre-defined distribution with equal probabilities for each treatment group (in the case of our motivating example, where there are four treatment arms, $A^s$ is generated from a multinomial with probability vector $p=(0.25, 0.25, 0.25, 0.25)$). Following that, $Z^s_1$ is generated. To do so, $Z^r_1$ is regressed on $\textbf{X}^r$ and $A^r$. Then, $\hat{Z}^s_1$ is predicted as a function of $\left( \textbf{X}^s,A^s \right)$ using the fitted regression model, and $Z^s_1$ is generated by sampling from the set of admissible values $\{ (\hat{Z}^s_1+r_{Z^r_1}) \geq0 \}$, where $r_{Z^r_1}$ denotes the residuals of the fitted regression model for $Z^r_1$. Note that a threshold of zero is set for our definition of admissibility due to features of the motivating example where the post-randomization variables are count variables and thus cannot be negative (see \autoref{subsec:datadescription} for details). However, a different threshold value can be used or indeed none may be needed depending on the data context. Then, following a similar procedure, $Z^s_2$ is generated. Here, the regression model includes $\textbf{X}^r$, $A^r$, and $Z^r_1$. The generation of $Z^s_2$ then proceeds much as $Z^s_1$. Finally, $Y^s$ is generated by first fitting a regression model with $Y^r$ as the outcome and all other variables as the predictors. In the case of our motivating example, where the outcome is binary, a logistic regression is used and then $\hat{\pi}^s_Y$, the probability of the outcome event occurring in the synthetic data, is predicted as a function of $\left( \textbf{X}^s,A^s,Z^s_1,Z^s_2 \right)$ and a Bernoulli$(\hat{\pi}^s_{Y})$ is drawn. The synthetic variables generated at each sequential step ($\textbf{X}^s,A^s,Z^s_1,Z^s_2,Y^s$) are merged to form the final generated synthetic data set. 

To determine whether and how best to extend this sequential framework such that missing data are incorporated, we compare the following approaches (both under non-monotone and monotone missingness): two CC methods, IPW, and MI. The first CC method (CC: All Stage) is how missing data are commonly treated in the data generation literature and is thus included as a baseline comparator. Individuals with any missing data are removed, and sequential data generation is then performed as described above. No missingness is imposed in the synthetic data. When the true missing data mechanism depends on covariates such that the data are MAR, this approach is expected to perform poorly. The second CC method (CC: By Stage) involves performing sequential data generation as previously described, removing only individuals with missing data in variables used in a given sequential step. Again, no missingness is imposed in the synthetic data. In this approach, the subsets of real data to which the generative models are fit at each sequential step may have different sample sizes. The motivation for this is to retain as much information as possible such that the synthetic data can better mimic the original data distribution. Hence, CC: By Stage differs from CC: All Stage in that the data pre-processing step is performed in a stage-wise fashion, rather than prior to any model fitting. 

In the IPW approach, the probability of being observed is modelled at each stage where missingness exists in the real data. Then, sequential data generation is performed as described before, however now, in any stage where data are missing, weighted regression is used to account for missingness. Note that when missingness is not monotone, this may create problems when a covariate is missing that is subsequently needed to model missingness of a variable at a later stage. We propose the following strategies to address this: (i) incorporate an indicator of being observed in the missingness model, or (ii) force the data to follow a monotone missing pattern. Once complete synthetic data have been generated according to these (possibly weighted) models, missingness is imposed in the synthetic data according to the missingness mechanism as determined by the estimated missingness model. For the MI approach that we propose, the first step is to create $m$ data sets. Next, sequential data generation is performed as previously described (without weights), for each of the $m$ data sets as detailed below; sampling is then used to select from the $m$ complete data sets so that only a single synthetic data set is ultimately obtained. In the real data, missingness is again modelled at each stage using the same IPW approach. Then, these missingness models are used to impose missingness in the single synthetic data set.

In the following sections, we describe in more detail how these proposed approaches can be naturally integrated into the sequential framework. We continue with $K=2$, though the proposed sequential frameworks can easily be adapted to other values of $K$. Since the methods differ depending on the pattern of missingness, we first present generative frameworks under the more general case of non-monotone missingness, followed by the special case of monotone missingness. Additionally, in order to perform comparisons between the synthetic data distribution and the completely-observed real data distribution to determine the data generation performance of each proposed framework, we also impose our own missingness mechanisms on the real data in order for the ``truth'' to be known in the subsequent simulations (see \autoref{subsec:missingnessscenarios} for details regarding this procedure).

\subsection{Methods for the General Case: Non-Monotone Missingness}
\label{subsubsec:generalcasenonmonotonemissingness}

Under non-monotone missingness, missingness was imposed in $Z^r_1$ and $Y^r$ such that $Y^r$ could be observed when $Z^r_1$ was missing and vice versa. $Z^r_2$ was completely observed. Missingness was imposed by defining a model and parameter values for the probability of being observed for $Z^r_1$ and for $Y^r$. For details regarding these models for imposing missingness in the real data, refer to the Supplementary Materials. In this setting, we present the modelling techniques for the five candidate frameworks to generate synthetic data, each with varying methods of accounting for missing data: CC: All Stage, CC: By Stage, IPW: Indicator Method, IPW: Force Monotonicity, and MI.

\subsubsection{CC: All Stage} 
\label{subsubsec:nonmonotoneccallstage}

The first step is to remove all observations in the real data with missing values, thus leaving only the real complete cases. Then, the data generation procedure follows as described in the beginning of \autoref{sec:sequentialdatageneration}.

\subsubsection{CC: By Stage} 
\label{subsubsec:nonmonotoneccbystage}

The first steps are to directly generate $X^s$ using an R-vine copula and $A^s$ sampling from the pre-defined probability distribution. Since no individuals have missing data at baseline or treatment assignment, all real participant data are included at these stages. Then, to generate $Z^s_1$, the same modelling procedure is implemented as before except that those with missing $Z^r_1$ are removed. To generate $Z^s_2$, again the same modelling procedure is used as explained above, but now those with missing $Z^r_1$ are removed since $Z^r_1$ is included as a predictor in the regression model (recall that $Z^r_2$ was fully observed in the non-monotone missingness scenario). Finally, $Y^s$ is generated using the same procedure described above (relying on logistic regression and prediction), but now individuals with missingness in $Z^r_1$ or $Y^r$ are omitted in the model fitting.

\subsubsection{IPW} 
\label{subsubsec:nonmonotoneipw}

In a first approach to addressing the missingness in a more sophisticated manner, rather than simply ignoring the missing values in the real data, a weighting procedure is incorporated into the data generation framework described in \autoref{sec:sequentialdatageneration}. As before, the first step is to generate complete $\textbf{X}^s$ using an R-vine copula, and the second step is to generate complete $A^s$ by sampling from the pre-defined probability distribution. As in CC: By Stage, data from all individuals are harnessed at both baseline and treatment data generation stages. Next, complete synthetic post-randomization variables are generated starting with $Z^s_1$. 

Unlike CC, before fitting the regression model directly, the probability of being observed for $Z^r_1$ is first estimated via $\hat{p}^r_{Z_1}=\text{Pr}( R^r_{Z_1}=1| \textbf{ X}^r,A^r )$, where $R^r_{Z_1}$ is the indicator variable representing whether $Z^r_1$ is observed (with 1 indicating observed and 0 missing). This model is fit to the real data using all real observations. Then, a \textit{weighted} regression model is fit (where $Z^r_1$ is regressed on $\textbf{X}^r$ and $A^r$) with weights equal to $1/\hat{p}^r_{Z_1}$. This weighted model is fit to the subset of data with observed $Z^r_1$. The final steps for generating (complete) $Z^s_1$ remain the same as in both CC methods: predict $\hat{Z}^s_1$ and then generate $Z^s_1$ by sampling from the set of admissible values.  

Next, generate $Z^s_2$. In the non-monotone scenarios presented in this work, $Z^r_2$ was completely observed. Hence, $Z^r_2$ is regressed on $\textbf{X}^r$, $A^r$, and $Z^r_1$. However, $Z^r_1$ was partially observed, so weights are also incorporated in this regression model, which is fit to the same real data observations as at the $Z^r_1$ data generation stage. These weights are to account for the missingness in $Z^r_1$ and thus are the same as the weights used in the weighted model for $Z^r_1$, i.e., $1/\hat{p}^r_{Z^r_1}$. The rest follows as before; predict $\hat{Z}^s_2$ and generate $Z^s_2$ by sampling from the set of admissible values.

Finally, $Y^s$ is generated. As $Y^r$ had missing values, the first step is to estimate the probability of being observed: $\hat{p}^r_Y=\text{Pr}( R^r_{Y}=1|\textbf{ X}^r,A^r,Z^r_1,Z^r_2 )$. However, this model cannot be fit directly to the data without removing observations due to $Z^r_1$ having missing values. One potential solution is to condition on whether $Z^r_1$ was observed:
\begin{equation}
\label{eq:nonmonotoneprobcondition}
    \begin{split}
        \hat{p}^r_Y = \text{Pr}\left(  R^r_{Y}=1|\textbf{ X}^r,A^r,Z^r_1,Z^r_2 \right) = &\text{Pr}\left(  R^r_{Y}=1|\textbf{ X}^r,A^r,Z^r_1,Z^r_2,R^r_{Z_1}=1 \right)\times\text{Pr}\left( R^r_{Z_1}=1|\textbf{ X}^r,A^r,Z^r_2 \right) \\
        & + \text{Pr}\left(  R^r_{Y}=1|\textbf{ X}^r,A^r,Z^r_1,Z^r_2,R^r_{Z_1}=0 \right)\times\text{Pr}\left( R^r_{Z_1}=0|\textbf{ X}^r,A^r,Z^r_2 \right).
    \end{split}
\end{equation}
Note that $\text{Pr}( R^r_{Z_1}=1|\textbf{ X}^r,A^r,Z^r_2 )=\text{Pr}( R^r_{Z_1}=1|\textbf{ X}^r,A^r,Z^r_1,Z^r_2 )$ since the probability of $Z^r_1$ being observed does not depend on the value of $Z^r_1$ under MAR. Similarly, $\text{Pr}( R^r_{Z_1}=0|\textbf{ X}^r,A^r,Z^r_2 )=\text{Pr}( R^r_{Z_1}=0|\textbf{ X}^r,A^r,Z^r_1,Z^r_2 )$.
The first term on the right-hand side, $\text{Pr}( R^r_{Y}=1|\textbf{ X}^r,A^r,Z^r_1,Z^r_2,R^r_{Z_1}=1 )$, can be fit directly to the real data for all individuals with observed $Z^r_1$. The second term, $\text{Pr}( R^r_{Z_1}=1|\textbf{ X}^r,A^r,Z^r_2 )$, can also be fit directly to the real data for all individuals as $\textbf{X}^r$, $A^r$, and $Z^r_2$ were completely observed. The fourth term, $\text{Pr}( R^r_{Z_1}=0|\textbf{ X}^r,A^r,Z^r_2 )$ can be computed simply by $1-\text{Pr}( R^r_{Z_1}=1|\textbf{ X}^r,A^r,Z^r_2 )$. However, the third term, $\text{Pr}( R^r_{Y}=1|\textbf{ X}^r,A^r,Z^r_1,Z^r_2,R^r_{Z_1}=0 )$, is problematic because this model cannot be fit directly to the data. We propose and compare two different ways to estimate $\hat{p}^r_Y$: either utilizing $R^r_{Z_1}$ as a predictor in the missingness model and incorporating the interaction term $R^r_{Z_1}\times Z^r_1$, or by forcing monotonicity in the data. Both methods are outlined below. Note, however, that no matter how $\hat{p}^r_Y$ is estimated, the inverse of this probability is then used to weight the outcome model, which is then used to predict $\hat{p}^s_Y$ and generate $Y^s$.

\medskip

\noindent
\textbf{\textit{IPW: Indicator Method.}} 
Here, $\hat{p}^r_Y$ is estimated by fitting $\text{Pr}( R^r_{Y}=1|\textbf{ X}^r,A^r,R^r_{Z_1},Z^r_2,R^r_{Z_1}\times Z^r_1 )$ rather than \autoref{eq:nonmonotoneprobcondition}. For those with $Z^r_1$ observed, the estimation of their probability of being observed at $Y^r$ includes their $Z^r_1$ value. Effectively, different models are fit for the subset of individuals with $Z^r_1$ observed and the subset of individuals with $Z^r_1$ missing. The rest of the data generation procedure follows as before. A weighted logistic regression model is fit with weights equal to the inverse of $\hat{p}^r_Y$, estimated using the model with $R^r_{Z_1}$ and $R^r_{Z_1}\times Z^r_1$ as covariates. Then, $Y^s$ is predicted from a Bernoulli distribution whose probability is found using the fitted weighted model and data $(\textbf{X}^s,A^s,Z^s_1,Z^s_2 )$. Note that the weighted logistic regression model to generate $Y^s$ in this framework differs from the other frameworks; here, rather than including the main effect of $Z^r_1$ as a covariate in the model, $R^r_{Z_1}$ and $R^r_{Z_1} \times Z^r_1$ are included instead.

\medskip

\noindent
\textbf{\textit{IPW: Force Monotonicity.}} Under this proposal, the original model to estimate $\hat{p}^r_Y$ is fit (i.e., \autoref{eq:nonmonotoneprobcondition}), but $Y^r$ is set to missing for those with missing values at $Z^r_1$. (Note that we do not set values to be missing for the fully-observed $Z^r_2$, even though this variable was collected after $Z^r_1$ in the RCT.) In other words, a monotone missing pattern is forced on the real data in order to fit the original missingness model. Forcing monotonicity allows for a decomposition of the conditional probability of being observed at $Y^r$ that avoids the difficulty with term three in Equation~\eqref{eq:nonmonotoneprobcondition}. Forcing monotonicity such that $Y^r$ is missing when $Z^r_1$ is missing ensures this term is equal to zero. This leaves the estimation of $\hat{p}^r_Y$ to be equal to $\text{Pr} ( R^r_{Y}=1|\textbf{ X}^r,A^r,Z^r_1,Z^r_2,R^r_{Z_1}=1 )\times\text{Pr}( R^r_{Z_1}=1|\textbf{ X}^r,A^r,Z^r_2 )$, both of which can be easily fit to the real data. Then, data generation continues as before by fitting a weighted logistic regression model with weights estimated by the original missingness model with monotone missingness imposed. Then, $\hat{p}^s_Y$ is predicted, and $Y^s$ is finally generated. Note that modelling the probability of being observed at $Z^r_1$ by conditioning on $Z^r_2$ does not follow the temporal ordering of the original data; this is discussed further in \autoref{sec:discussion}. Alternatively, one could simply impose the assumption that the model does not depend on $Z^r_2$ due to the known temporal ordering.

The last step is to generate synthetic missingness. Starting with $Z^s_1$, $\hat{p}^s_{Z_1}$ is predicted for $( \textbf{X}^s,A^s )$ using the fitted missingness model (note that this model is the same for both IPW: Indicator Method and IPW: Force Monotonicity; it does not include an indicator of being observed as a covariate and does not impose monotonicity in the data at this stage). Then, $R^s_{Z_1}$ is generated by drawing from a Bernoulli distribution with probability $\hat{p}^s_{Z_1}$ (a similar procedure to generating binary $Y^s$). Finally, the $i^{\text{th}}$ value of $Z^s_1$ is set to missing whenever $R^s_{Z_1i}=0$. The same procedure is implemented to generate synthetic missingness in $Y^s$, however here, the missingness model used to predict $\hat{p}^s_{Z_1}$ differs between the two IPW methods. Then, the merged data $\{\textbf{X}^s,A^s,\tilde{Z}^s_1,Z^s_2,\tilde{Y}^s\}$ form the generated synthetic data set, where $\tilde{Z}^s_1$ and $\tilde{Y}^s$ represent the versions of these synthetic variables that also include missing values. (If completely-observed synthetic data are preferred, the last step of imposing synthetic missingness can be omitted and the complete version of the synthetic data $\{\textbf{X}^s,A^s,Z^s_1,Z^s_2,Y^s\}$ can instead be returned.)

\subsubsection{MI}
\label{subsubsec:nonmonotonemi}

With an MI strategy to account for missingness in the real data when generating a completely synthetic data set, the first step is to apply MI to the set of real data $\{ \mathbf{X}^r,A^r,Z^r_1,Z^r_2,Y^r \}$, thus creating $m$ complete data sets. In this work, $m$ is set to be equal to the largest proportion of missingness in the real data \citep{White2011}; e.g., if $Z^r_1$ was missing $24\%$ and $Y^r$ was missing $26\%$, $m=26$. The maximum number of iterations is set to be 10. Generating $\mathbf{X}^s,A^s$ follows the same procedure as for CC and IPW, since $\mathbf{X}^r$ and $A^r$ were completely observed (i.e., generating $\mathbf{X}^s,A^s$ does not involve the imputed data). This procedure also ensures that all real data samples are included at each data generation stage, since missing values are imputed at the beginning.

Next, the post-randomization variables are generated in the sequential generation framework. The general strategy implemented in this work to generate these variables using $m$ imputed data sets is to fit regression models to each imputed data set, generate a vector of synthetic data for the given post-randomization variable, and then sample from the $m$ vectors of generated data for the given variable to result in a final vector of synthetic observations. Other strategies could be applied instead, such as harnessing ``pooled'' regression models, however this raises additional complications; further discussion can be found in the Supplementary Materials. We now present the generation procedure with MI, using the sampling strategy.

Starting with the first post-randomization variable, a model is fit regressing $Z^r_1$ on $\mathbf{X}^r,A^r$ using each of the $m$ imputed data sets, thus resulting in $m$ sets of parameter estimates, $\{\hat{\boldsymbol{\beta}}_1,\hat{\alpha}_1\}, \{\hat{\boldsymbol{\beta}}_2,\hat{\alpha}_2\},...,\{\hat{\boldsymbol{\beta}}_m,\hat{\alpha}_m\}$, and $m$ sets of model residuals, $r_1, r_2,...,r_m$ (we suppress the $Z^r_1$ subscript for notational brevity). Next, $m$ sets of $Z^s_1$ are generated via the usual process: 
\begin{equation*}
    \begin{split}
    &\hat{Z}^{s_1}_1=\hat{\boldsymbol{\beta}}_1\mathbf{X}^s+\hat{\alpha}_1A^s \hspace{1.5cm}\hat{Z}^{s_2}_1=\hat{\boldsymbol{\beta}}_2\mathbf{X}^s+\hat{\alpha}_2A^s \hspace{1.5cm} ... \hspace{1cm} \hat{Z}^{s_m}_1=\hat{\boldsymbol{\beta}}_m\mathbf{X}^s+\hat{\alpha}_mA^s \\
    &Z^{s_1}_1\sim \Bigl\{ \left( \hat{Z}^{s_1}_1+r_1 \right) \geq 0 \Bigr\} \hspace{0.75cm} Z^{s_2}_1\sim \Bigl\{ \left( \hat{Z}^{s_2}_1+r_2 \right) \geq 0\Bigr\} \hspace{0.78cm} ... \hspace{0.98cm} Z^{s_m}_m\sim \Bigl\{ \left( \hat{Z}^{s_m}_1+r_m \right) \geq 0 \Bigr\}
    \end{split}
\end{equation*}
where the superscript $s_j$ indicates the vector of synthetic data for the $j^{\text{th}}$ imputed data set, $j=1,...,m$. Then, the final vector of $Z^s_1$ observations is generated via
\begin{gather*}
Z^s_1 = \left( Z^s_{11},Z^s_{12},...,Z^s_{1n} \right)\nonumber\\[1ex]
\text{where } Z^s_{1i} \sim \bigl\{ Z^{s_1}_{1i},Z^{s_2}_{1i},...,Z^{s_m}_{1i} \bigr\} \text{ for } i=1,...,n.\\
\end{gather*}
In other words, the $i^{\text{th}}$ observation of $Z^s_1$ is randomly sampled from the collection of the $i^{\text{th}}$ observations of $Z^s_1$ across all $m$ imputed data sets. This procedure (fit a regression model to all $m$ data sets, generate $m$ sets of the synthetic variable, then sample from the $m$ sets to generate the final version of the synthetic variable) is repeated to generate the subsequent post-randomization variables, $Z^s_2$ and $Y^s$. Note that the generation of $Z^s_2$ involved the imputed data despite $Z^r_2$ being completely observed, since $Z^r_1$ is included as a predictor in the generative regression model and originally had missing values.

This MI procedure results in completely observed synthetic data. Thus, in order to generate synthetic data with missing values similar to those of the real data, the same strategy from the IPW: Indicator Method approach is employed to generate synthetic missingness. The choice to implement the indicator method here, and not the method that forced monotone missingness, was based on preliminary investigations into the performance of both IPW frameworks under non-monotone missing. Hence, the MI strategy presented here does not entirely side-step the assumptions required to model the probability of being observed in the real data, as this model is needed to predict the probability of being observed in the synthetic data. As in both IPW methods, the merged data $\{\textbf{X}^s,A^s,\tilde{Z}^s_1,Z^s_2,\tilde{Y}^s\}$ result in a final synthetic data set that includes missing values, but it is also possible to return the completely-observed synthetic data $\{\textbf{X}^s,A^s,Z^s_1,Z^s_2,Y^s\}$ if desired.

\subsection{Methods for the Special Case: Monotone Missingness}
\label{subsec:specialcasemonotonemissingness}

In the special case of monotone missingness, missingness was imposed in $Z^r_1$, $Z^r_2$, and $Y^r$ such that missingness in $Z^r_1$ implied missingness in $Z^r_2$, and missingness in $Z^r_2$ implied missingness in $Y^r$. In the following subsections, we present the same modelling techniques and data generation frameworks as before, but now in the case of monotone missingness. In particular, we compare CC: All Stage, CC: By Stage, IPW, and MI. Only one version of IPW is considered here.

\subsubsection{CC: All Stage}
\label{subsubsec:monotoneccallstage}

The CC: All Stage method under monotone missingness is the same procedure as the CC: All Stage method under non-monotone missingness, as the missing values are removed during data pre-processing and hence ignored irregardless of the missingness pattern for the remainder of the data generation procedure. As before, this method cannot generate synthetic missingness.

\subsubsection{CC: By Stage}
\label{subsubsec:monotoneccbystage}

This framework also mimics that of CC: By Stage under non-monotone missingness. At baseline and treatment randomization, all individuals are included, and at subsequent stages, the available real data sample for fitting generative models becomes smaller (and are nested). At the final stage to generate the outcome, the available data are the same as the real complete cases that are used in all sequential data generation stages in CC: All Stage. This framework cannot generate synthetic missingness. 

\subsubsection{IPW}
\label{subsubsec:monotoneipw}

The IPW method for monotone missingness begins the same as that of non-monotone missingness: complete $\textbf{X}^s$ is generated using an R-vine copula and $A^s$ is generated by sampling from the pre-defined probability distribution. Then, the post-randomization variables are generated starting with $Z^s_1$. For $Z^s_1$, the procedure follows that of IPW for non-monotone missingness since in both cases, $\textbf{X}^r$ and $A^r$ are completely observed.

Next, when generating $Z^r_2$, missing values must be taken into account. Recall that with a monotone missingness pattern, if $Z^r_1$ is missing, then $Z^r_2$ must also be missing by definition (i.e., $R^r_{Z_1}=0$ implies $R^r_{Z_2}=0$). First, the probability of being observed is estimated for $Z^r_2$, denoted $\hat{p}^r_{Z_2}$, by further conditioning on $Z^r_1$ being observed:
\begin{equation*}
\begin{split}
    \hat{p}^r_{Z_2}=\text{Pr}\left( R^r_{Z_2}=1|\textbf{ X}^r,A^r,Z^r_1 \right) = & \text{Pr}\left( R^r_{Z_2}=1|\textbf{ X}^r,A^r,Z^r_1,R^r_{Z_1}=1 \right)\times\text{Pr}\left( R^r_{Z_1}=1|\textbf{ X}^r,A^r \right) \\
    & + \text{Pr}\left( R^r_{Z_2}=1|\textbf{ X}^r,A^r,Z^r_1,R^r_{Z_1}=0 \right)\times\text{Pr}\left( R^r_{Z_1}=0|\textbf{ X}^r,A^r \right).
\end{split}
\end{equation*}
Again, $\text{Pr}( R^r_{Z_1}=1|\textbf{ X}^r,A^r)=\text{Pr}( R^r_{Z_1}=1|\textbf{ X}^r,A^r,Z^r_1)$ and $\text{Pr}( R^r_{Z_1}=0|\textbf{ X}^r,A^r)=\text{Pr}( R^r_{Z_1}=0|\textbf{ X}^r,A^r,Z^r_1)$ under the MAR working assumption. But $\text{Pr}( R^r_{Z_2}=1|\textbf{ X}^r,A^r,Z^r_1,R^r_{Z_1}=0 )=0$ by definition of monotone missingness. Thus,
\begin{equation*}
    \begin{split}
        \hat{p}^r_{Z_2}&=\text{Pr}\left( R^r_{Z_2}=1|\textbf{ X}^r,A^r,Z^r_1,R^r_{Z_1}=1 \right)\times\text{Pr}\left( R^r_{Z_1}=1|\textbf{ X}^r,A^r \right) \\
        &=:\hat{p}^r_{Z_2|R^r_{Z_1}=1}\times\hat{p}^r_{Z_1}.
    \end{split}
\end{equation*}
The first term on the right-hand side, $\hat{p}^r_{Z_2|R^r_{Z_1}=1}$, can be estimated directly from the data, and the second term, $\hat{p}^r_{Z_1}$, was already estimated in the first step when generating $Z^s_1$. For the remaining individuals with missing $Z^r_1$, their probability of being observed at $Z^r_2$ is set to zero, again due to the definition of monotone missingness. Unlike in the non-monotone missingness case, conditioning on observed $Z^r_1$ now leads to a helpful simplification in order to estimate $\hat{p}^r_{Z_2}$. As before, the next steps involve fitting a weighted regression model, regressing $Z^r_2$ on $\textbf{X}^r$, $A^r$, and $Z^r_1$ with weights $1/\hat{p}^r_{Z_2}$ using data from participants with observed $Z^r_1$ and $Z^r_2$. Then, $\hat{Z}^s_2$ is predicted for fully-observed $( \textbf{X}^s,A^s,Z^s_1 )$ using the fitted weighted model, and finally $Z^s_2$ is generated by sampling from the set of admissible values.

To generate $Y^s$, the same four steps are repeated: estimate the probability of being observed $\hat{p}^r_Y$, fit a weighted regression model, predict, and sample. For the first step, the similar logic based on the monotonic missingness is used to simplify the estimation of $\hat{p}^r_Y$ by conditioning on $Z^r_2$ being observed. (Note that this differs from the decomposition that is used in the IPW: Force Monotonicity framework under non-monotone missingness, where the model involves conditioning on $Z^r_1$ being observed instead of $Z^r_2$.) Details of the derivation can be found in the Supplementary Materials. The estimation of $\hat{p}^r_Y$ is thus equivalent to
\begin{equation*}
    \hat{p}^r_Y=\text{Pr} \left( R^r_{Y}=1|\textbf{ X}^r,A^r,Z^r_1,Z^r_2,R^r_{Z_2}=1 \right)\times\text{Pr}\left( R^r_{Z_2}=1|\textbf{ X}^r,A^r,Z^r_1 \right).
\end{equation*}
Again, $\text{Pr}( R^r_{Z_2}=1|\textbf{ X}^r,A^r,Z^r_1 )=\text{Pr}( R^r_{Z_2}=1|\textbf{ X}^r,A^r,Z^r_1,Z^r_2 )$ since here, $Z^r_2$ cannot influence the probability of being observed at $Z^r_2$ under the assumption of MAR. Further, under monotone missingness, $$\text{Pr} \left( R^r_{Y}=1|\textbf{ X}^r,A^r,Z^r_1,Z^r_2,R^r_{Z_2}=1 \right)=\text{Pr} \left( R^r_{Y}=1|\textbf{ X}^r,A^r,Z^r_1,Z^r_2,R^r_{Z_2}=1,R^r_{Z_1}=1 \right).$$ This can also be shown mathematically by additionally conditioning on $Z^r_1$ being observed (refer to Supplementary Materials for details). 
Thus, we have
\begin{equation*}
    \begin{split}
        \hat{p}^r_Y &= \text{Pr}\left( R^r_{Y}=1|\textbf{X}^r,A^r,Z^r_1,Z^r_2,R^r_{Z_2}=1,R^r_{Z_1}=1 \right)\times\text{Pr}\left( R^r_{Z_2}=1|\textbf{ X}^r,A^r,Z^r_1 \right) \\
        &=: \hat{p}^r_{Y|R^r_{Z_2}=1,R^r_{Z_1}=1}\times\hat{p}^r_{Z_2}.
    \end{split}
\end{equation*}
The first term on the right-hand side, $\hat{p}^r_{Y|R^r_{Z_2}=1,R^r_{Z_1}=1}$, can be estimated directly from the data, and the second term, $\hat{p}^r_{Z_2}$, was previously estimated in the first step when generating $Z^s_2$. For the remaining individuals with missing $Z^r_1$ and missing $Z^r_2$, the probability of being observed at $Y^r$ is set to be zero, again due to the definition of monotone missingness. Then, a weighted (logistic) regression model is fit to those with available data at this final stage (i.e., the complete cases) to generate the (binary) outcome, same as before. Finally, the same procedure for generating synthetic missingness as described for the IPW methods under non-monotone missingness is implemented, resulting in $\tilde{Z}^s_1,\tilde{Z}^s_2,\tilde{Y}^s$. Then, the final generated synthetic data set with missing values is the merged data set $\{\textbf{X}^s,A^s,\tilde{Z}^s_1,\tilde{Z}^s_2,\tilde{Y}^s\}$. Of course, the completely-observed synthetic data set $\{\textbf{X}^s,A^s,Z^s_1,Z^s_2,Y^s\}$ could be returned as well.

\subsubsection{MI}
\label{subsubsecmi}

The data generation framework incorporating MI to handle missing values is the same under monotone missingness and non-monotone missingness. Here, the missingness models used to generate synthetic missingness are as described for IPW under monotone missingness.

\section{Experimental Methods}
\label{sec:experimentalmethods}

Next, the real data set used in the empirical comparisons of the sequential data generation frameworks is described, as well as the multiple missingness scenarios considered in simulation and metrics used to quantify synthetic data generation performance.

\subsection{Data Description}
\label{subsec:datadescription}

Publicly-available data are from the AIDS Clinical Trials Group Study 175 (ACTG 175), which studied the effect of four HIV treatment regimens on a composite outcome defined as either having a drop in CD4 cell count of at least 50\%, a progression of HIV to AIDS, or death \citep{Hammer1996}. The baseline treatment comparator was zidovudine only, and the three other treatments of interest were zidovudine and didanozine, zidovudine and zalcitabine, or diadnozine only. The trial included people living with HIV who had CD4 cell counts between 200 and 500 per cubic millimeter, and it enrolled 2139 patients. At baseline, several variables of mixed types were measured, including age, weight, sex, race, hemophilia status, homosexual identity (participant-reported), injection drug use status, Karnofsky score (a health score that measures a participant's functional status, with a maximum score of 100), history of prior antiretroviral therapy (ART), symptomatic HIV infection status, and CD4 cell count. Participants were followed for a median of 33 months with follow-up visits at weeks 2, 4, 8, and then every 12 weeks thereafter. However, in the available data set, only data at week 20 and week 96 were included. Hence, for the data used in our empirical experiments, $K=2$. In this work, we consider only the CD4 cell count measurements at week 20 and week 96 for simplicity, though other variables were measured as well. The real data are thus comprised of $\textbf{X}^r$ being the baseline variables listed above, $A^r$ comprised of one of four possible treatment arms per participant, $Z^r_1$ being CD4 count at week 20, $Z^r_2$ being CD4 count at week 96, and $Y^r$ the composite binary outcome.

Though the outcome was fully-observed in the real data, CD4 count at week 96 exhibited approximately 37\% missingness. For the purposes of our empirical simulations, we imposed our own missingness mechanisms on the real data. This was done by first excluding all rows in the original RCT data that had a missing value for CD4 count at week 96; this led to a total sample size of 1342. Then, using the sample of 1342 participants with completely-observed data, additional missingness was imposed on the real data to create a real data set with missing values where the missingness mechanism was known. More details regarding the models used to impose missingness mechanisms on the real data can be found in the Supplementary Materials. Thus, we aim to account for this (non-negligible) missingness as well as produce synthetic data with missingness similar to that of the real data set. Additionally, note that the sample sizes of the real data subsets available at each sequential data generation step differed depending on the simulated missingness scenario (as described in \autoref{subsec:missingnessscenarios} and \autoref{tbl:simulationscenarios}); these sample sizes can be found in the Supplementary Materials.

\subsection{Missingness Scenarios}
\label{subsec:missingnessscenarios}

Twelve scenarios with varying missingness mechanisms were explored to investigate which methods may perform better under different circumstances. In particular, three main factors were considered, for both non-monotone and monotone missingness: the missingness mechanism (MCAR or MAR), the proportion of missingness (10\%, 25\%, or 50\%), and the strength of the missingness mechanism (strong or weak). The proportion of missingness refers to the proportion missing for a particular variable, not the complement of the proportion of complete cases. The strength of the missingness mechanism can be thought of as the degree of impact that certain variables may have on the probability of being missing. Six different combinations of mechanism, proportion, and strength were defined, and these six combinations were investigated for both non-monotone and monotone missingness (thus resulting in twelve distinct simulation scenarios). \autoref{tbl:simulationscenarios} shows all scenarios that were explored. The primary scenario was set to be one that we believe to realistically mimic what often occurs in real life RCTs; we chose non-monotone missingness in this primary scenario as it is the more general case. The other scenarios (2-6) were defined by changing one factor at a time (MCAR/MAR, 10\%/25\%/50\% missing, strong/weak mechanism), with the last scenario defined by changing two factors. The results presented in \autoref{sec:results} largely focus on scenarios 1A and 1B; results from the remaining scenarios can be found in the Supplementary Materials. In the simulation scenarios involving monotone missingness (1B - 6B), the proportion of missingness was set to be approximately the same for all variables with missing values. We also experimented with varying the proportion of missingness across variables, however this did not reveal new insights and so these scenarios are omitted. 

\begin{table}[ht!]
\begin{center}
 \caption{All missingness scenarios investigated via simulation. The factors that differ from those in 1A and 1B are presented in bold text. Strength refers to the impact that observed variables have on the probability of being missing; strong means the observed data have a large effect on the missingness probability whereas weak means the observed data do not have much of an impact.}
 \label{tbl:simulationscenarios}
\begin{tabular}{||c c c c c ||} 
 \hline
 \textbf{Scenario} & \textbf{Pattern} & \textbf{Mechanism} & \textbf{Proportion} & \textbf{Strength} \\
 \hline\hline
 \hline
 1A (Primary) & Non-Monotone & MAR & 25\% & Strong \\
 \hline
 1B & Monotone & MAR & 25\% & Strong \\
 \hline\hline
 2A & Non-Monotone & \textbf{MCAR} & 25\% & Strong \\ 
 \hline
 2B & Monotone & \textbf{MCAR} & 25\% & Strong \\ 
 \hline
 3A & Non-Monotone & MAR & \textbf{10\%} & Strong \\ 
 \hline
 3B & Monotone & MAR & \textbf{10\%} & Strong \\ 
 \hline
 4A & Non-Monotone & MAR & \textbf{50\%} & Strong \\ 
 \hline
 4B & Monotone & MAR & \textbf{50\%} & Strong \\ 
 \hline
 5A & Non-Monotone & MAR & 25\% & \textbf{Weak} \\ 
 \hline
 5B & Monotone & MAR & 25\% & \textbf{Weak} \\ 
 \hline
 6A & Non-Monotone & MAR & \textbf{50\%} & \textbf{Weak} \\ 
 \hline
 6B & Monotone & MAR & \textbf{50\%} & \textbf{Weak} \\ 
 \hline\hline
\end{tabular}
\end{center}
\end{table}

\subsection{Metrics and Evaluation}
\label{subsec:metrics}

To ensure data generation performance results were robust to the type of metric employed, several metrics were used to quantify how well each framework generated synthetic data with univariate and multivariate distributions that were close to those in the real data. 

\subsubsection{Distributional Similarity}
\label{subsubsec:distributionalsimilaritymetrics}

First, similarity metrics that captured the ``closeness'' between univariate and bivariate distributions were calculated, comparing synthetic to real data, and the choice of these similarity metrics was based on common practice in the ML data generation literature \citep{Patki2016,sdmetrics_misc}. For univariate continuous comparisons, the complement of the Kolmogorov Smirnov (KS) statistic was used. The KS statistic captures the largest distance between two cumulative distribution functions (CDFs), and hence the complement of the KS statistic represents the closeness of two CDFs. To demonstrate data generation performance, where the goal is for the synthetic distribution to be close to the real distribution, a higher value of the complement of the KS statistic implies better generative performance. The complement of the KS statistic is defined as $1 - \sup_{x \subset X}\left|F^{\text{r}}_{n}(x)-F^{\text{s}}_{n}(x)\right|$ for a continuous variable $X$ with realization $x$, where $F^{\text{r}}_{n}(x)$ is the real data CDF and $F^{\text{s}}_{n}(x)$ is the synthetic data CDF. For univariate discrete comparisons, the complement of the total variation distance (TVD) was calculated. The TVD is defined as $1 - \frac{1}{2}\sum_{a \subset A}\left|\pi^{\text{r}}_{a}-\pi^{\text{s}}_{a}\right|$ for a discrete variable $A$ with stratum $a$, where $\pi^{\text{r}}_{a}$ is the proportion of stratum $a$ in the real data and $\pi^{\text{s}}_{a}$ is the proportion of stratum $a$ in the synthetic data. A higher value of the complement of the TVD implies better generative performance. These univariate comparisons were first plotted for all variables comparing observed real data to observed synthetic data, and then were plotted only for variables with missingness, comparing both observed real to observed synthetic data and complete real to complete synthetic data, as the focus of this work was to determine how well the real data distribution was captured when missingness is an important feature of the data to be mimicked. Note that for the two CC frameworks, the complete and observed synthetic data were the same because these frameworks do not generate synthetic missing values.

For bivariate continuous comparisons, Spearman correlation was calculated for a given pair of continuous variables in the synthetic and real data, and the complement of the normalized difference between the two correlations was plotted. Again, a higher value indicated better performance. This similarity metric is defined as $1 - 1/2\left|\rho^{\text{r}}_{XY}-\rho^{\text{s}}_{XY}\right|$ for continuous variables $X$ and $Y$, where $\rho^{\text{r}}_{XY}$ is the Spearman correlation in the real data and $\rho^{\text{s}}_{XY}$ is the Spearman correlation in the synthetic data. In prior work, there was little difference in performance results when calculating the Pearson correlation as opposed to the Spearman correlation \citep{Petrakos2025}, so we report only the Spearman correlation here. For bivariate discrete comparisons, a similarity score representing differences in two-way contingency tables was used. Note that this score was also calculated for pairs of variables where one was continuous and one was discrete; in this case, continuous variables were discretized by quartiles. This contingency similarity score is defined as $1 - 1/2\sum_{a \subset A}\sum_{b \subset B}\left|\pi^{\text{r}}_{ab} - \pi^{\text{s}}_{ab}\right|$ for some discrete variables $A$ and $B$ with strata $a$ and $b$, respectively, where $\pi^{\text{r}}_{ab}$ represents the proportion of observations in the real data contained in both strata $a$ and $b$, and $\pi^{\text{s}}_{ab}$ represents that in the synthetic data. A higher value indicates better generative performance.

To capture higher-order multivariate distributional comparisons, we also investigated principal component analysis (PCA) plots from a simulation run selected at random, as was done in \cite{Wang2023}. To do so, PCA was performed on the real and synthetic data sets, and the first two principal components were plotted with the amount of variation in the real data captured by each principal component also displayed. If the synthetic data are similar to the real data, the plots between the real and synthetic data should look very similar and the proportions of variation should also be similar.

\subsubsection{ML Efficacy}
\label{subsubsec:mlefficacymetrics}

Another way to quantify data generation performance that is often presented in the data generation literature is to use what are called ML efficacy metrics \citep{Xu2019,Zhou2020}. In this approach, an ML classifier is trained on real data, validated on the real data test set, and metrics such as accuracy ($\frac{\text{True Positive + True Negative}}{\text{Total}}$), precision ($\frac{\text{True Positive}}{\text{True Positive + False Positive}}$), recall ($\frac{\text{True Positive}}{\text{True Positive + False Negative}}$), and F-1 score ($2\cdot\frac{\text{Precision}\cdot\text{Recall}}{\text{Precision} + \text{Recall}}$) are computed, where a ``positive'' generally refers to the (binary) outcome event occurring. Another ML classifier is then trained on synthetic data, validated on real data, and the same metrics are computed. Finally, the metric values for the two classifiers are compared such that if the values are close (e.g., the accuracy of the classifier trained on real data is close to the accuracy of the classifier trained on synthetic data), then it is concluded that the synthetic data are similar to the real data. We have harnessed this method to evaluate whether the ML classifier could detect a real data observation versus a synthetic data observation. To do so, the real and synthetic data sets were merged and an additional column was added that indicated from which data set the row/observation originated -- real or synthetic. Then, only one classifier was trained and tested, and the classification task was to identify the additional data label of real or synthetic. Finally, the same metrics were computed (accuracy, recall, precision, and F-1 score). As the goal was to generate synthetic data that were close to real data, better generative performance is indicated by the classifier not being able to distinguish between real data rows and synthetic data rows, and hence by low metric values. However, in order to remain consistent by reporting metrics such that higher values indicate better performance, we plotted and reported on the complement of these ML efficacy metrics. Moreover, to ensure results were robust to the type of classifier, we calculated these results using two different classifiers -- Extreme Gradient Boosting, or XGBoost, and $k$ Nearest Neighbors, or KNN where $k=5$. XGBoost was able to handle missing values directly, whereas KNN involved an extra step of imputing missing values in the real data using a KNN-based imputation scheme \citep{Hastie2001}.

\subsubsection{Trial Inference}
\label{subsubsec:trialinferencemetrics}

Trial inference results between synthetic and real data were also compared to determine generative performance. We emphasize that we do \textbf{\textit{not}} suggest that synthetic data should be used for inference. Rather, our purpose is only to quantify generative performance using yet another metric, and one that is pertinent to RCTs, as synthetic data generation of RCTs may be used, for example, to simulate sample size or power and thus accurate reflection of inference is valuable. Since our context involved RCTs, it was important to consider that one of the main focuses of RCTs is to determine the effect of treatment on the outcome. Hence, another way to investigate how well the data generation frameworks captured the real data distribution is to compare the estimated treatment effect from the real data and the synthetic data. Ideally, the estimated effects should be similar. 

To do so, we first dichotomized treatment by keeping zidovudine only as the baseline comparator and grouping together the three other treatment arms to form the other stratum. This was done for the sake of simplicity in capturing differences in trial inference results between the real and synthetic data, and was implemented only after synthetic data were generated. This comparison was performed utilizing a) complete synthetic data and complete real data, as well as b) observed synthetic data and observed real data, to ensure both the complete and observed data distributions were well captured. The effect of treatment on outcome was estimated by fitting a logistic regression model and reporting the odds ratio (OR) and associated 95\% confidence interval (CI). This resulted in 1000 OR and CI estimates across all simulation runs, which were plotted. Good data generation performance was evidenced by synthetic data OR estimates being close to that of the real data and synthetic 95\% CIs overlapping with the 95\% CI estimated from the real data. 

After missingness was imposed on the real data, each approach was employed to generate synthetic data, and the closeness of the synthetic data to the real was quantified using these metrics. To capture the variation in the data generation procedure, 1000 simulation runs were implemented. One simulation run involved fitting models to the real data, generating one synthetic data set per candidate approach, and then evaluating data generation performance. All simulations were implemented in R version 4.3.1 using a machine with 16 GB RAM, and simulations were parallelized across 10 cores. Computing time for generating a synthetic data set and calculating metrics was also recorded to determine computational load.

\section{Results}
\label{sec:results}

Unless otherwise specified, results and visualizations presented here are from scenarios 1A and 1B, as outlined in \autoref{subsec:missingnessscenarios} and \autoref{tbl:simulationscenarios}. 

\subsection{Distributional Similarity Metrics: Univariate and Bivariate}
\label{subsec:results_simmetrics}

\begin{figure}[h!]
    \centering
    \includegraphics[width=\linewidth]{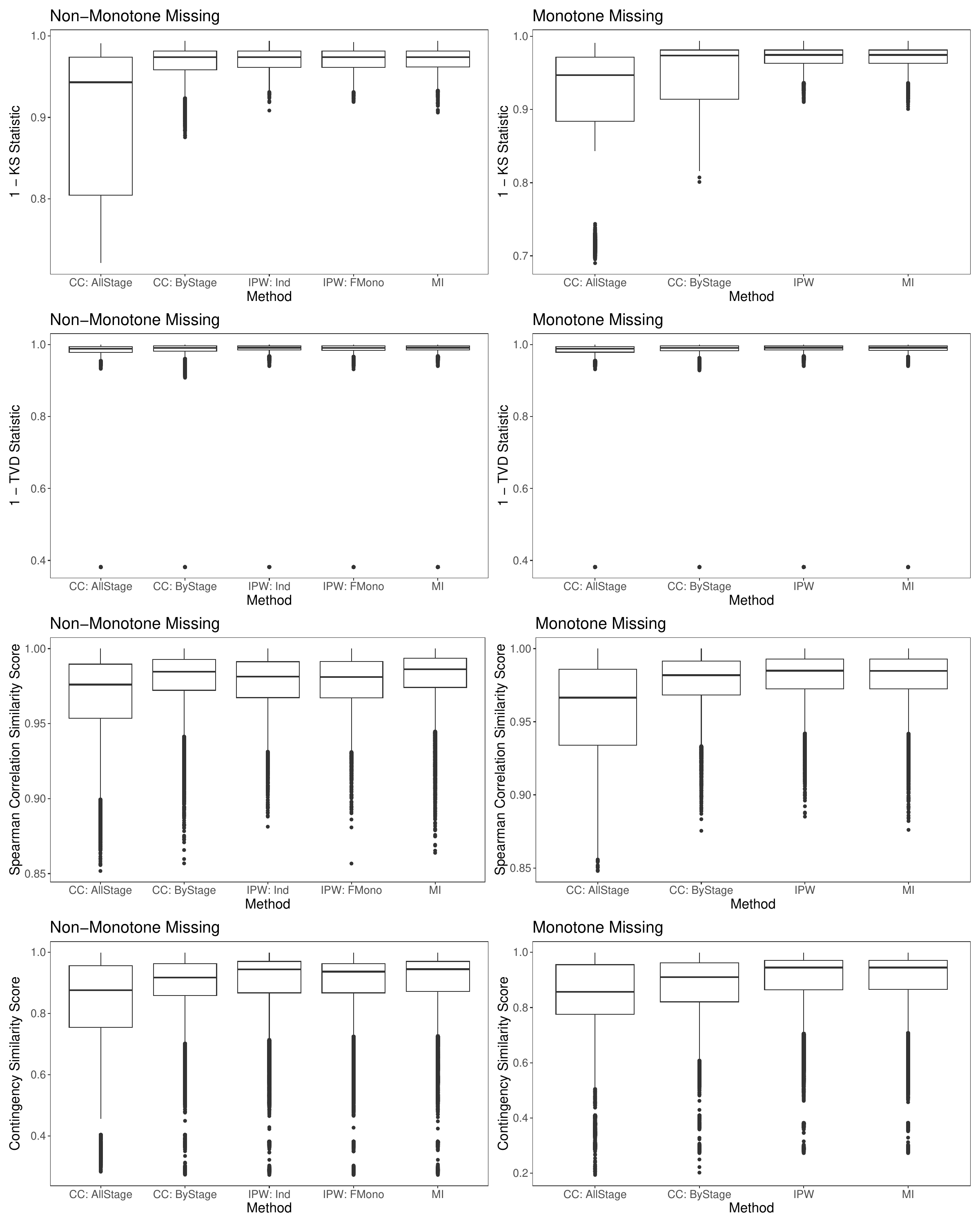}
    \caption{Plot of similarity metrics for capturing univariate continuous (first row), univariate discrete (second row), bivariate continuous (third row), and bivariate discrete (fourth row) distributions, across all variables and all 1000 simulation runs for Scenarios 1A (left column) and 1B (right column). These metrics compare the observed synthetic data to the observed real data (note: observed synthetic data is the same as complete synthetic data for the two CC methods because these do not generate synthetic missingness).}
    \label{fig:univarbivarcontdisc}
\end{figure}

We first present results for the univariate and bivariate similarity metrics, considering all variables including those at baseline which did not have missing values (\autoref{fig:univarbivarcontdisc}). In all scenarios, CC: All Stage demonstrated the worst generative performance, with the most stark contrast in performance evident in capturing univariate continuous distributions (though, all frameworks performed similarly well at capturing univariate discrete distributions, under both missingness patterns). IPW methods and MI performed similarly well in capturing univariate and bivariate relationships, no matter the type of variable(s) or missingness pattern. Under non-monotone missingness, both IPW methods performed slightly worse on average than MI, however the box plot representing MI results had a longer tail than those for both IPW methods, thus indicating that it was more likely for synthetic data to be distributionally less similar than the real data when using the MI framework rather than either IPW framework. CC: By Stage generally showed an improvement over CC: All Stage but the long tails in the box plots particularly for univariate continuous distributions suggest that there were several instances where the synthetic distributions were far from those of the real data. Hence, simply removing everyone with missing data as a pre-processing step was a hindrance. Indeed, synthetic data were closer to the real when as much information as possible was included at each sequential step, even at baseline when there was no missingness. Additionally, all frameworks had an instance of poorly capturing the discrete distribution of Karnofsky score, as shown by the points close to 0.4 in the second row of \autoref{fig:univarbivarcontdisc}. This was not surprising since CC: By Stage and all IPW and MI methods generated baseline data in the same way. Though CC: All Stage used only complete cases, it was possible for this framework to perform similarly to the other frameworks when generating baseline covariates, likely when the variable did not play a role in the missingness mechanism (and indeed, Karnofsky score was not included in the missingness model imposed on the real data when simulating the known missingness mechanism).

\FloatBarrier

For the other scenarios, the same general pattern was observed. CC: All Stage performed the worst whereas IPW and MI methods performed the best (refer to the Supplementary Materials for visualizations). In Scenario 2 (MCAR), there was less of a difference in performance but CC: All Stage still showed the worst performance. In Scenario 3 (10\% missing), there was a stark contrast in generative performance between CC: All Stage and the rest of the frameworks, even despite the lower proportion of missingness. With a higher missingness proportion, like in Scenario 4 (50\% missing), the performance of CC: By Stage suffered especially in capturing the univariate continuous distributions and bivariate discrete distributions. With a weak missingness mechanism such as in Scenario 5 (25\% missing and weak) and Scenario 6 (50\% missing and weak), IPW and MI frameworks demonstrated best performance, particularly in replicating the univariate and bivariate continuous distributions and the bivariate discrete distributions.

\subsection{Multivariate Relationships: PCA Plots}
\label{subsec:results_pca}

\begin{figure}[h!]
    \centering
    \includegraphics[width=\linewidth]{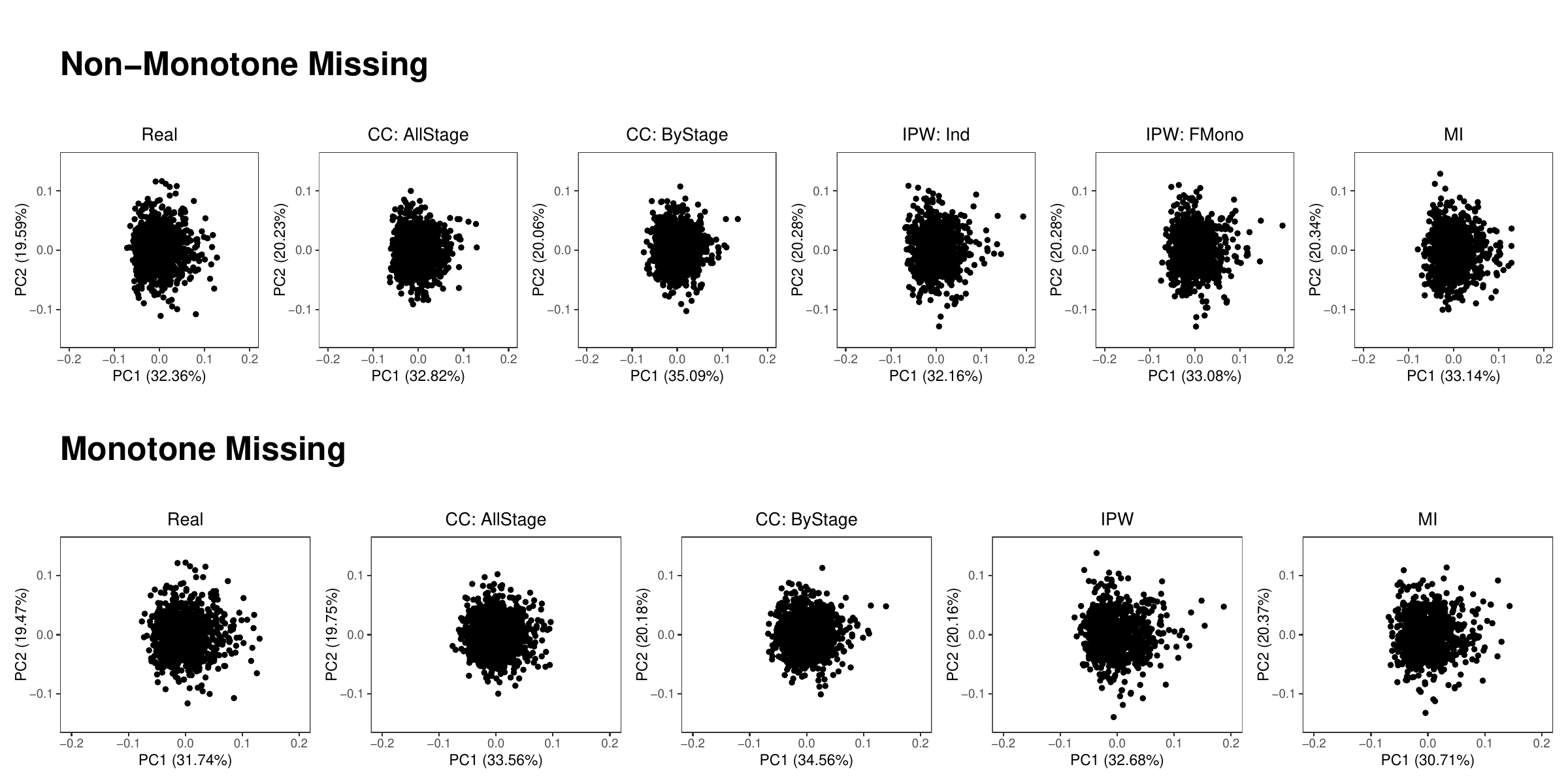}
    \caption{PCA plots of the first two principal components for the real data and for each synthetic data generation framework, taken from one simulation run chosen at random under Scenarios 1A (top row) and 1B (bottom row). The horizontal axis represents the first principal component and the vertical axis represents the second principal component; the proportion of variation from the data captured by the principal component is indicated along each axis.}
    \label{fig:pca}
\end{figure}

In the PCA plots from one simulation run chosen at random (\autoref{fig:pca}), both IPW methods and MI performed better at capturing multivariate (continuous) distributions than either CC method under non-monotone missingness, as evidenced by the spread of points being visually more similar to that of the real. All frameworks resulted in the first and second principal components capturing a similar proportion of variation in the data as compared to the principal components in the real data, though IPW: Indicator Method had the closest proportion to that of the real for the first principal component. CC: By Stage had the closest proportion to that of the real for the second principal component. Under monotone missingness, MI and IPW displayed a relationship between the first two principal components that was the most visually similar to that of the real data. The first principal component in the MI-generated synthetic data captured a proportion of variation that was most similar to that of the real data. CC: All Stage had a proportion most similar to that of the real data when considering the second principal component analysis. For the other scenarios, the general pattern observed was that both CC methods resulted in tighter PCA plots whereas points in the PCA plots under IPW and MI frameworks were more spread out and better reflected the shape of the real data PCA plots.

\FloatBarrier

\subsection{ML Efficacy Metrics}
\label{subsec:results_MLefficacy}

\begin{figure}[h!]
    \centering
    \includegraphics[width=\linewidth]{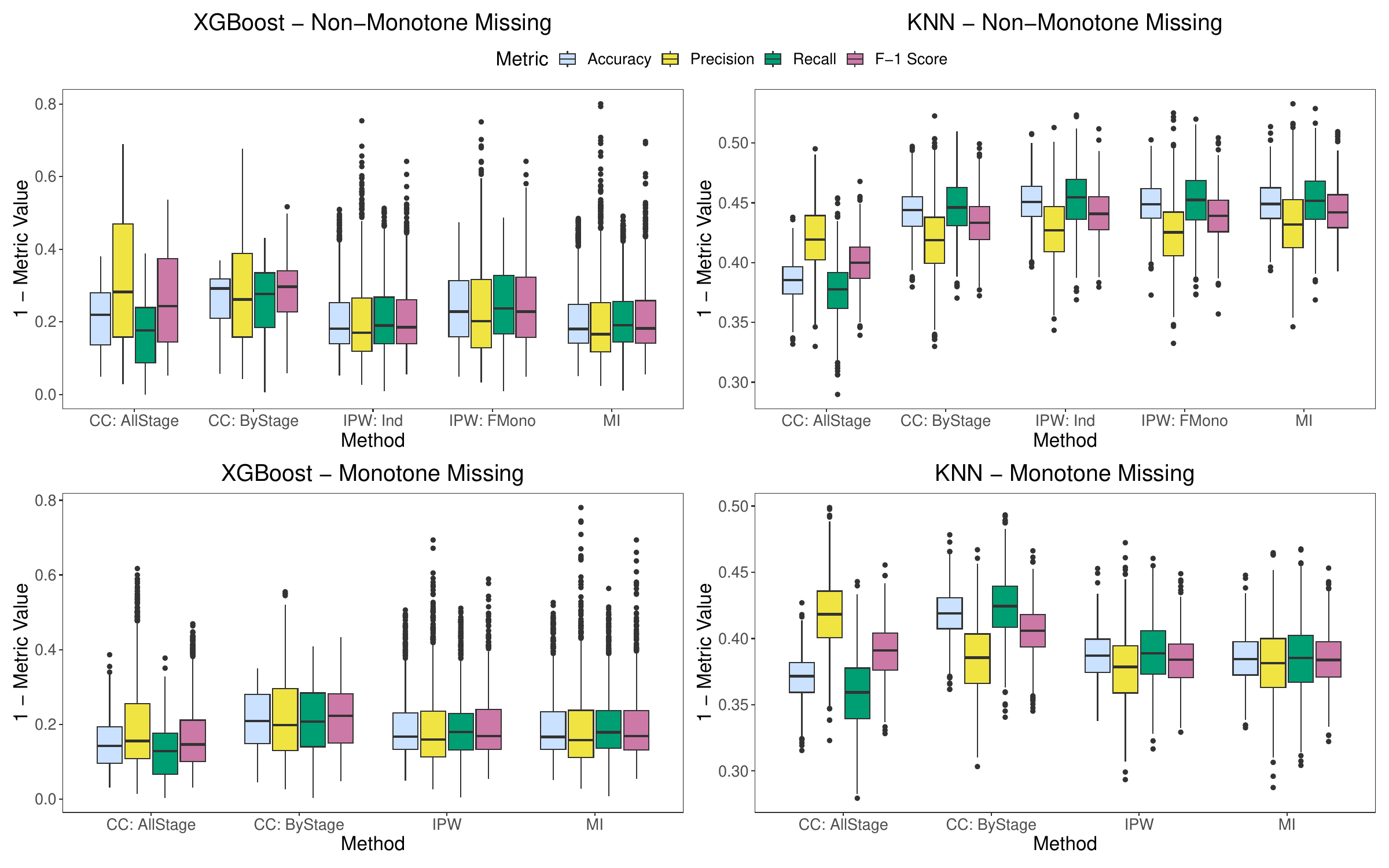}
    \caption{Plot of ML efficacy metrics (accuracy, precision, recall, and F-1 score) for both the XGBoost classifier (left column) and KNN classifier (right column) across all frameworks and 1000 simulation runs, for Scenarios 1A (top row) and 1B (bottom row). The classification task was to correctly identify whether a row was from the real data or the synthetic data.}
    \label{fig:MLefficacy}
\end{figure}

Next we consider ML efficacy metrics, as shown in \autoref{fig:MLefficacy}. Under non-monotone missingness, and defining a KNN classifier, both IPW frameworks and the MI framework performed the best across all metrics. However for the XGBoost classifier, CC: All Stage outperformed the other frameworks when considering precision and F-1 score. In contrast, CC: All Stage performed the worst when considering accuracy and recall. Again, CC: By Stage showed an improvement over CC: All Stage but did not perform as well as the frameworks that handled missingness (IPW, MI). Under monotone missingness with the XGBoost classifier, CC: By Stage slightly outperformed the other frameworks across all metrics on average, though IPW and MI also performed similarly well and had several instances of better performance as evidenced by the long tails in the box plots. For the KNN classifier, both CC methods outperformed IPW and MI when considering recall, with CC: By Stage showing better performance than IPW and MI for accuracy and F-1 score as well. Hence, the results from the ML efficacy plots were less straightforward to interpret, as they were highly dependent on the type of ML classifier (i.e., XGBoost or KNN) as well as the metric (accuracy, precision, recall, or F-1 score). This may also be related in part to the handling of missing data in XGBoost versus KNN, as the latter required an additional imputation step.

\FloatBarrier

Under MCAR (Scenario 2), all results were quite similar, which was to be expected. With a lower proportion of missingness (Scenario 3, 10\% missing), using an XGBoost classifier led to results showing that CC: By Stage outperformed the other frameworks under both non-monotone and monotone missing. However, the results using a KNN classifier were less obvious. With a higher proportion of missingness (Scenario 4, 50\% missing), a much clearer pattern emerged -- in almost all cases, the IPW and MI frameworks outperformed both CC frameworks. The sole exception was that CC: All Stage showed best ML efficacy results based on precision with an XGBoost classifier under non-monotone missingness. However, this framework showed the worst performance when considering the other three metrics in the same setting. For scenarios with weak missingness mechanisms (Scenarios 5 and 6), CC: All Stage still showed poor performance. Specifically in Scenario 6 (50\% missing and weak mechanism), the MI framework slightly outperformed IPW: Indicator Method and IPW: Force Monotonicity in the non-monotone missing setting and IPW in the monotone missing setting when a KNN classifier was employed. With an XGBoost classifier, the difference in performance was negligible. In general, the same interpretability issue also applied for the ML efficacy metric plots of the remaining scenarios, except when the proportion of missingness was high (i.e., Scenario 4, 50\% missing). With a high missingness proportion, results clearly showed that IPW and MI frameworks had best generative performance.

\subsection{Trial Inference}
\label{subsec:results_trialinf}

\begin{figure}[h!]
    \centering
    \includegraphics[width=\linewidth]{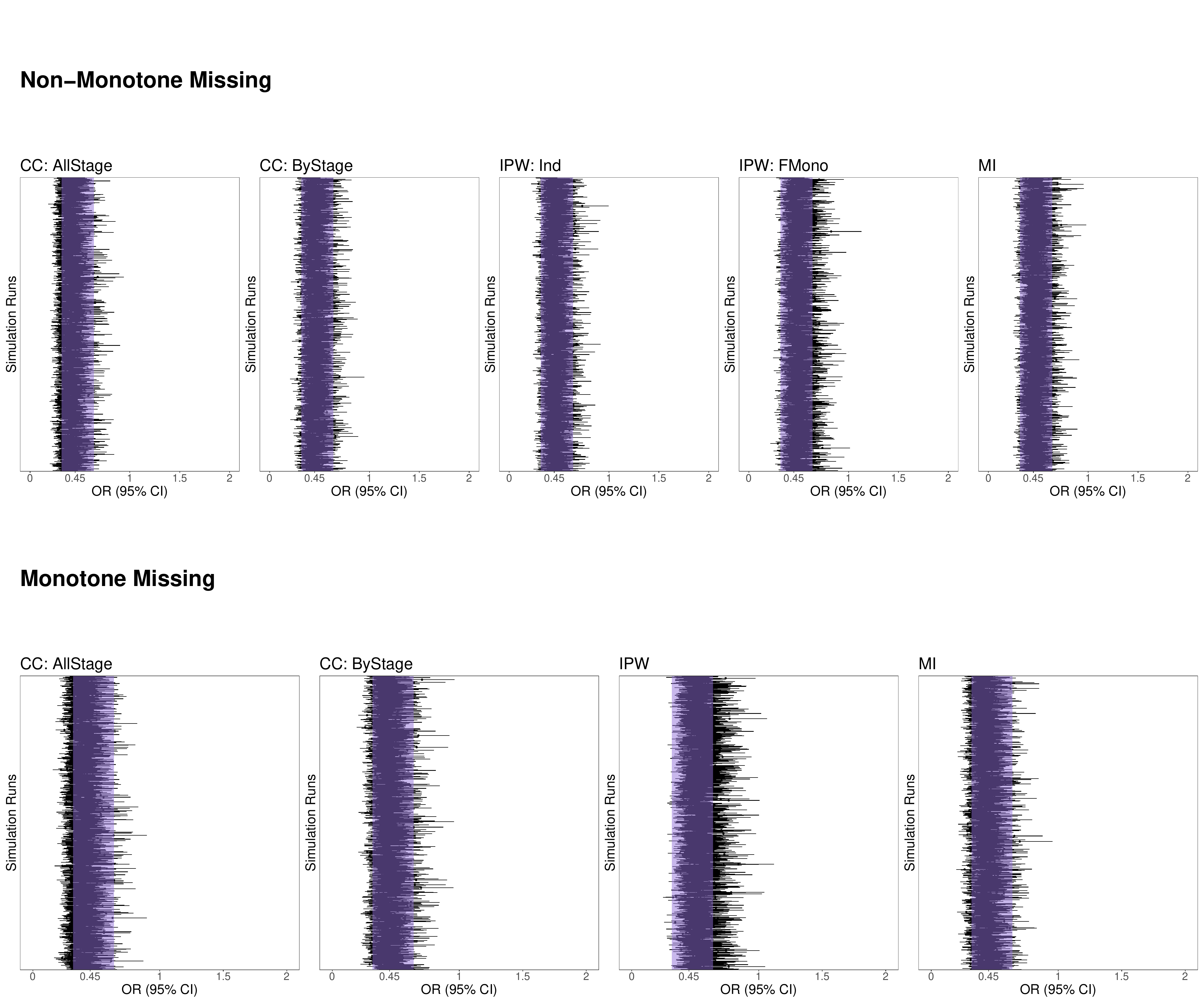}
    \caption{Plot of trial inference results utilizing complete real data and complete synthetic data, for all frameworks under both non-monotone missingness and monotone missingness (Scenario 1A, top row, and Scenario 1B, bottom row). The real (complete) data OR estimate was approximately 0.45, and the real (complete) data 95\% CI is shown by the purple shaded region. Each horizontal bar represents the synthetic data 95\% CI for one simulation run.}
    \label{fig:trialinfall}
\end{figure}

\autoref{fig:trialinfall} shows the trial inference results, utilizing complete real data and complete synthetic data. (Visualizations of results comparing observed real data and observed synthetic data are omitted from the main paper for brevity but can be found in the Supplementary Materials.) Under non-monotone missingness, OR estimates using synthetic data generated by the IPW: Indicator Method framework were the closest to the (complete) real data OR estimate, on average, though the MI framework also resulted in close OR estimates. The CC: All Stage framework resulted in OR estimates that were furthest from the real data OR estimate on average, thus showing that again, simply removing all observations with missing data as a data pre-processing step led to worse generative performance. Though, the CC: By Stage framework showed comparable results to the other IPW and MI frameworks, even outperforming IPW: Force Monotonicity on average. Under monotone missingness, IPW performed rather poorly, whereas MI showed best performance with synthetic data OR estimates being closest to the real data OR estimate, on average. Interestingly, comparing the performance of each framework across missingness patterns, all frameworks performed slightly worse under monotone missingness as compared to non-monotone missingness. For the remaining scenarios, the general pattern observed was that MI performed as well as, and often outperformed, the IPW frameworks under both non-monotone and monotone missingness. Even under MCAR (Scenario 2), where both CC frameworks resulted in synthetic data with estimated ORs that were very close to the real data OR (as expected), MI still showed slightly better results than both CC frameworks. The scenario where all frameworks seemed to suffer the most in capturing the complete real data OR was Scenario 6, which had a higher proportion of missingness and a weak missingness mechanism.

\FloatBarrier

Comparing observed synthetic data OR estimates to the observed real data OR estimate (for which the visualizations are provided in the Supplementary Materials), it was unsurprising to find that the CC frameworks performed rather well at recapturing the observed data OR estimates, thus further supporting our claim that omitting missing values during the data generation procedure allows for the analyst to capture the \textit{observed} real data distribution, but not the \textit{complete} real data distribution (as demonstrated in \autoref{fig:trialinfall}). Under non-monotone missingness, the IPW: Indicator Method was most successful at recovering the observed real data OR estimate out of the other non-CC frameworks, whereas both IPW and MI frameworks performed similarly well under a monotone missingness setting. Note that comparisons regarding lengths of CIs are not meaningful in this context since the CC approaches utilized the full synthetic data (as no missingness could be simulated in these frameworks) whereas the IPW and MI approaches utilized the observed synthetic data to mimic the real data analysis; hence, it is to be expected that the CC approaches will, on average, result in shorter estimated CI lengths. As for the other scenarios, the IPW and MI frameworks still outperformed both CC frameworks, though results were very similar under MCAR, as to be expected. On average, in a setting with a higher proportion of missingness, IPW: Indicator Method synthetic data ORs were closest to the real data OR under non-monotone missingness, and MI synthetic data ORs were closest to the real data OR under monotone missingness. Surprisingly, under monotone missingness, the synthetic data ORs resulting from the IPW framework were the furthest from the real data OR, on average; MI was the closest. With a weak missingness mechanism and a higher proportion of missingness (Scenario 6), all IPW and MI frameworks led to synthetic data ORs that were close to the real data OR, on average, under both non-monotone and monotone missingness. Here, it seemed that the strength of the missingness mechanism had a larger impact on generative performance than the proportion of missing data. The IPW and MI frameworks still fared well under a weak mechanism with a high proportion of missingness, whereas a strong mechanism paired with a higher proportion of missingness (as in Scenario 4) led to some instances of poorer performance for these frameworks.

\subsection{Variables with Missing Data}
\label{subsec:results_missmetrics}

In terms of proportions of missing values generated, IPW: Indicator Method and MI generated data with proportions similar to that of the real data for both CD4 count at week 20 and the outcome under the non-monotone missingness scenario. However, IPW: Force Monotonicity generated the synthetic outcome with a proportion of missing values that was much larger than existed in the real data for this variable. Under monotone missingness, all frameworks generated proportions of missingness that were close to that of the real data, for all three variables (CD4 count week 20, CD4 count week 96, and the outcome). A plot of these results can be found in the Supplementary Materials. The other scenarios also showed similar results, with an inflated proportion of missingness for the synthetic outcome using the IPW: Force Monotonicity framework.

\begin{figure}[h!]
    \centering
    \includegraphics[width=\linewidth]{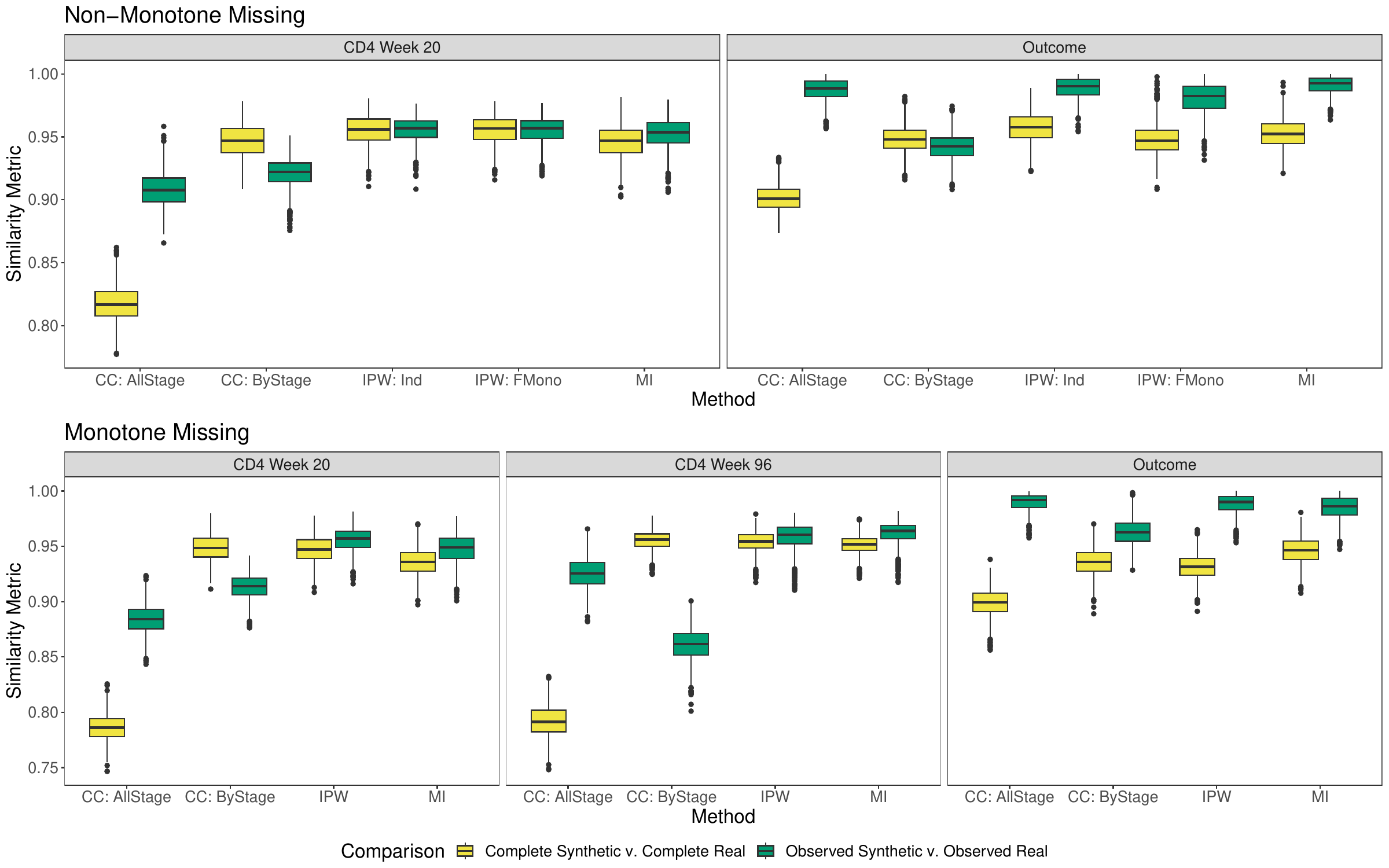}
    \caption{Scenarios 1A and 1B (MAR x 25\% Missing x Strong Mechanism) with 1A shown in the top row and 1B shown in the bottom row; plot of similarity metrics for variables with missing values: 1-KS statistic for continuous variables CD4 count week 20 ($Z_1$) and week 96 ($Z_2$), 1-TVD for the discrete outcome ($Y$). In yellow: comparison of complete synthetic data to complete real data; in green: comparison of observed synthetic data to observed real data. Recall that for CC: All Stage and CC: By Stage, the complete synthetic data and observed synthetic data were the same, as these frameworks did not generate any missing values.}
    \label{fig:missmetrics}
\end{figure}

\autoref{fig:missmetrics} shows the univariate similarity metrics for only those variables with missing data, comparing complete synthetic data to complete real data and comparing observed synthetic data to observed real data. Under both non-monotone and monotone missingness, CC: All Stage clearly showed the worst performance in capturing the complete real data distribution, though for certain variables, it outperformed CC: By Stage in capturing the observed real data distribution. Another pattern that was apparent under both non-monotone missingness and monotone missingness was that the IPW and MI frameworks never performed worse, and often outperformed, both CC frameworks in generating the observed data distribution, thus indicating that introducing more models and hence modelling assumptions to the data generation framework did not negatively impact the ability to generate synthetic data with distributions close to the real data. Indeed, accounting for the missingness rather than omitting observations with missing data at the start benefited data generation performance. For capturing the complete data distribution, IPW: Force Monotonicity under non-monotone missingness and MI under both missingness patterns showed worse performance than the other weighting method and even CC: By Stage in certain cases. For IPW: Force Monotonicity, one potential explanation may be that forcing certain values to be missing in order to fit the missingness model with CD4 count at week 20 as a predictor could have led to the missingness model to be misspecified for the outcome variable (in addition to noisier parameter estimates). IPW: Force Monotonicity also resulted in slightly worse performance when capturing the observed data distribution, compared to IPW: Indicator Method and MI, thus further suggesting that the issue may be linked to misspecification of the missingness model when forcing monotonicity in the real data. Note also that forcing monotonicity only impacted the data at the outcome stage and not at CD4 count at week 20, so it was to be expected that any drop in performance would occur at the outcome and not at CD4 count at week 20. For MI, as the real data were MAR (in Scenarios 1A and 1B), it was anticipated that the complete data distribution would be recovered appropriately. However, it is possible that this may not always be the case due to model misspecification of the imputation model. This would lead to poorly-imputed values and thus not a good match between the real and synthetic complete data. Overall, IPW: Indicator Method and MI under non-monotone missing, and IPW and MI under monotone missing, performed similarly well, though there were a few stages where MI performed slightly worse in capturing the complete data distribution.  

\FloatBarrier

For Scenario 2 (MCAR), all frameworks performed similarly in capturing both the complete and observed data distributions (though, IPW: Force Monotonicity performed slightly worse at capturing the outcome observed distribution). With only 10\% missing (Scenario 3), CC: All Stage was clearly the worst at capturing the complete data distribution and CC: By Stage did not perform well in capturing the observed data distributions at subsequent stages (i.e., for CD4 count at week 96 and the final outcome). The IPW and MI frameworks performed similarly well, with IPW: Indicator Method under non-monotone missing and IPW under monotone missing performing slightly better at the first post-randomization stage (i.e., for CD4 count at week 20). With 50\% missing (Scenario 4), both CC frameworks clearly showed the worst generative performance. Depending on the post-randomization time point (CD4 count at week 20 or week 96, or for the outcome), IPW: Indicator Method (non-monotone), IPW (monotone), and MI (non-monotone and monotone) frameworks performed much better at capturing the complete data distributions. With a weak missingness mechanism (Scenario 5), CC: All Stage drastically underperformed in capturing the complete data distributions for all post-randomization variables (including the outcome) and under both missingness patterns. The MI framework performed slightly worse than both IPW frameworks under non-monotone and monotone missingness when mimicking the complete data distributions. With both a high missingness proportion and a weak missingness mechanism (Scenario 6), IPW methods on average outperformed the CC frameworks, but there were a few instances (i.e., simulation runs) where the IPW frameworks did not work as well. Under this scenario, MI performed consistently well for all post-randomization variables and under both missingness patterns.

\subsection{Computation Time}
Over 1000 simulations, the following computation times were recorded for the generation of synthetic data and calculation of metrics, ordered from quickest to slowest (in format hh:mm:ss): CC: All Stage (01:13:39), CC: By Stage (01:31:15), IPW: Indicator Method (01:32:45), IPW: Force Monotonicity (01:33:21), MI (01:59:18). Similarly, for Scenario 1B: CC: All Stage (01:17:31), CC: By Stage (01:29:02), IPW (01:31:46), MI (02:12:39). Though the IPW and MI frameworks required more computational resources than the CC methods (with MI having the largest computational time), the difference was not so much greater, and computation time did not detract from the feasibility of using any of the frameworks presented here. The ordering was typically the same for the other scenarios (2-6) as well; these computation times can be found in the Supplementary Materials.

\section{Discussion}
\label{sec:discussion}

We have demonstrated how missing values can be effectively handled when generating realistic, complex synthetic tabular data in an RCT context, as well as how to generate synthetic missing values. Rather than simply discarding individuals with missing data, additional models (either to model the probability of being observed, or to model the conditional distribution of the missing data given the observed) can be added with relative ease to our previously-proposed framework for generating synthetic data. Having already shown in previous work that a sequential framework works best, where the data generation task is split into multiple time-dependent steps that follow the natural temporal ordering of an RCT, we have shown in this paper that we can further improve upon data generation performance by properly handling missing values in the real data. If real data are MAR, then data generation performance in terms of fidelity to the original data source improves greatly by employing either IPW or MI. In particular, if data follow a non-monotone missingness pattern, then implementing IPW by including the indicator of being observed as a main effect and in an interaction term with the variable with missingness as covariates in the missingness model was a better strategy than forcing monotonicity and using IPW. Even when data are MCAR, data generation performance is not lost (and sometimes even gained) if we account for the missingness mechanism rather than executing CC.

Under non-monotone missingness, IPW using the indicator method allowed for missingness in $Y^r$ to be modeled using as much information as possible and thus may have led to potentially less noisy results. However, the missingness model was misspecified for those with missing $Z^r_1$ and hence estimating the probability of being observed at $Y^r$ for these individuals introduced some bias. In our simulations, the impact of the misspecification appeared to be minimal, and the approach still yielded superior performance as compared to CC. Implementing IPW by forcing monotonicity was also subject to potential misspecification of the missingness model. In the real data, we know that missingness in the outcome was simulated such that people with lower CD4 counts at week 96 were more likely to have a missing outcome value (as shown in the Supplementary Materials). Thus, the subset of data included when modelling missingness in the outcome was a non-random subgroup of patients. This method of forcing monotonicity would also likely be inefficient and would lead to noisier results due to the loss of information when imposing missingness on $Y^r$ for individuals with missing $Z^r_1$. In a real data setting, we would expect that any (or every) model could be misspecified and it can never be known which are or are not. Additionally, though the frameworks that incorporated IPW and MI performed similarly in generating synthetic data that were close to the real data under non-monotone missingness, there were instances when the framework IPW: Indicator Method showed better performance in capturing complete univariate distributions of variables with missing data, perhaps due to it being an easier task to correctly specify the missingness model rather than the imputation model. However, the MI framework slightly outperformed in capturing bivariate continuous relationships. In a monotone missingness setting, both IPW and MI performed similarly well, though IPW was noticeably more successful at capturing the complete data distribution whereas MI was slightly better at capturing the observed data distribution. However, across almost all scenarios, the MI framework showed the greatest success at recovering the complete real data estimated effect of the treatment on the trial outcome. Hence, though IPW and MI methods showed different strengths, they both still resulted in data generation performance that was superior to that of either CC framework. Thus, a global recommendation based on the results presented in this work would be to use either the framework IPW: Indicator Method or the MI framework in a non-monotone missing data setting, and either IPW or MI under a monotone missing data setting. Choosing between these two would further involve which assumptions are less likely to be violated in the given context, as well as whether it is more desirable to produce synthetic data with univariate and bivariate distributions that closely match those of the real data or if it is more important to generate synthetic data that capture the ``true'' treatment effect. If the former, then IPW may be more suitable, and if the latter, then MI may be preferable.

We pursued R-vine copulas to generate the baseline cohort rather than deep learning approaches in this work since previous work showed it to be an effective method, however the conclusion that missing data cannot be ignored is very likely to generalize to other ML approaches to synthetic data generation. If the data generation methods are biased by the selection procedure that leads to the missingness in the real data, then it does not matter which method is used to generate (completely-observed) baseline data, deep learning or otherwise. Additionally, though our real data example in this paper involved four treatment arms and a binary outcome, our proposed data generation frameworks accounting for missingness can easily be adapted to handle other trial designs and other types of outcomes, such as continuous or time-to-event \citep{Demeulemeester2023}. Moreover, while we considered the context of RCT data, these methods for handling missing data in the data generation procedure would be relevant in any real data situation with missing data that are MAR, especially when longitudinal aspects are also involved.

An interesting challenge that arose when generating post-randomization variables was that fitting missingness or generative regression models to the observed real complete cases at a given stage sometimes caused rarely-observed strata for real baseline data to drop out entirely. Then, the fitted models could not produce predictions using the complete synthetic data since these data included observations of these strata. (Recall that all frameworks except CC: All Stage fit generative models to all real baseline data, which included observations of these rare strata.) When this occurred, we manually set the probabilities of being observed to zero and the post-randomization variable values to be missing for synthetic participants with these strata observed. However, other ways to account for this would be interesting to investigate, as this challenge could occur in real world applications, especially with higher proportions of missingness.

While the work proposed here demonstrated many strengths, in particular by providing a user-friendly manner of accounting for missing values and quantifying generation performance using a large array of metrics, there were also some limitations. One limitation of IPW was that for the parameter estimators to be unbiased in the analysis model, the missingness model must be correctly-specified. In the missing data literature, there have been developments of IPW that have the advantage of being doubly robust, requiring only one of two auxiliary models to be correctly specified in order for parameter estimators to be (asymptotically) unbiased; one such method is Augmented IPW, or AIPW \citep{Robins1994,Rotnitzky1998}. AIPW includes an extra term in the score equations derived from the original analysis model that incorporates the expected value of the first derivative of the log likelihood function given the observed data. This extra term is estimated via an imputation model, i.e., by predicting the values that are missing, conditional on a set of observed covariates. Hence, AIPW requires only one of the missingness model or the imputation model to be correctly specified in order for parameter estimates to be unbiased. Though not currently implemented, this extension is possible using our proposed weighting framework. However, recall that there were certain instances in the results presented here when MI did not perform as well as IPW, likely due to misspecification of the imputation model. Hence, it is unclear if implementing AIPW rather than IPW would be of great benefit -- at worst, if both the missingness model and imputation model are misspecified, then AIPW may result in larger variance than IPW \citep{McIsaac2017}, but at best, if both models are correctly specified, then AIPW would be more efficient than IPW \citep{Seaman2013}. If one of the two models is correctly specified, then AIPW would still be preferable over IPW due to its double robustness property. This extension would be an interesting investigation for future work. 

Another limitation of the presented work was that baseline and treatment data were assumed to be completely observed. However, it is possible for real RCT data to involve missingness at baseline or treatment assignment, either due to human error in data collection, patient refusal on specific (perhaps sensitive) questions, inability to collect measurements (e.g., participants unable to tolerate a particular procedure), or otherwise. In such cases, the generative model at baseline must be adapted to handle missing values. It would be interesting to determine whether copulas could still be harnessed at baseline with missing data, as this is currently an active area of research, though mainly focused on the MCAR setting \citep{Wang2014,DiLascio2015,Liebscher2024} with few studies considering MAR \citep{Hamori2019,Kertel2022}. However, a simpler method may be sufficient at baseline, as the missing data literature suggests that rather simple approaches such as mean imputation or conditional mean imputation are appropriate for handling missingness for baseline variables in trials \citep{White2005,Groenwold2012,Sullivan2018}. We also did not consider MNAR in this work, as strong assumptions and domain-specific knowledge are often needed to attempt to model the unknown missingness mechanisms, thus making the task highly context dependent. It would be of interest to further extend our proposed framework to the MNAR setting, thus lending greater utility to the method.

The framework IPW: Force Monotonicity, where $Y^r$ was set to missing whenever $Z^r_1$ was missing but $Z^r_2$ was left to be fully observed, led to a model that was temporally illogical, i.e., $\text{Pr}( R^r_{Z_1}=1|\textbf{ X}^r,A^r,Z^r_2 )$, where the probability of $Z^r_1$ being observed or missing was conditional on $Z^r_2$ even though $Z^r_1$ was measured before $Z^r_2$. The impact of this, if any, remains unclear. Perhaps it is of little consequence that the definition of this model does not make sense temporally; the model may not necessarily be misspecified since the coefficient for $Z^r_2$ would be equal to zero if $Z^r_2$ did not impact the probability of being observed at $Z^r_1$. Another option could be to invoke the knowledge of temporal ordering and assume that the probability does not depend on $Z^r_2$. Moreover, though the sample size of synthetic data could be larger or smaller than that of the real data, we did not consider either option as we did not believe this to be a scientific question of interest.

In this work, we have proposed a user-friendly way to incorporate additional models (either missingness models via IPW or imputation models via MI) in our sequential data generation framework that account for missing data and generate synthetic missingness. Incorporating additional models within the data generation framework allows for missingness to be accounted for at the same time as generating synthetic data, thus removing the need for a data pre-processing or post-processing step. We have shown that it is necessary to take into account missing values in order to generate synthetic data that adequately capture the real data distribution; it is not enough to simply omit these observations at the beginning of the data generation procedure and only use the real data complete cases to fit generative models. We have effectively bridged the gap in the literature in how best to handle missing values within the broader goal of generating complex and realistic synthetic tabular data, particularly in the context of RCTs, thus opening the door for data generation to involve more complex aspects like missing data and hence become more realistic.

\section*{Author contributions}

NZP contributed to the experimental design, coding of simulations, and writing and editing of the original manuscript. EEMM and NS contributed to the experimental design and editing of the original manuscript.

\section*{Financial disclosure}

EEMM is a Canada Research Chair (Tier 1, Canadian Institutes of Health Research) in Statistical Methods for Precision Medicine. This work is supported by a Discovery Grant from the Natural Sciences and Engineering Research Council of Canada.

\section*{Supporting information}

Additional details regarding methods and simulation results can be found in the appendix at the end of this article. Data are publicly available from the R library \texttt{speff2trial}, for which documentation can be found here: \url{https://cran.r-project.org/package=speff2trial}.

\bibliographystyle{apalike}  
\bibliography{references}  %%% Remove comment to use the external .bib file (using bibtex).

@article{Hammer1996,
      author = {Hammer, Scott M and Katzenstein, David A and Hughes, Michael D and Gundacker, Holly and Schooley, Robert T and Haubrich, Richard H and Henry, W Keith and Lederman, Michael M and Phair, John P and Niu, Manette and Hirsch, Martin S and Merigan, Thomas C},
       title = {A Trial Comparing Nucleoside Monotherapy with Combination Therapy in {HIV}-Infected Adults with {CD4} Cell Counts from 200 to 500 per Cubic Millimeter},
        year = {1996},
     journal = "{The New England Journal of Medicine}",
      volume = {335},
      number = {15},
       pages = {1081--1090}
}

@article{Petrakos2025,
      author = {Petrakos, Niki Z and Moodie, Erica E M and Savy, Nicolas},
       title = {A Framework for Generating Realistic Synthetic Tabular Data in a Randomized Controlled Trial Setting},
        year = {2025},
     journal = "{Statistics in Medicine}",
      volume = {44},
      number = {18-19},
       pages = {e70227}
}

@article{Little2024,
      author = {Little, Roderick J A and Carpenter, James R and Lee, Katherine J},
       title = {A Comparison of Three Popular Methods for Handling Missing Data: {C}omplete-Case Analysis, Inverse Probability Weighting, and Multiple Imputation},
        year = {2024},
     journal = "{Sociological Methods \& Research}",
      volume = {53},
      number = {3},
       pages = {1105--1135}
}

@article{Seaman2013,
      author = {Seaman, Shaun R and White, Ian R},
       title = {Review of Inverse Probability Weighting for Dealing with Missing Data},
        year = {2013},
     journal = "{Statistical Methods in Medical Research}",
      volume = {2},
      number = {3},
       pages = {278--295}
}

@article{Liu2023,
      author = {Liu, Mingxuan and Li, Siqi and Yuan, Han and Ong, Marcus E H and Ning, Yilin and Xie, Feng and Saffari, Seyed E and Shang, Yuqing and Volovici, Victor and Chakraborty, Bibhas and Liu, Nan},
       title = {Handling Missing Values in Healthcare Data: {A} Systematic Review of Deep Learning-Based Imputation Techniques},
        year = {2023},
     journal = "{Artificial Intelligence in Medicine}",
      volume = {142},
       pages = {102587}
}

@article{Pingi2024,
      author = {Pingi, Sharon T and Nayak, Richi and Bashar, Md A},
       title = {Conditional Generative Adversarial Network for Early Classification of Longitudinal Datasets Using an Imputation Approach},
        year = {2024},
     journal = "{ACM Transactions on Knowledge Discovery from Data}",
      volume = {18},
      number = {5}
}

@InProceedings{Yoon2018,
     author = {Yoon, Jinsung and Jordon, James and van der Schaar, Mihaela},
      title = {{GAIN}: {M}issing Data Imputation using Generative Adversarial Nets},
  booktitle = {Proceedings of the 35th International Conference on Machine Learning},
      pages = {5689--5698},
       year = {2018},
     editor = {Dy, Jennifer and Krause, Andreas},
      month = {10--15 Jul},
  publisher = {PMLR}
}

@InProceedings{Ma2020,
     author = {Ma, Fenglong and Wang, Yaqing and Gao, Jing and Houping Xiao, Houping and Zhou, Jing},
      title = {Rare Disease Prediction by Generating Quality-Assured Electronic Health Records},
  booktitle = {Proceedings of the 2020 SIAM International Conference on Data Mining},
      pages = {514--522},
       year = {2020},
     editor = {Demeniconi, Carlotta and Chawla, Nitesh},
      month = {7--9 May},
  publisher = {Society for Industrial and Applied Mathematics}
}

@article{Spinelli2020,
      author = {Spinelli, Indro and Scardapane, Simone and Uncini, Aurelio},
       title = {Missing data imputation with adversarially-trained graph convolutional networks},
        year = {2020},
     journal = "{Neural Networks}",
      volume = {129},
       pages = {249--260}
}

@InProceedings{Jarrett2022,
     author = {Jarrett, Daniel and Cebere, Bogdan C and Liu, Tennison and Curth, Alicia and van der Schaar, Mihaela},
      title = {{H}yper{I}mpute: {G}eneralized Iterative Imputation with Automatic Model Selection},
  booktitle = {Proceedings of the 39th International Conference on Machine Learning},
      pages = {9916--9937},
       year = {2022},
     editor = {Chaudhuri, Kamalika and Jegelka, Stefanie and Song, Le and Szepesvari, Csaba and Niu, Gang and Sabato, Sivan},
      month = {17--23 Jul},
  publisher = {PMLR}
}

@article{Rubin1976,
      author = {Rubin, Donald B},
       title = {Inference and missing data},
        year = {1976},
     journal = "{Biometrika}",
      volume = {63},
      number = {3},
       pages = {581--592}
}

@article{Dong2013,
      author = {Dong, Yiran and Peng, Chao-Ying J},
       title = {Principled missing data methods for researchers},
        year = {2013},
     journal = "{SpringerPlus}",
      volume = {2},
      number = {1},
       pages = {222}
}

@book{Rubin1987,
  author    = {Rubin, Donald B},
  title     = {Multiple imputation for nonresponse in surveys},
  year      = {1987},
  publisher = "{John Wiley \& Sons Inc}",
  address   = "{New York}"
}

@article{Raghunathan2001,
      author = {Raghunathan, Trivellore E and Lepkowski, James M and Van Hoewyk, John and Solenberger, Peter and others},
       title = {A multivariate technique for multiply imputing missing values using a sequence of regression models},
        year = {2001},
     journal = "{Survey Methodology}",
      volume = {27},
      number = {1},
       pages = {85--96}
}

@article{vanBuuren2007,
      author = {van Buuren, Stef},
       title = {Multiple imputation of discrete and continuous data by fully conditional specification},
        year = {2007},
     journal = "{Statistical Methods in Medical Research}",
      volume = {16},
      number = {3},
       pages = {219--242}
}

@article{White2011,
      author = {White, Ian R and Royston, Patrick and Wood, Angela M},
       title = {Multiple imputation using chained equations: {I}ssues and guidance for practice},
        year = {2011},
     journal = "{Statistics in Medicine}",
      volume = {30},
      number = {4},
       pages = {377--399}
}

@article{Austin2021,
      author = {Austin, Peter C and White, Ian R and Lee, Douglas S and {van Buuren}, Stef},
       title = {Missing Data in Clinical Research: {A} Tutorial on Multiple Imputation},
        year = {2021},
     journal = "{Canadian Journal of Cardiology}",
      volume = {37},
      number = {9},
       pages = {1322--1331}
}

@article{Sullivan2018,
      author = {Sullivan, Thomas R and White, Ian R and Salter, Amy B and Ryan, Philip and Lee, Katherine J},
       title = {Should multiple imputation be the method of choice for handling missing data in randomized trials?},
        year = {2018},
     journal = "{Statistical Methods in Medical Research}",
      volume = {27},
      number = {9},
       pages = {2610--2626}
}

@article{White2005,
      author = {White, Ian R and Thompson, Simon G},
       title = {Adjusting for partially missing baseline measurements in randomized trials},
        year = {2005},
     journal = "{Statistics in Medicine}",
      volume = {24},
      number = {7},
       pages = {993--1007}
}

@article {Groenwold2012,
      author = {Groenwold, Rolf H H and White, Ian R and Donders, A Rogier T and Carpenter, James R and Altman, Douglas G and Moons, Karel G M},
       title = {Missing covariate data in clinical research: {W}hen and when not to use the missing-indicator method for analysis},
        year = {2012},
     journal = "{Canadian Medical Association Journal}",
      volume = {184},
      number = {11},
       pages = {1265--1269}
}

@InProceedings{Patki2016,
     author = {Patki, Neha and Wedge, Roy and Veeramachaneni, Kalyan},
      title = {The Synthetic Data Vault},
  booktitle = {2016 IEEE International Conference on Data Science and Advanced Analytics},
      pages = {399--410},
       year = {2016},
     editor = {Bilof, Randall},
      month = {17--19 Oct},
  publisher = {IEEE}
}

@misc{sdmetrics_misc,
      author = "{DataCebo Inc}",
       title = {Synthetic data metrics},
        year = {2024},
        month = {07},
        note = "{Version 0.15.0. \url{https://docs.sdv.dev/sdmetrics/}}"
}

@inproceedings{Xu2019,
     author = {Xu, Lei and Skoularidou, Maria and Cuesta-Infante, Alfredo and Veeramachaneni, Kalyan},
      title = {Modeling tabular data using conditional {GAN}},
  booktitle = {Advances in Neural Information Processing Systems},
      pages = {1--11},
       year = {2019},
     editor = {Wallach, H and Larochelle, H and Beygelzimer, A and d\textquotesingle Alch\'{e}-Buc, F and Fox, E and Garnett, R},
      month = {8--14 Dec},
  publisher = {Curran Associates, Inc}
}

@article{Zhou2020,
      author = {Zhou, Y and Dong, F and Liu, Y and Li, Z and Du, J and Zhang, L},
       title = {Forecasting emerging technologies using data augmentation and deep learning},
        year = {2020},
     journal = "{Scientometrics}",
      volume = {123},
       pages = {1--29}
}

@article{Wang2023,
      author = {Wang, Winston and Pai, Tun-Wen},
       title = {Enhancing Small Tabular Clinical Trial Dataset through Hybrid Data Augmentation: {C}ombining {SMOTE} and {WCGAN-GP}},
        year = {2023},
     journal = "{Data}",
      volume = {8},
      number = {9},
       pages = {135}
}

@article{Hastie2001,
      author = {Hastie, T and Tibshirani, R and Sherlock, G and Eisen, M and Brown, P and Botstein, D},
       title = {Imputing missing data for gene expression arrays},
        year = {2001},
     journal = "{Technical report, Stanford Statistics Department}",
      volume = {17},
      number = {6},
       pages = {520--525}
}

@article{Robins1994,
      author = {Robins, James M and Rotnitzky, Andrea and Zhao, Lue Ping},
       title = {Estimation of regression coefficients when some regressors are not always observed},
        year = {1994},
     journal = "{Journal of the American Statistical Association}",
      volume = {89},
      number = {427},
       pages = {846--866}
}

@article{Rotnitzky1998,
      author = {Rotnitzky, Andrea and Robins, James M and Scharfstein, Daniel O},
       title = {Semiparametric regression for repeated outcomes with nonignorable nonresponse},
        year = {1998},
     journal = "{Journal of the American Statistical Association}",
      volume = {93},
      number = {444},
       pages = {1321--1339}
}

@article{McIsaac2017,
      author = {McIsaac, Michael and Cook, RJ},
       title = {Statistical methods for incomplete data: {S}ome results on model misspecification},
        year = {2017},
     journal = "{Statistical Methods in Medical Research}",
      volume = {26},
      number = {1},
       pages = {248--267}
}

@article{Liebscher2024,
      author = {Liebscher, Eckhard},
       title = {Fitting copulas in the case of missing data},
        year = {2024},
     journal = "{Statistical Papers}",
      volume = {65},
      number = {6},
       pages = {3681--3711}
}

@article{Kertel2022,
      author = {Kertel, Maximilian and Pauly, Markus},
       title = {Estimating {Gaussian} Copulas with Missing Data with and without Expert Knowledge},
        year = {2022},
     journal = "{Entropy}",
      volume = {24},
      number = {12},
       pages = {3681--3711}
}

@article{DiLascio2015,
      author = {Di Lascio, F Marta L and Giannerini, Simone and Reale, Alessandra},
       title = {Exploring copulas for the imputation of complex dependent data},
        year = {2015},
     journal = "{Statistical Methods \& Applications}",
      volume = {24},
      number = {1},
       pages = {159--175}
}

@article{Hamori2019,
      author = {Hamori, Shigeyuki and Motegi, Kaiji and Zhang, Zheng},
       title = {Calibration estimation of semiparametric copula models with data missing at random},
        year = {2019},
     journal = "{Journal of Multivariate Analysis}",
      volume = {173},
       pages = {85--109}
}

@InProceedings{Wang2014,
     author = {Wang, Huahua and Fazayeli, Farideh and Chatterjee, Soumyadeep and Banerjee, Arindam},
      title = {Gaussian Copula Precision Estimation with Missing Values},
  booktitle = {Proceedings of the 17th International Conference on Artificial Intelligence and Statistics},
      pages = {978--986},
       year = {2014},
     editor = {Kaski, Samuel and Corander, Jukka},
      month = {22--25 Apr},
  publisher = {PMLR}
}

@article{Friedrich2024,
      author = {Friedrich, S and Friede, T},
       title = {On the role of benchmarking data sets and simulations in method comparison studies},
        year = {2024},
     journal = "{Biometrical Journal}",
      volume = {66},
      number = {1},
       pages = {2200212}
}

@article{Chen2021,
      author = {Chen, Z and Zhang, H and Guo, Y and George, T J and Prosperi, M and Hogan, W R and He, Z and Shenkman, E A and Wang, F and Bian, J},
       title = {Exploring the feasibility of using real-world data from a large clinical data research network to simulate clinical trials of {A}lzheimer’s disease},
        year = {2021},
     journal = "{NPJ Digital Medicine}",
      volume = {4},
      number = {1},
       pages = {84}
}

@article{SarramiForoushani2021,
      author = {Sarrami-Foroushani, A and Lassila, T and MacRaild, M and Asquith, J and Roes, K C B and Byrne, J V and Frangi, A F},
       title = {In-silico trial of intracranial flow diverters replicates and expands insights from conventional clinical trials},
        year = {2021},
     journal = "{Nature Communications}",
      volume = {12},
      number = {3861}
}

@article{Zwep2024,
      author = {Zwep, L B and Guo, T and Nagler, T and Knibbe, C A J and Meulman, J J and {van Hasselt}, J G C},
       title = {Virtual patient simulation using copula modeling},
        year = {2024},
     journal = "{Clinical Pharmacology \& Therapeutics}",
      volume = {115},
      number = {4},
       pages = {795--804}
}

@InProceedings{Zhao2021,
     author = {Zhao, Zilong and Kunar, Aditya and Birke, Robert and Chen, Lydia Y},
      title = {{CTAB-GAN}: {E}ffective Table Data Synthesizing},
  booktitle = {Proceedings of the 13th Asian Conference on Machine Learning},
      pages = {97--112},
       year = {2021},
     editor = {Balasubramanian, Vineeth N. and Tsang, Ivor},
      month = {17--19 Nov},
  publisher = {PMLR}
}

@InProceedings{Mendikowski2023,
     author = {Mendikowski, Melle and Schindler, Benjamin and Schmid, Thomas and M{\" o}ller, Ralf and Hartwig, Mattis},
      title = {Improved Techniques for Training Tabular {GANs} Using {Cramer}\textquoteright{}s {V} Statistics},
  booktitle = {Proceedings of the 36th Canadian Conference on Artificial Intelligence},
       year = {2023},
     editor = {Soares, Amilcar and Zulkernine, Farhana and Dividino, Renata and Rabbany, Reihaneh and Ye, Qiang and Beach, David and Ali, Karim},
      month = {5--9 Jun},
  publisher = {Canadian Artificial Intelligence Association}
}

@article{Stekhoven2011,
      author = {Stekhoven, Daniel J and Bühlmann, Peter},
       title = {Miss{F}orest -- {N}on-parametric missing value imputation for mixed-type data},
        year = {2011},
     journal = "{Bioinformatics}",
      volume = {28},
      number = {1},
       pages = {112--118}
}

@InProceedings{Choi2017,
     author = {Choi, Edward and Biswal, Siddharth and Malin, Bradley and Duke, Jon and Stewart, Walter F and Sun, Jimeng},
      title = {Generating multi-label discrete patient records using generative adversarial networks},
  booktitle = {Proceedings of the 2nd Machine Learning for Healthcare Conference},
      pages = {286--305},
       year = {2017},
     editor = {Doshi-Velez, Finale and Fackler, Jim and Kale, David and Ranganath, Rajesh and Wallace, Byron and Wiens, Jenna},
      month = {18--19 Aug},
  publisher = {PMLR}
}

@InProceedings{Hyun2020,
     author = {Hyun, Jayun and Lee, Seo Hu and Son, Ha Min and Park, Ji-Ung and Chung, Tai-Myoung},
      title = {A Synthetic Data Generation Model for Diabetic Foot Treatment},
  booktitle = {Future Data and Security Engineering. Big Data, Security and Privacy, Smart City and Industry 4.0 Applications},
      pages = {249--264},
       year = {2020},
     editor = {Dang, Tran Khanh and K{\"u}ng, Josef and Takizawa, Makoto and Chung, Tai M},
  publisher = {Springer Singapore}
}

@InProceedings{Koloi2023,
     author = {Koloi, Angela and Loukas, Vasileios S and Sakellarios, Antonis and Bosch, Jos A and Quax, Rick and Nowakowska, Karina and Tachos, Nikolaos and Kaźmierski, Jakub and Papaloukas, Costas and Fotiadis, Dimitrios},
      title = {A comparison study on creating simulated patient data for individuals suffering from chronic coronary disorders},
  booktitle = {2023 45th Annual International Conference of the IEEE Engineering in Medicine \& Biology Society (EMBC)},
      pages = {1--4},
       year = {2023},
     editor = {Barbieri, Riccardo},
      month = {24--27 Jul},
  publisher = {IEEE}
}

@InProceedings{Walia2020,
     author = {Walia, Manhar and Tierney, Brendan and McKeever, Susan},
      title = {Synthesising tabular data using {Wasserstein} conditional {GANs} with gradient penalty {(WCGAN-GP)}},
  booktitle = {Irish Conference on Artificial Intelligence and Cognitive Science},
       year = {2020},
     editor = {Longo, Luca and Rizzo, Lucas and Hunter, Elizabeth and Pakrashi, Arjun},
      month = {7--8 Dec},
  publisher = {CEUR-WS.org}
}

@InProceedings{Gondara2018,
     author = {Gondara, Lovedeep and Wang, Ke},
      title = {{MIDA}: {M}ultiple Imputation Using Denoising Autoencoders},
  booktitle = {Advances in Knowledge Discovery and Data Mining},
      pages = {260--272},
       year = {2018},
     editor = {Phung, Dinh and Tseng, Vincent S and Webb, Geoffrey I and Ho, Bao and Ganji, Mohadeseh and Rashidi, Lida},
      month = {3--6 Jun},
  publisher = {Springer International Publishing}
}

@article{Nazabal2020,
      author = {Nazábal, Alfredo and Olmos, Pablo M and Ghahramani, Zoubin and Valera, Isabel},
       title = {Handling incomplete heterogeneous data using {VAEs}},
        year = {2020},
     journal = "{Pattern Recognition}",
      volume = {107},
       pages = {107501}
}

@article{Li2023,
      author = {Li, Jin and Cairns, Benjamin J and Li, Jingsong and Zhu, Tingting},
       title = {Generating synthetic mixed-type longitudinal electronic health records for artificial intelligent applications},
        year = {2023},
     journal = "{NPJ Digital Medicine}",
      volume = {6},
      number = {98}
}

@article{Lim2025,
      author = {Lim, David K and Rashid, Naim U and Oliva, Junier B and Ibrahim, Joseph G},
       title = {Unsupervised Imputation of Non-Ignorably Missing Data Using Importance-Weighted Autoencoders},
        year = {2025},
     journal = "{Statistics in Biopharmaceutical Research}",
      volume = {17},
      number = {2},
       pages = {222--234}
}

@article{Kazijevs2023,
      author = {Kazijevs, Maksims and Samad, Manar D},
       title = {Deep imputation of missing values in time series health data: {A} review with benchmarking},
        year = {2023},
     journal = "{Journal of Biomedical Informatics}",
      volume = {144},
       pages = {104440}
}

@article{Neves2022,
      author = {Neves, Diogo Telmo and Alves, João and Naik, Marcel Ganesh and Proença, Alberto José and Prasser, Fabian},
       title = {From Missing Data Imputation to Data Generation},
        year = {2022},
     journal = "{Journal of Computational Science}",
      volume = {61},
       pages = {101640}
}

@article{Farhadyar2024,
      author = {Farhadyar, Kiana and Bonofiglio, Federico and Hackenberg, Maren and Behrens, Max and Z{\"o}ller, Daniela and Binder, Harald},
       title = {Combining propensity score methods with variational autoencoders for generating synthetic data in presence of latent sub-groups},
        year = {2024},
     journal = "{BMC Medical Research Methodology}",
      volume = {24},
      number = {198}
}

@article{Gomez2025,
      author = {Gomez, Louis A and Toye, Adedolapo Aishat and Hum, R Stanley and Kleinberg, Samantha},
       title = {Simulating Realistic Continuous Glucose Monitor Time Series By Data Augmentation},
        year = {2025},
     journal = "{Journal of Diabetes Science and Technology}",
      volume = {19},
      number = {1},
       pages = {114--122}
}

@article{Bianchi2019,
      author = {Bianchi, Filippo Maria and Livi, Lorenzo and Mikalsen, Karl {\O}yvind and Kampffmeyer, Michael and Jenssen, Robert},
       title = {Learning representations of multivariate time series with missing data},
        year = {2019},
     journal = "{Pattern Recognition}",
      volume = {96},
       pages = {106973}
}

@article{Mosquera2023,
      author = {Mosquera, Lucy and El Emam, Khaled and Ding, Lei and Sharma, Vishal and Zhang, Xue Hua and Kababji, Samer El and Carvalho, Chris and Hamilton, Brian and Palfrey, Dan and Kong, Linglong and Jiang, Bei and Eurich, Dean T},
       title = {A method for generating synthetic longitudinal health data},
        year = {2023},
     journal = "{BMC Medical Research Methodology}",
      volume = {23},
      number = {67}
}

@article{Weng2024,
      author = {Weng, Xutao and Song, Hong and Lin, Yucong and Wu, You and Zhang, Xi and Liu, Bowen and Yang, Jian},
       title = {A joint learning method for incomplete and imbalanced data in electronic health record based on generative adversarial networks},
        year = {2024},
     journal = "{Computers in Biology and Medicine}",
      volume = {168},
       pages = {107687}
}

@article{Eckardt2024,
      author = {Eckardt, Jan-Niklas and Hahn, Waldemar and R{\"o}llig, Christoph and Stasik, Sebastian and Platzbecker, Uwe and M{\"u}ller-Tidow, Carsten and Serve, Hubert and Baldus, Claudia D and Schliemann, Christoph and Sch{\"a}fer-Eckart, Kerstin and Hanoun, Maher and Kaufmann, Martin and Burchert, Andreas and Thiede, Christian and Schetelig, Johannes and Sedlmayr, Martin and Bornh{\"a}user, Martin and Wolfien, Markus and Middeke, Jan Moritz},
       title = {Mimicking clinical trials with synthetic acute myeloid leukemia patients using generative artificial intelligence},
        year = {2024},
     journal = "{NPJ Digital Medicine}",
      volume = {7},
      number = {1}
}

@InProceedings{Denton2015,
     author = {Denton, Emily L and Chintala, Soumith and Szlam, Arthur and Fergus, Rob},
      title = {Deep generative image models using a {Laplacian} pyramid of adversarial networks},
  booktitle = {Advances in Neural Information Processing Systems},
       year = {2015},
     editor = {Cortes, C and Lawrence, N and Lee, D and Sugiyama, M and Garnett, R},
      month = {7--10 Dec},
  publisher = {Curran Associates, Inc.}
}

@misc{Radford2016,
      author = {Radford, Alec and Metz, Luke and Chintala, Soumith},
       title = {Unsupervised representation learning with deep convolutional generative adversarial networks},
        year = {2016},
        note = "{Accessed November 27, 2025. \url{https://arxiv.org/abs/1511.06434}}"
}

@inbook{LittleRubinBook2002,
      author = {Little, Roderick J A and Rubin, Donald B},
   publisher = {John Wiley \& Sons, Ltd},
        isbn = {9781119013563},
       title = {Complete-Case and Available-Case Analysis, Including Weighting Methods},
   booktitle = {Statistical Analysis with Missing Data},
     chapter = {3},
       pages = {41--58},
        year = {2002}
}

@article{Demeulemeester2023,
      author = {Demeulemeester, Romain and Savy, Nicolas and Grosclaude, Pascale and Costa, Nadège and Saint-Pierre, Philippe},
       title = {Agent based modeling in health care economics: {E}xamples in the field of thyroid cancer},
        year = {2023},
     journal = "{The International Journal of Biostatistics}",
      volume = {19},
      number = {2},
       pages = {351--368}
}

@article{Little2012,
      author = {Little, Roderick J and D'Agostino, Ralph and Cohen, Michael L and Dickersin, Kay and Emerson, Scott S and Farrar, John T and Frangakis, Constantine and Hogan, Joseph W and Molenberghs, Geert and Murphy, Susan A and Neaton, James D and Rotnitzky, Andrea and Scharfstein, Daniel and Shih, Weichung J and Siegel, Jay P and Stern, Hal},
       title = {The prevention and treatment of missing data in clinical trials},
        year = {2012},
     journal = "{The New England Journal of Medicine}",
      volume = {367},
      number = {14},
       pages = {1355--1360}
}

@article{White2010,
      author = {White, Ian R and Carlin, John },
       title = {Bias and efficiency of multiple imputation compared with complete-case analysis for missing covariate values},
        year = {2010},
     journal = "{Statistics in Medicine}",
      volume = {29},
      number = {28},
       pages = {2920--2931}
}
%%% and comment out the ``thebibliography'' section.

%%% Comment out this section when you \bibliography{references} is enabled.
% \begin{thebibliography}{1}

% \bibitem{kour2014real}
% George Kour and Raid Saabne.
% \newblock Real-time segmentation of on-line handwritten arabic script.
% \newblock In {\em Frontiers in Handwriting Recognition (ICFHR), 2014 14th
%   International Conference on}, pages 417--422. IEEE, 2014.

% \bibitem{kour2014fast}
% George Kour and Raid Saabne.
% \newblock Fast classification of handwritten on-line arabic characters.
% \newblock In {\em Soft Computing and Pattern Recognition (SoCPaR), 2014 6th
%   International Conference of}, pages 312--318. IEEE, 2014.

% \bibitem{hadash2018estimate}
% Guy Hadash, Einat Kermany, Boaz Carmeli, Ofer Lavi, George Kour, and Alon
%   Jacovi.
% \newblock Estimate and replace: A novel approach to integrating deep neural
%   networks with existing applications.
% \newblock {\em arXiv preprint arXiv:1804.09028}, 2018.

% \end{thebibliography}

\newpage

\appendix

\section*{Appendix}

\section{Additional Methods Details}

\subsection{Models for Imposing Missingness in Real Data}

Here we define the models that were used to impose missingness in the real data for simulation purposes. For the treatment variable, 0 refers to the baseline arm (zidovudine only), 1 refers to zidovudine and didanosine, 2 refers to zidovudine and zalcitabine, and 3 refers to didanosine only. Continuous variables (CD4 count at baseline, week 20, and week 96) were standardized by subtracting the mean and dividing by the standard deviation. In the monotone missing setting, $Z^r_2$ was set to missing when $Z^r_1$ was missing, and $Y^r$ was set to missing when $Z^r_2$ was missing in addition to the models presented below. A modest number of decimal places are shown in the data generating models; for reproducibility, please see code for the exact values.

\subsubsection{Scenario 1 (MAR x 25\% Missing x Strong Mechanism)}

\noindent
\textit{A: Non-Monotone Setting}

\smallskip

\begin{table}[H]
    \centering
    \begin{minipage}[t]{0.48\textwidth} % First minipage for the first row
        \centering
        \caption*{Model for $R^r_{Z_1}$}
        \begin{tabular}{|| c c ||}
            \hline
            Variable & Coefficient \\
            \hline
            Intercept & 2.562 \\
            \hline
            ART History (1-52 wks v.~naive) & 3 \\ 
            \hline
            ART History (52+ wks v.~naive) & 5 \\ 
            \hline
            CD4 Baseline & 10 \\ 
            \hline
            Treatment (1 v.~0) & 3.5 \\ 
            \hline
            Treatment (2 v.~0) & 2 \\ 
            \hline
            Treatment (3 v.~0) & 2.5 \\ 
            \hline
            \end{tabular}
    \end{minipage}
    \hfill % Adds horizontal space between minipages
    \begin{minipage}[t]{0.48\textwidth} % Second minipage for the first row
        \centering
        \caption*{Model for $R^r_{Y}$}
        \begin{tabular}{|| c c ||}
            \hline
            Variable & Coefficient \\
            \hline
            Intercept & 27.546 \\
            \hline
            ART History (1-52 wks v.~naive) & 5 \\ 
            \hline
            ART History (52+ wks v.~naive) & 7 \\ 
            \hline
            CD4 Baseline & 20 \\ 
            \hline
            CD4 Week 96 & 28 \\ 
            \hline
            \end{tabular}
    \end{minipage}
\end{table}

\noindent
\textit{B: Monotone Setting}

\smallskip

\begin{table}[H]
    \centering
    \begin{minipage}[t]{0.48\textwidth} % First minipage for the first row
        \centering
        \caption*{Model for $R^r_{Z_1}$}
        \begin{tabular}{|| c c ||}
            \hline
            Variable & Coefficient \\
            \hline
            Intercept & 2.562 \\
            \hline
            ART History (1-52 wks v.~naive) & 3 \\ 
            \hline
            ART History (52+ wks v.~naive) & 5 \\ 
            \hline            
            CD4 Baseline & 10 \\ 
            \hline
            Treatment (1 v.~0) & 3.5 \\ 
            \hline
            Treatment (2 v.~0) & 2 \\ 
            \hline
            Treatment (3 v.~0) & 2.5 \\ 
            \hline
            \end{tabular}
    \end{minipage}
    \hfill % Adds horizontal space between minipages
    \begin{minipage}[t]{0.48\textwidth} % Second minipage for the first row
        \centering
        \caption*{Model for $R^r_{Z_2}$}
        \begin{tabular}{|| c c ||}
            \hline
            Variable & Coefficient \\
            \hline
            Intercept & 14.357 \\
            \hline
            ART History (1-52 wks v.~naive) & 5 \\ 
            \hline
            ART History (52+ wks v.~naive) & 7 \\ 
            \hline
            CD4 Baseline & 15 \\ 
            \hline
            Treatment (2 v.~0) & 3 \\ 
            \hline
            CD4 Week 20 & 15 \\ 
            \hline
            \end{tabular}
    \end{minipage}

    \bigskip % Adds vertical space between rows of tables

    \begin{minipage}[t]{0.48\textwidth} % Minipage for the third table (second row)
        \centering
        \caption*{Model for $R^r_{Y}$}
        \begin{tabular}{|| c c ||}
            \hline
            Variable & Coefficient \\
            \hline
            Intercept & 26.878 \\
            \hline
            ART History (1-52 wks v.~naive) & 6 \\ 
            \hline
            ART History (52+ wks v.~naive) & 8 \\ 
            \hline
            CD4 Baseline & 20 \\ 
            \hline
            CD4 Week 96 & 28 \\ 
            \hline
            \end{tabular}
    \end{minipage}
    % If a fourth table is desired in the second row, add another minipage here
\end{table}

\newpage

\subsubsection{Scenario 2 (MCAR x 25\% Missing x Strong Mechanism)}

\textit{A: Non-Monotone Setting}

\smallskip

\noindent
Model for $R^r_{Z_1}$: $R^r_{Z_1i}$ was sampled from a Bernoulli distribution with $p=0.75$, for $i=1,...,n$ where $n=1342$ was the total sample size. \\
Model for $R^r_{Y}$: $R^r_{Yi}$ was sampled from a Bernoulli distribution with $p=0.75$, for $i=1,...,n$. 

\medskip

\noindent
\textit{B: Monotone Setting}

\smallskip

\noindent
Model for $R^r_{Z_1}$: $R^r_{Z_1i}$ was sampled from a Bernoulli distribution with $p=0.75$, for $i=1,...,n$ where $n=1342$ was the total sample size. \\
Model for $R^r_{Z_2}$: $R^r_{Z_2i}$ was sampled from a Bernoulli distribution with $p=0.75$, for $i=1,...,n$. \\
Model for $R^r_{Y}$: $R^r_{Yi}$ was sampled from a Bernoulli distribution with $p=0.75$, for $i=1,...,n$. 

\subsubsection{Scenario 3 (MAR x 10\% Missing x Strong Mechanism)}

\noindent
\textit{A: Non-Monotone Setting}

\smallskip

\begin{table}[H]
    \centering
    \begin{minipage}[t]{0.48\textwidth} % First minipage for the first row
        \centering
        \caption*{Model for $R^r_{Z_1}$}
        \begin{tabular}{|| c c ||}
            \hline
            Variable & Coefficient \\
            \hline
            Intercept & 7.643 \\
            \hline
            ART History (1-52 wks v.~naive) & 3 \\ 
            \hline
            ART History (52+ wks v.~naive) & 5 \\ 
            \hline
            CD4 Baseline & 10 \\ 
            \hline
            Treatment (1 v.~0) & 3.5 \\ 
            \hline
            Treatment (2 v.~0) & 2 \\ 
            \hline
            Treatment (3 v.~0) & 2.5 \\ 
            \hline
            \end{tabular}
    \end{minipage}
    \hfill % Adds horizontal space between minipages
    \begin{minipage}[t]{0.48\textwidth} % Second minipage for the first row
        \centering
        \caption*{Model for $R^r_{Y}$}
        \begin{tabular}{|| c c ||}
            \hline
            Variable & Coefficient \\
            \hline
            Intercept & 47.949 \\
            \hline
            ART History (1-52 wks v.~naive) & 5 \\ 
            \hline
            ART History (52+ wks v.~naive) & 7 \\ 
            \hline
            CD4 Baseline & 20 \\ 
            \hline
            CD4 Week 96 & 28 \\ 
            \hline
            \end{tabular}
    \end{minipage}
\end{table}

\noindent
\textit{B: Monotone Setting}

\smallskip

\begin{table}[H]
    \centering
    \begin{minipage}[t]{0.48\textwidth} % First minipage for the first row
        \centering
        \caption*{Model for $R^r_{Z_1}$}
        \begin{tabular}{|| c c ||}
            \hline
            Variable & Coefficient \\
            \hline
            Intercept & 7.643 \\
            \hline
            ART History (1-52 wks v.~naive) & 3 \\ 
            \hline
            ART History (52+ wks v.~naive) & 5 \\ 
            \hline
            CD4 Baseline & 10 \\ 
            \hline
            Treatment (1 v.~0) & 3.5 \\ 
            \hline
            Treatment (2 v.~0) & 2 \\ 
            \hline
            Treatment (3 v.~0) & 2.5 \\ 
            \hline
            \end{tabular}
    \end{minipage}
    \hfill % Adds horizontal space between minipages
    \begin{minipage}[t]{0.48\textwidth} % Second minipage for the first row
        \centering
        \caption*{Model for $R^r_{Z_2}$}
        \begin{tabular}{|| c c ||}
            \hline
            Variable & Coefficient \\
            \hline
            Intercept & 27.717 \\
            \hline
            ART History (1-52 wks v.~naive) & 5 \\ 
            \hline
            ART History (52+ wks v.~naive) & 7 \\ 
            \hline
            CD4 Baseline & 15 \\ 
            \hline
            Treatment (2 v.~0) & 3 \\ 
            \hline
            CD4 Week 20 & 15 \\ 
            \hline
            \end{tabular}
    \end{minipage}

    \bigskip % Adds vertical space between rows of tables

    \begin{minipage}[t]{0.48\textwidth} % Minipage for the third table (second row)
        \centering
        \caption*{Model for $R^r_{Y}$}
        \begin{tabular}{|| c c ||}
            \hline
            Variable & Coefficient \\
            \hline
            Intercept & 47.236 \\
            \hline
            ART History (1-52 wks v.~naive) & 6 \\ 
            \hline
            ART History (52+ wks v.~naive) & 8 \\ 
            \hline
            CD4 Baseline & 20 \\ 
            \hline
            CD4 Week 96 & 28 \\ 
            \hline
            \end{tabular}
    \end{minipage}
    % If a fourth table is desired in the second row, add another minipage here
\end{table}

\newpage

\subsubsection{Scenario 4 (MAR x 50\% Missing x Strong Mechanism)}

\noindent
\textit{A: Non-Monotone Setting}

\smallskip

\begin{table}[H]
    \centering
    \begin{minipage}[t]{0.48\textwidth} % First minipage for the first row
        \centering
        \caption*{Model for $R^r_{Z_1}$}
        \begin{tabular}{|| c c ||}
            \hline
            Variable & Coefficient \\
            \hline
            Intercept & -3.801 \\
            \hline
            ART History (1-52 wks v.~naive) & 3 \\ 
            \hline
            ART History (52+ wks v.~naive) & 5 \\ 
            \hline
            CD4 Baseline & 10 \\ 
            \hline
            Treatment (1 v.~0) & 3.5 \\ 
            \hline
            Treatment (2 v.~0) & 2 \\ 
            \hline
            Treatment (3 v.~0) & 2.5 \\ 
            \hline
            \end{tabular}
    \end{minipage}
    \hfill % Adds horizontal space between minipages
    \begin{minipage}[t]{0.48\textwidth} % Second minipage for the first row
        \centering
        \caption*{Model for $R^r_{Y}$}
        \begin{tabular}{|| c c ||}
            \hline
            Variable & Coefficient \\
            \hline
            Intercept & -1.626 \\
            \hline
            ART History (1-52 wks v.~naive) & 5 \\ 
            \hline
            ART History (52+ wks v.~naive) & 7 \\ 
            \hline
            CD4 Baseline & 20 \\ 
            \hline
            CD4 Week 96 & 28 \\ 
            \hline
            \end{tabular}
    \end{minipage}
\end{table}

\medskip

\noindent
\textit{B: Monotone Setting}

\smallskip

\begin{table}[H]
    \centering
    \begin{minipage}[t]{0.48\textwidth} % First minipage for the first row
        \centering
        \caption*{Model for $R^r_{Z_1}$}
        \begin{tabular}{|| c c ||}
            \hline
            Variable & Coefficient \\
            \hline
            Intercept & -3.801 \\
            \hline
            ART History (1-52 wks v.~naive) & 3 \\ 
            \hline
            ART History (52+ wks v.~naive) & 5 \\ 
            \hline
            CD4 Baseline & 10 \\ 
            \hline
            Treatment (1 v.~0) & 3.5 \\ 
            \hline
            Treatment (2 v.~0) & 2 \\ 
            \hline
            Treatment (3 v.~0) & 2.5 \\ 
            \hline
            \end{tabular}
    \end{minipage}
    \hfill % Adds horizontal space between minipages
    \begin{minipage}[t]{0.48\textwidth} % Second minipage for the first row
        \centering
        \caption*{Model for $R^r_{Z_2}$}
        \begin{tabular}{|| c c ||}
            \hline
            Variable & Coefficient \\
            \hline
            Intercept & -2.430 \\
            \hline
            ART History (1-52 wks v.~naive) & 5 \\ 
            \hline
            ART History (52+ wks v.~naive) & 7 \\ 
            \hline
            CD4 Baseline & 15 \\ 
            \hline
            Treatment (2 v.~0) & 3 \\ 
            \hline
            CD4 Week 20 & 15 \\ 
            \hline
            \end{tabular}
    \end{minipage}

    \bigskip % Adds vertical space between rows of tables

    \begin{minipage}[t]{0.48\textwidth} % Minipage for the third table (second row)
        \centering
        \caption*{Model for $R^r_{Y}$}
        \begin{tabular}{|| c c ||}
            \hline
            Variable & Coefficient \\
            \hline
            Intercept & -2.185 \\
            \hline
            ART History (1-52 wks v.~naive) & 6 \\ 
            \hline
            ART History (52+ wks v.~naive) & 8 \\ 
            \hline
            CD4 Baseline & 20 \\ 
            \hline
            CD4 Week 96 & 28 \\ 
            \hline
            \end{tabular}
    \end{minipage}
    % If a fourth table is desired in the second row, add another minipage here
\end{table}

\subsubsection{Scenario 5 (MAR x 25\% Missing x Weak Mechanism)}

\noindent
\textit{A: Non-Monotone Setting}

\smallskip

\begin{table}[H]
    \centering
    \begin{minipage}[t]{0.48\textwidth} % First minipage for the first row
        \centering
        \caption*{Model for $R^r_{Z_1}$}
        \begin{tabular}{|| c c ||}
            \hline
            Variable & Coefficient \\
            \hline
            Intercept & 3.440 \\
            \hline
            ART History (1-52 wks v.~naive) & 0.3 \\ 
            \hline
            ART History (52+ wks v.~naive) & 0.5 \\ 
            \hline
            CD4 Baseline & 5 \\ 
            \hline
            Treatment (1 v.~0) & 0.7 \\ 
            \hline
            Treatment (2 v.~0) & 0.3 \\ 
            \hline
            Treatment (3 v.~0) & 0.4 \\ 
            \hline
            \end{tabular}
    \end{minipage}
    \hfill % Adds horizontal space between minipages
    \begin{minipage}[t]{0.48\textwidth} % Second minipage for the first row
        \centering
        \caption*{Model for $R^r_{Y}$}
        \begin{tabular}{|| c c ||}
            \hline
            Variable & Coefficient \\
            \hline
            Intercept & 15.821 \\
            \hline
            ART History (1-52 wks v.~naive) & 0.5 \\ 
            \hline
            ART History (52+ wks v.~naive) & 0.7 \\ 
            \hline
            CD4 Baseline & 10 \\ 
            \hline
            CD4 Week 96 & 14 \\ 
            \hline
            \end{tabular}
    \end{minipage}
\end{table}

\newpage

\noindent
\textit{B: Monotone Setting}

\smallskip

\begin{table}[H]
    \centering
    \begin{minipage}[t]{0.48\textwidth} % First minipage for the first row
        \centering
        \caption*{Model for $R^r_{Z_1}$}
        \begin{tabular}{|| c c ||}
            \hline
            Variable & Coefficient \\
            \hline
            Intercept & 3.155 \\
            \hline
            ART History (1-52 wks v.~naive) & 0.3 \\ 
            \hline
            ART History (52+ wks v.~naive) & 0.5 \\ 
            \hline
            CD4 Baseline & 5 \\ 
            \hline
            Treatment (1 v.~0) & 0.7 \\ 
            \hline
            Treatment (2 v.~0) & 0.3 \\ 
            \hline
            Treatment (3 v.~0) & 0.4 \\ 
            \hline
            \end{tabular}
    \end{minipage}
    \hfill % Adds horizontal space between minipages
    \begin{minipage}[t]{0.48\textwidth} % Second minipage for the first row
        \centering
        \caption*{Model for $R^r_{Z_2}$}
        \begin{tabular}{|| c c ||}
            \hline
            Variable & Coefficient \\
            \hline
            Intercept & 8.947 \\
            \hline
            ART History (1-52 wks v.~naive) & 0.5 \\ 
            \hline
            ART History (52+ wks v.~naive) & 0.7 \\ 
            \hline
            CD4 Baseline & 7 \\ 
            \hline
            Treatment (2 v.~0) & 0.3 \\ 
            \hline
            CD4 Week 20 & 7.5 \\ 
            \hline
            \end{tabular}
    \end{minipage}

    \bigskip % Adds vertical space between rows of tables

    \begin{minipage}[t]{0.48\textwidth} % Minipage for the third table (second row)
        \centering
        \caption*{Model for $R^r_{Y}$}
        \begin{tabular}{|| c c ||}
            \hline
            Variable & Coefficient \\
            \hline
            Intercept & 7.549 \\
            \hline
            ART History (1-52 wks v.~naive) & 0.5 \\ 
            \hline
            ART History (52+ wks v.~naive) & 0.7 \\ 
            \hline
            CD4 Baseline & 5 \\ 
            \hline
            CD4 Week 96 & 7 \\ 
            \hline
            \end{tabular}
    \end{minipage}
    % If a fourth table is desired in the second row, add another minipage here
\end{table}

\subsubsection{Scenario 6}

\noindent
\textit{A: Non-Monotone Setting}

\smallskip

\begin{table}[H]
    \centering
    \begin{minipage}[t]{0.48\textwidth} % First minipage for the first row
        \centering
        \caption*{Model for $R^r_{Z_1}$}
        \begin{tabular}{|| c c ||}
            \hline
            Variable & Coefficient \\
            \hline
            Intercept & -0.230 \\
            \hline
            ART History (1-52 wks v.~naive) & 0.3 \\ 
            \hline
            ART History (52+ wks v.~naive) & 0.5 \\ 
            \hline
            CD4 Baseline & 5 \\ 
            \hline
            Treatment (1 v.~0) & 0.7 \\ 
            \hline
            Treatment (2 v.~0) & 0.3 \\ 
            \hline
            Treatment (3 v.~0) & 0.4 \\ 
            \hline
            \end{tabular}
    \end{minipage}
    \hfill % Adds horizontal space between minipages
    \begin{minipage}[t]{0.48\textwidth} % Second minipage for the first row
        \centering
        \caption*{Model for $R^r_{Y}$}
        \begin{tabular}{|| c c ||}
            \hline
            Variable & Coefficient \\
            \hline
            Intercept & 0.590 \\
            \hline
            ART History (1-52 wks v.~naive) & 0.5 \\ 
            \hline
            ART History (52+ wks v.~naive) & 0.7 \\ 
            \hline
            CD4 Baseline & 10 \\ 
            \hline
            CD4 Week 96 & 14 \\ 
            \hline
            \end{tabular}
    \end{minipage}
\end{table}

\noindent
\textit{B: Monotone Setting}

\smallskip

\begin{table}[H]
    \centering
    \begin{minipage}[t]{0.48\textwidth} % First minipage for the first row
        \centering
        \caption*{Model for $R^r_{Z_1}$}
        \begin{tabular}{|| c c ||}
            \hline
            Variable & Coefficient \\
            \hline
            Intercept & -0.230 \\
            \hline
            ART History (1-52 wks v.~naive) & 0.3 \\ 
            \hline
            ART History (52+ wks v.~naive) & 0.5 \\ 
            \hline
            CD4 Baseline & 5 \\ 
            \hline
            Treatment (1 v.~0) & 0.7 \\ 
            \hline
            Treatment (2 v.~0) & 0.3 \\ 
            \hline
            Treatment (3 v.~0) & 0.4 \\ 
            \hline
            \end{tabular}
    \end{minipage}
    \hfill % Adds horizontal space between minipages
    \begin{minipage}[t]{0.48\textwidth} % Second minipage for the first row
        \centering
        \caption*{Model for $R^r_{Z_2}$}
        \begin{tabular}{|| c c ||}
            \hline
            Variable & Coefficient \\
            \hline
            Intercept & 0.534 \\
            \hline
            ART History (1-52 wks v.~naive) & 0.5 \\ 
            \hline
            ART History (52+ wks v.~naive) & 0.7 \\ 
            \hline
            CD4 Baseline & 7 \\ 
            \hline
            Treatment (2 v.~0) & 0.3 \\ 
            \hline
            CD4 Week 20 & 7.5 \\ 
            \hline
            \end{tabular}
    \end{minipage}

    \bigskip % Adds vertical space between rows of tables

    \begin{minipage}[t]{0.48\textwidth} % Minipage for the third table (second row)
        \centering
        \caption*{Model for $R^r_{Y}$}
        \begin{tabular}{|| c c ||}
            \hline
            Variable & Coefficient \\
            \hline
            Intercept & 0.131 \\
            \hline
            ART History (1-52 wks v.~naive) & 0.5 \\ 
            \hline
            ART History (52+ wks v.~naive) & 0.7 \\ 
            \hline
            CD4 Baseline & 5 \\ 
            \hline
            CD4 Week 96 & 7 \\ 
            \hline
            \end{tabular}
    \end{minipage}
    % If a fourth table is desired in the second row, add another minipage here
\end{table}

\subsection{Sample Size at Each Stage}

Below are the sample sizes used to fit the \textbf{generative} models in each framework, for all stages and across all scenarios.

\subsubsection{Scenario 1 (MAR x 25\% Missing x Strong Mechanism)}

\noindent
\textit{A: Non-Monotone Setting}

\vspace{-1cm}
\begin{table}[H]
\caption*{}
\begin{tabular}{|| c c c c c ||}
\hline
\textbf{Framework} & \textbf{Baseline ($\mathbf{X}$)} & \textbf{Week 20 ($\mathbf{Z_1}$)} & \textbf{Week 96 ($\mathbf{Z_2}$)} & \textbf{Outcome ($\mathbf{Y}$)} \\
\hline
CC: All Stage           & 900  & 900  & 900  & 900 \\
\hline
CC: By Stage            & 1342 & 1015 & 1015 & 900 \\ 
\hline
IPW: Indicator Method   & 1342 & 1015 & 1015 & 1010 \\ 
\hline
IPW: Force Monotonicity & 1342 & 1015 & 1015 & 900 \\ 
\hline
MI                      & 1342 & 1342 & 1342 & 1342 \\ 
\hline
\end{tabular}
\end{table}

\noindent
\textit{B: Monotone Setting}

\vspace{-1cm}
\begin{table}[H]
\caption*{}
\begin{tabular}{|| c c c c c ||}
\hline
\textbf{Framework} & \textbf{Baseline ($\mathbf{X}$)} & \textbf{Week 20 ($\mathbf{Z_1}$)} & \textbf{Week 96 ($\mathbf{Z_2}$)} & \textbf{Outcome ($\mathbf{Y}$)} \\
\hline
CC: All Stage & 859  & 859  & 859 & 859 \\
\hline
CC: By Stage  & 1342 & 1015 & 916 & 859 \\ 
\hline
IPW           & 1342 & 1015 & 916  & 859 \\ 
\hline
MI            & 1342 & 1342 & 1342 & 1342 \\ 
\hline
\end{tabular}
\end{table}

\subsubsection{Scenario 2 (MCAR x 25\% Missing x Strong Mechanism)}

\noindent
\textit{A: Non-Monotone Setting}

\vspace{-1cm}
\begin{table}[H]
\caption*{}
\begin{tabular}{|| c c c c c ||}
\hline
\textbf{Framework} & \textbf{Baseline ($\mathbf{X}$)} & \textbf{Week 20 ($\mathbf{Z_1}$)} & \textbf{Week 96 ($\mathbf{Z_2}$)} & \textbf{Outcome ($\mathbf{Y}$)} \\
\hline
CC: All Stage           & 770  & 770  & 770  & 770 \\
\hline
CC: By Stage            & 1342 & 1018 & 1018 & 770 \\ 
\hline
IPW: Indicator Method   & 1342 & 1018 & 1018 & 1011 \\ 
\hline
IPW: Force Monotonicity & 1342 & 1018 & 1018 & 770 \\ 
\hline
MI                      & 1342 & 1342 & 1342 & 1342 \\ 
\hline
\end{tabular}
\end{table}

\noindent
\textit{B: Monotone Setting}

\vspace{-1cm}
\begin{table}[H]
\caption*{}
\begin{tabular}{|| c c c c c ||}
\hline
\textbf{Framework} & \textbf{Baseline ($\mathbf{X}$)} & \textbf{Week 20 ($\mathbf{Z_1}$)} & \textbf{Week 96 ($\mathbf{Z_2}$)} & \textbf{Outcome ($\mathbf{Y}$)} \\
\hline
CC: All Stage & 843  & 843  & 843  & 843 \\
\hline
CC: By Stage  & 1342 & 1018 & 918  & 843 \\ 
\hline
IPW           & 1342 & 1018 & 918  & 843 \\ 
\hline
MI            & 1342 & 1342 & 1342 & 1342 \\ 
\hline
\end{tabular}
\end{table}

\subsubsection{Scenario 3 (MAR x 10\% Missing x Strong Mechanism)}

\noindent
\textit{A: Non-Monotone Setting}

\vspace{-1cm}
\begin{table}[H]
\caption*{}
\begin{tabular}{|| c c c c c ||}
\hline
\textbf{Framework} & \textbf{Baseline ($\mathbf{X}$)} & \textbf{Week 20 ($\mathbf{Z_1}$)} & \textbf{Week 96 ($\mathbf{Z_2}$)} & \textbf{Outcome ($\mathbf{Y}$)} \\
\hline
CC: All Stage           & 1148 & 1148 & 1148 & 1148 \\
\hline
CC: By Stage            & 1342 & 1221 & 1221 & 1148 \\ 
\hline
IPW: Indicator Method   & 1342 & 1221 & 1221 & 1209 \\ 
\hline
IPW: Force Monotonicity & 1342 & 1221 & 1221 & 1148 \\ 
\hline
MI                      & 1342 & 1342 & 1342 & 1342 \\ 
\hline
\end{tabular}
\end{table}

\noindent
\textit{B: Monotone Setting}

\vspace{-1cm}
\begin{table}[H]
\caption*{}
\begin{tabular}{|| c c c c c ||}
\hline
\textbf{Framework} & \textbf{Baseline ($\mathbf{X}$)} & \textbf{Week 20 ($\mathbf{Z_1}$)} & \textbf{Week 96 ($\mathbf{Z_2}$)} & \textbf{Outcome ($\mathbf{Y}$)} \\
\hline
CC: All Stage & 1125 & 1125 & 1125 & 1125 \\
\hline
CC: By Stage  & 1342 & 1221 & 1168 & 1125 \\ 
\hline
IPW           & 1342 & 1221 & 1168 & 1125 \\ 
\hline
MI            & 1342 & 1342 & 1342 & 1342 \\ 
\hline
\end{tabular}
\end{table}

\subsubsection{Scenario 4 (MAR x 50\% Missing x Strong Mechanism)}

\noindent
\textit{A: Non-Monotone Setting}

\vspace{-1cm}
\begin{table}[H]
\caption*{}
\begin{tabular}{|| c c c c c ||}
\hline
\textbf{Framework} & \textbf{Baseline ($\mathbf{X}$)} & \textbf{Week 20 ($\mathbf{Z_1}$)} & \textbf{Week 96 ($\mathbf{Z_2}$)} & \textbf{Outcome ($\mathbf{Y}$)} \\
\hline
CC: All Stage           & 557 & 557 & 557 & 557 \\
\hline
CC: By Stage            & 1342 & 681 & 681 & 557 \\ 
\hline
IPW: Indicator Method   & 1342 & 681 & 681 & 678 \\ 
\hline
IPW: Force Monotonicity & 1342 & 681 & 681 & 557 \\ 
\hline
MI                      & 1342 & 1342 & 1342 & 1342 \\ 
\hline
\end{tabular}
\end{table}

\noindent
\textit{B: Monotone Setting}

\vspace{-1cm}
\begin{table}[H]
\caption*{}
\begin{tabular}{|| c c c c c ||}
\hline
\textbf{Framework} & \textbf{Baseline ($\mathbf{X}$)} & \textbf{Week 20 ($\mathbf{Z_1}$)} & \textbf{Week 96 ($\mathbf{Z_2}$)} & \textbf{Outcome ($\mathbf{Y}$)} \\
\hline
CC: All Stage & 503  & 503  & 503  & 503 \\
\hline
CC: By Stage  & 1342 & 681  & 570  & 503 \\ 
\hline
IPW           & 1342 & 681  & 570  & 503 \\ 
\hline
MI            & 1342 & 1342 & 1342 & 1342 \\ 
\hline
\end{tabular}
\end{table}

\subsubsection{Scenario 5 (MAR x 25\% Missing x Weak Mechanism)}

\noindent
\textit{A: Non-Monotone Setting}

\vspace{-1cm}
\begin{table}[H]
\caption*{}
\begin{tabular}{|| c c c c c ||}
\hline
\textbf{Framework} & \textbf{Baseline ($\mathbf{X}$)} & \textbf{Week 20 ($\mathbf{Z_1}$)} & \textbf{Week 96 ($\mathbf{Z_2}$)} & \textbf{Outcome ($\mathbf{Y}$)} \\
\hline
CC: All Stage           & 906  & 906  & 906  & 906 \\
\hline
CC: By Stage            & 1342 & 1032 & 1032 & 906 \\ 
\hline
IPW: Indicator Method   & 1342 & 1032 & 1032 & 1016 \\ 
\hline
IPW: Force Monotonicity & 1342 & 1032 & 1032 & 906 \\ 
\hline
MI                      & 1342 & 1342 & 1342 & 1342 \\ 
\hline
\end{tabular}
\end{table}

\noindent
\textit{B: Monotone Setting}

\vspace{-1cm}
\begin{table}[H]
\caption*{}
\begin{tabular}{|| c c c c c ||}
\hline
\textbf{Framework} & \textbf{Baseline ($\mathbf{X}$)} & \textbf{Week 20 ($\mathbf{Z_1}$)} & \textbf{Week 96 ($\mathbf{Z_2}$)} & \textbf{Outcome ($\mathbf{Y}$)} \\
\hline
CC: All Stage & 850  & 850  & 850  & 850 \\
\hline
CC: By Stage  & 1342 & 1031 & 921  & 850 \\ 
\hline
IPW           & 1342 & 1031 & 921  & 850 \\ 
\hline
MI            & 1342 & 1342 & 1342 & 1342 \\ 
\hline
\end{tabular}
\end{table}

\subsubsection{Scenario 6}

\noindent
\textit{A: Non-Monotone Setting}

\vspace{-1cm}
\begin{table}[H]
\caption*{}
\begin{tabular}{|| c c c c c ||}
\hline
\textbf{Framework} & \textbf{Baseline ($\mathbf{X}$)} & \textbf{Week 20 ($\mathbf{Z_1}$)} & \textbf{Week 96 ($\mathbf{Z_2}$)} & \textbf{Outcome ($\mathbf{Y}$)} \\
\hline
CC: All Stage           & 525 & 525 & 525 & 525 \\
\hline
CC: By Stage            & 1342 & 663 & 663 & 525 \\ 
\hline
IPW: Indicator Method   & 1342 & 663 & 663 & 663 \\ 
\hline
IPW: Force Monotonicity & 1342 & 663 & 663 & 525 \\ 
\hline
MI                      & 1342 & 1342 & 1342 & 1342 \\ 
\hline
\end{tabular}
\end{table}

\noindent
\textit{B: Monotone Setting}

\vspace{-1cm}
\begin{table}[H]
\caption*{}
\begin{tabular}{|| c c c c c ||}
\hline
\textbf{Framework} & \textbf{Baseline ($\mathbf{X}$)} & \textbf{Week 20 ($\mathbf{Z_1}$)} & \textbf{Week 96 ($\mathbf{Z_2}$)} & \textbf{Outcome ($\mathbf{Y}$)} \\
\hline
CC: All Stage & 477 & 477  & 477  & 477 \\
\hline
CC: By Stage  & 1342 & 663  & 533  & 477 \\ 
\hline
IPW           & 1342 & 663  & 533  & 477 \\ 
\hline
MI            & 1342 & 1342 & 1342 & 1342 \\ 
\hline
\end{tabular}
\end{table}

\newpage

\subsection{Missingness and Generative Models}

Below are the definitions of the models that were fit in each framework for generating the post-randomization variables. The vector of baseline covariates, $\textbf{X}^r$, included age, weight, hemophilia status (yes/no), homosexual identity (yes/no), history of intravenous drug use (yes/no), Karnofsky score (categorized as 70/80/90/100), history of non-zidovudine antiretroviral therapy use (yes/no), zidovudine use 30 days before RCT treatment (yes/no), number of days of previous ART, race (white/non-white), gender (woman/man), history of ART (naive/1-52 weeks/more than 52 weeks), symptomatic HIV (symptomatic/asymptomatic), and CD4 count. 

\subsubsection{A: Non-Monotone Setting}

\noindent
CC: All Stage
\vspace{-0.75cm}
\begin{table}[H]
\caption*{}
\begin{tabular}{|| c c c c ||}
\hline
\textbf{Variable to Generate} & \textbf{Model} & \textbf{Covariates} & \textbf{Response}\\
\hline
CD4 week 20 ($Z_1$) & Generative (linear regression) & $\textbf{X}, A$ & $Z_1$ \\
\hline
CD4 week 96 ($Z_2$) & Generative (linear regression) & $\textbf{X}, A, Z_1$ & $Z_2$ \\ 
\hline
Outcome ($Y$) & Generative (logistic regression) & $\textbf{X}, A, Z_1, Z_2$ & $Y$ \\ 
\hline
\end{tabular}
\end{table}

\noindent
CC: By Stage
\vspace{-0.75cm}
\begin{table}[H]
\caption*{}
\begin{tabular}{|| c c c c ||}
\hline
\textbf{Variable to Generate} & \textbf{Model} & \textbf{Covariates} & \textbf{Response} \\
\hline
CD4 week 20 ($Z_1$) & Generative (linear regression) & $\textbf{X}, A$ & $Z_1$ \\
\hline
CD4 week 96 ($Z_2$) & Generative (linear regression) & $\textbf{X}, A, Z_1$ & $Z_2$ \\ 
\hline
Outcome ($Y$) & Generative (logistic regression) & $\textbf{X}, A, Z_1, Z_2$ & $Y$ \\ 
\hline
\end{tabular}
\end{table}

\noindent
IPW: Indicator Method
\vspace{-0.75cm}
\begin{table}[H]
\caption*{}
\begin{tabular}{|| c c c c ||}
\hline
\textbf{Variable to Generate} & \textbf{Model} & \textbf{Covariates} & \textbf{Response} \\
\hline
CD4 week 20 ($Z_1$) & Missingness (logistic regression) & $\textbf{X}, A$ & $R_{Z_1}$\\
\hline
CD4 week 20 ($Z_1$) & Generative (weighted linear regression) & $\textbf{X}, A$ & $Z_1$\\
\hline
CD4 week 96 ($Z_2$) & Generative (weighted linear regression) & $\textbf{X}, A, Z_1$ & $Z_2$\\ 
\hline
Outcome ($Y$) & Missingness (logistic regression) & $\textbf{X}, A, Z_2, R_{Z_1},Z_1 \times R_{Z_1}$ & $R_Y$\\ 
\hline
Outcome ($Y$) & Generative (weighted logistic regression) & $\textbf{X}, A, Z_2, R_{Z_1}, Z_1 \times R_{Z_1}$ & $Y$ \\ 
\hline
\end{tabular}
\end{table}

\noindent
IPW: Force Monotonicity
\vspace{-0.75cm}
\begin{table}[H]
\caption*{}
\begin{tabular}{|| c c c c ||}
\hline
\textbf{Variable to Generate} & \textbf{Model} & \textbf{Covariates} & \textbf{Response} \\
\hline
CD4 week 20 ($Z_1$) & Missingness (logistic regression) & $\textbf{X}, A$ & $R_{Z_1}$\\
\hline
CD4 week 20 ($Z_1$) & Generative (weighted linear regression) & $\textbf{X}, A$ & $Z_1$\\
\hline
CD4 week 96 ($Z_2$) & Generative (weighted linear regression) & $\textbf{X}, A, Z_1$ & $Z_2$\\ 
\hline
Outcome ($Y$) & Missingness, conditional on $R_{Z_1}=1$ (logistic regression) & $\textbf{X}, A, Z_1, Z_2$ & $R_Y$\\ 
\hline
Outcome ($Y$) & Missingness, in $Z_1$ (logistic regression) & $\textbf{X}, A, Z_2$ & $R_{Z_1}$\\ 
\hline
Outcome ($Y$) & Generative (weighted logistic regression) & $\textbf{X}, A, Z_1, Z_2$ & $Y$ \\ 
\hline
\end{tabular}
\end{table}

\noindent
MI (Recall: real data missing values were imputed via MICE and PMM, with all variables in the original data)
\vspace{-1cm}
\begin{table}[H]
\caption*{}
\begin{tabular}{|| c c c c ||}
\hline
\textbf{Variable to Generate} & \textbf{Model} & \textbf{Covariates} & \textbf{Response} \\
\hline
CD4 week 20 ($Z_1$) & Generative (linear regression) & $\textbf{X}, A$ & $Z_1$\\
\hline
CD4 week 20 ($Z_1$) & Missingness (logistic regression) & $\textbf{X}, A$ & $R_{Z_1}$\\
\hline
CD4 week 96 ($Z_2$) & Generative (linear regression) & $\textbf{X}, A, Z_1$ & $Z_2$\\ 
\hline
Outcome ($Y$) & Generative (logistic regression) & $\textbf{X}, A, Z_1, Z_2$ & $Y$ \\ 
\hline
Outcome ($Y$) & Missingness (logistic regression) & $\textbf{X}, A, Z_2, R_{Z_1},Z_1 \times R_{Z_1}$ & $R_Y$\\ 
\hline
\end{tabular}
\end{table}

\newpage

\subsubsection{B: Monotone Setting}

\noindent
CC: All Stage
\vspace{-0.75cm}
\begin{table}[H]
\caption*{}
\begin{tabular}{|| c c c c ||}
\hline
\textbf{Variable to Generate} & \textbf{Model} & \textbf{Covariates} & \textbf{Response}\\
\hline
CD4 week 20 ($Z_1$) & Generative (linear regression) & $\textbf{X}, A$ & $Z_1$ \\
\hline
CD4 week 96 ($Z_2$) & Generative (linear regression) & $\textbf{X}, A, Z_1$ & $Z_2$ \\ 
\hline
Outcome ($Y$) & Generative (logistic regression) & $\textbf{X}, A, Z_1, Z_2$ & $Y$ \\ 
\hline
\end{tabular}
\end{table}

\noindent
CC: By Stage
\vspace{-0.75cm}
\begin{table}[H]
\caption*{}
\begin{tabular}{|| c c c c ||}
\hline
\textbf{Variable to Generate} & \textbf{Model} & \textbf{Covariates} & \textbf{Response} \\
\hline
CD4 week 20 ($Z_1$) & Generative (linear regression) & $\textbf{X}, A$ & $Z_1$ \\
\hline
CD4 week 96 ($Z_2$) & Generative (linear regression) & $\textbf{X}, A, Z_1$ & $Z_2$ \\ 
\hline
Outcome ($Y$) & Generative (logistic regression) & $\textbf{X}, A, Z_1, Z_2$ & $Y$ \\ 
\hline
\end{tabular}
\end{table}

\noindent
IPW
\vspace{-0.75cm}
\begin{table}[H]
\caption*{}
\begin{tabular}{|| c c c c ||}
\hline
\textbf{Variable to Generate} & \textbf{Model} & \textbf{Covariates} & \textbf{Response} \\
\hline
CD4 week 20 ($Z_1$) & Missingness (logistic regression) & $\textbf{X}, A$ & $R_{Z_1}$\\
\hline
CD4 week 20 ($Z_1$) & Generative (weighted linear regression) & $\textbf{X}, A$ & $Z_1$\\
\hline
CD4 week 96 ($Z_2$) & Missingness, conditional on $R_{Z_1}=1$ (logistic regression) & $\textbf{X}, A, Z_1$ & $R_{Z_2}$\\
\hline
CD4 week 96 ($Z_2$) & Missingness, in $Z_1$ (logistic regression) & $\textbf{X}, A$ & $R_{Z_1}$\\ 
\hline
CD4 week 96 ($Z_2$) & Generative (weighted linear regression) & $\textbf{X}, A, Z_1$ & $Z_2$\\ 
\hline
Outcome ($Y$) & Missingness, conditional on $R_{Z_2}=1$ (logistic regression) & $\textbf{X}, A, Z_1, Z_2$ & $R_Y$\\ 
\hline
Outcome ($Y$) & Missingness, in $Z_2$ (logistic regression) & $\textbf{X}, A, Z_1$ & $R_{Z_2}$\\ 
\hline
Outcome ($Y$) & Generative (weighted logistic regression) & $\textbf{X}, A, Z_1, Z_2$ & $Y$ \\ 
\hline
\end{tabular}
\end{table}

\noindent
MI (Recall: real data missing values were imputed via MICE and PMM, with all variables in the original data)
\vspace{-1cm}
\begin{table}[H]
\caption*{}
\begin{tabular}{|| c c c c ||}
\hline
\textbf{Variable to Generate} & \textbf{Model} & \textbf{Covariates} & \textbf{Response} \\
\hline
CD4 week 20 ($Z_1$) & Generative (linear regression) & $\textbf{X}, A$ & $Z_1$\\
\hline
CD4 week 20 ($Z_1$) & Missingness (logistic regression) & $\textbf{X}, A$ & $R_{Z_1}$\\
\hline
CD4 week 96 ($Z_2$) & Generative (linear regression) & $\textbf{X}, A, Z_1$ & $Z_2$\\ 
\hline
CD4 week 96 ($Z_2$) & Missingness, conditional on $R_{Z_1}=1$ (logistic regression) & $\textbf{X}, A, Z_1$ & $R_{Z_2}$\\
\hline
CD4 week 96 ($Z_2$) & Missingness, in $Z_1$ (logistic regression) & $\textbf{X}, A$ & $R_{Z_1}$\\ 
\hline
Outcome ($Y$) & Generative (logistic regression) & $\textbf{X}, A, Z_1, Z_2$ & $Y$ \\ 
\hline
Outcome ($Y$) & Missingness, conditional on $R_{Z_2}=1$ (logistic regression) & $\textbf{X}, A, Z_1, Z_2$ & $R_Y$\\ 
\hline
Outcome ($Y$) & Missingness, in $Z_2$ (logistic regression) & $\textbf{X}, A, Z_1$ & $R_{Z_2}$\\ 
\hline
\end{tabular}
\end{table}

\subsection{IPW: Force Monotonicity -- Conditional Distribution Decomposition}

The decomposition of the distribution of $Y^r$ being observed given all other variables involves conditioning on $Z^r_1$ being observed. Note that this differs from the decomposition shown under the monotone missing setting, where the decomposition required conditioning on $Z^r_2$ being observed.
\begin{equation*}
    \begin{split}
        \hat{p}^r_Y=\text{Pr}\left( R^r_{Y}=1|\textbf{ X}^r,A^r,Z^r_1,Z^r_2 \right)&=\text{Pr} \left( R^r_{Y}=1|\textbf{ X}^r,A^r,Z^r_1,Z^r_2,R^r_{Z_1}=1 \right)\times\text{Pr}\left( R^r_{Z_1}=1|\textbf{ X}^r,A^r,Z^r_2 \right) \\
        &\hspace{0.5cm} + \text{Pr}\left( R^r_{Y}=1|\textbf{ X}^r,A^r,Z^r_1,Z^r_2,R^r_{Z_1}=0 \right)\times\text{Pr}\left( R^r_{Z_1}=0|\textbf{ X}^r,A^r,Z^r_2 \right). 
    \end{split}
\end{equation*}
Note that the second and fourth terms, $\text{Pr}\left( R^r_{Z_1}=1|\textbf{ X}^r,A^r,Z^r_2 \right)$ and $\text{Pr}\left( R^r_{Z_1}=1|\textbf{ X}^r,A^r,Z^r_2 \right)$, can be further simplified as $\text{Pr}\left( R^r_{Z_1}=1|\textbf{ X}^r,A^r \right)$ and $\text{Pr}\left( R^r_{Z_1}=1|\textbf{ X}^r,A^r \right)$ due to the temporal ordering of the data. As before, the third term, $\text{Pr}\left( R^r_{Y}=1|\textbf{ X}^r,A^r,Z^r_1,Z^r_2,R^r_{Z_1}=0 \right)$, poses an issue in the non-monotone missing setting, so forcing monotonicity such that $Y^r$ is missing whenever $Z^r_1$ is missing would lead to this term being equal to zero, and thus canceling out. This leaves the estimation of $\hat{p}^r_Y$ to be equal to $\text{Pr} \left( R^r_{Y}=1|\textbf{ X}^r,A^r,Z^r_1,Z^r_2,R^r_{Z_1}=1 \right)\times\text{Pr}\left( R^r_{Z_1}=1|\textbf{ X}^r,A^r,Z^r_2 \right)$, both of which can be easily fit to the real data.

\subsection{IPW under Monotone Missingness - Derivation Details}

We present here the details for the results shown in the main paper in Section 4.2.3.

\begin{equation*}
    \begin{split}
        \hat{p}^r_Y=\text{Pr}\left( R^r_{Y}=1|\textbf{ X}^r,A^r,Z^r_1,Z^r_2 \right)&=\text{Pr} \left( R^r_{Y}=1|\textbf{ X}^r,A^r,Z^r_1,Z^r_2,R^r_{Z_2}=1 \right)\times\text{Pr}\left( R^r_{Z_2}=1|\textbf{ X}^r,A^r,Z^r_1 \right) \\
        &\hspace{0.5cm} + \text{Pr}\left( R^r_{Y}=1|\textbf{ X}^r,A^r,Z^r_1,Z^r_2,R^r_{Z_2}=0 \right)\times\text{Pr}\left( R^r_{Z_2}=0|\textbf{ X}^r,A^r,Z^r_1 \right) \\
        &= \text{Pr} \left( R^r_{Y}=1|\textbf{ X}^r,A^r,Z^r_1,Z^r_2,R^r_{Z_2}=1 \right)\times\text{Pr}\left( R^r_{Z_2}=1|\textbf{ X}^r,A^r,Z^r_1 \right) \\
        &\hspace{0.5cm} + 0\times\text{Pr}\left( R^r_{Z_2}=0|\textbf{ X}^r,A^r,Z^r_1 \right)\\
        &=\text{Pr} \left( R^r_{Y}=1|\textbf{ X}^r,A^r,Z^r_1,Z^r_2,R^r_{Z_2}=1 \right)\times\text{Pr}\left( R^r_{Z_2}=1|\textbf{ X}^r,A^r,Z^r_1 \right).
    \end{split}
\end{equation*}
Again, $\text{Pr}\left( R^r_{Z_2}=1|\textbf{ X}^r,A^r,Z^r_1 \right)=\text{Pr}\left( R^r_{Z_2}=1|\textbf{ X}^r,A^r,Z^r_1,Z^r_2 \right)$ since here, $Z^r_2$ did not influence the probability of being observed at $Z^r_2$ and similarly $\text{Pr}\left( R^r_{Z_2}=0|\textbf{ X}^r,A^r,Z^r_1 \right)=\text{Pr}\left( R^r_{Z_2}=0|\textbf{ X}^r,A^r,Z^r_1,Z^r_2 \right)$. Intuitively, we also know that under monotone missingness, $$\text{Pr} \left( R^r_{Y}=1|\textbf{ X}^r,A^r,Z^r_1,Z^r_2,R^r_{Z_2}=1 \right)=\text{Pr} \left( R^r_{Y}=1|\textbf{ X}^r,A^r,Z^r_1,Z^r_2,R^r_{Z_2}=1,R^r_{Z_1}=1 \right).$$ This can be seen via the following decomposition, additionally conditioning on $Z^r_1$ being observed:
\begin{equation*}
    \begin{split}
        \text{Pr} \left(R^r_{Y}=1|\textbf{ X}^r,A^r,Z^r_1,Z^r_2,R^r_{Z_2}=1 \right)&=\{\text{Pr}\left( R^r_{Y}=1|\textbf{X}^r,A^r,Z^r_1,Z^r_2,R^r_{Z_2}=1,R^r_{Z_1}=1 \right) \\
        &\hspace{0.5cm} \times\text{Pr}\left( R^r_{Z_1}=1|\textbf{ X}^r,A^r,Z^r_1,Z^r_2,R^r_{Z_2}=1 \right)\} \\
        &+ \{\text{Pr}\left( R^r_{Y}=1|\textbf{ X}^r,A^r,Z^r_1,Z^r_2,R^r_{Z_2}=1,R^r_{Z_1}=0 \right) \\
        &\hspace{0.5cm} \times\text{Pr}\left( R^r_{Z_1}=0|\textbf{ X}^r,A^r,Z^r_1,Z^r_2,R^r_{Z_2}=1 \right)\} \\
        &= \text{Pr}\left( R^r_{Y}=1|\textbf{X}^r,A^r,Z^r_1,Z^r_2,R^r_{Z_2}=1,R^r_{Z_1}=1 \right)\times 1 \\
        & \hspace{0.5cm} + 0 \times 0 \\
        &= \text{Pr}\left( R^r_{Y}=1|\textbf{X}^r,A^r,Z^r_1,Z^r_2,R^r_{Z_2}=1,R^r_{Z_1}=1 \right).
    \end{split}
\end{equation*}
Thus, we have
\begin{equation*}
    \begin{split}
        \hat{p}^r_Y &= \text{Pr}\left( R^r_{Y}=1|\textbf{X}^r,A^r,Z^r_1,Z^r_2,R^r_{Z_2}=1,R^r_{Z_1}=1 \right)\times\text{Pr}\left( R^r_{Z_2}=1|\textbf{ X}^r,A^r,Z^r_1 \right) \\
        &=: \hat{p}^r_{Y|R^r_{Z_2}=1,R^r_{Z_1}=1}\times\hat{p}^r_{Z_2}.
    \end{split}
\end{equation*}

\subsection{MI Implementation -- Alternative Method}
\label{appendix:miimplementation}

Another strategy for incorporating MI in the sequential data generation framework as proposed in \cite{Petrakos2025} and harnessed in this work could be to impute, pool, and then generate. 
\footnote{Petrakos, N. Z., Moodie, E. E. M., and Savy, N. (2025). A framework for generating realistic synthetic tabular data in a randomized controlled trial setting. \textit{Statistics in Medicine}, 44(18-19):e70227.}
(Recall that in the main paper, we instead imputed, generated, and sampled.) For illustrative purposes, generating $Z_1$ is used as an example. First, $m$ data sets are imputed. A model regressing $Z^r_1$ on $\mathbf{X}^r,A^r$ is fit to each of the $m$ data sets resulting in $m$ sets of parameter estimates, $\{\hat{\boldsymbol{\beta}}_1,\hat{\alpha}_1\}, \{\hat{\boldsymbol{\beta}}_2,\hat{\alpha}_2\},...,\{\hat{\boldsymbol{\beta}}_m,\hat{\alpha}_m\}$, and model residuals. Then, the parameter estimates are pooled using Rubin's rules, and the pooled estimates, $\{\hat{\boldsymbol{\beta}}_p,\hat{\alpha}_p\}$, are used to predict $\hat{Z}^s_1$. Following the usual sequential steps, the final step would be to generate $Z^s_1$ by sampling from the set of admissible values, $\{ \left( \hat{Z}^s_1+r \right) \geq 0 \}$. Recall that this restriction to admissible values is required since $Z_1$ represents a count a count and hence must be non-negative.

However, there is no ``pooled'' version of model residuals. Potential solutions include letting $r$ be the collection of all $m$ model residuals, or to define $r=Z^r_1-\left( \hat{\boldsymbol{\beta}}_p\mathbf{X}^r+\hat{\alpha}_pA^r \right)$, restricted to the observed real data $\left( \mathbf{X}^r,A^r,Z^r_1 \right)$. A potential drawback of this strategy is that defining $r$ in either of these ways may not incorporate enough randomness in the data generation process and hence would lead to $Z^s_1$ exhibiting less variation than desired. The same applies for $Z^s_2$ and $Y^s$.

\section{Additional Results}

\subsection{Distributional Similarity Metrics: Univariate and Bivariate}

\subsubsection{Scenarios 2A and 2B (MCAR x 25\% Missing x Strong Mechanism)}

\begin{figure}[H]
    \centering
    \includegraphics[width=0.91\linewidth]{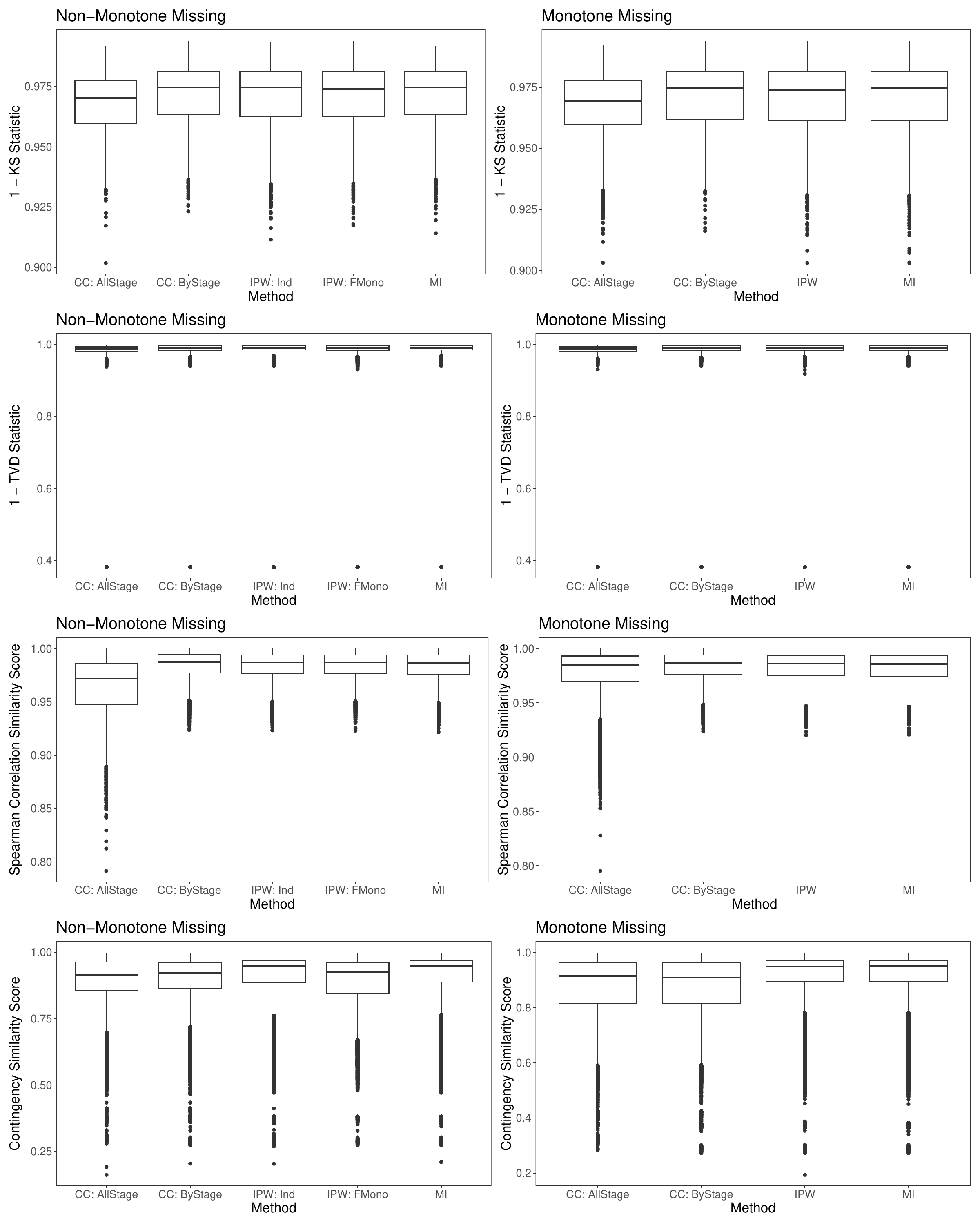}
    \caption{Plot of similarity metrics capturing univariate continuous (first row), univariate discrete (second row), bivariate continuous (third row), and bivariate discrete (fourth row) distributions, across all variables and all 1000 simulation runs for Scenarios 2A and 2B (MCAR x 25\% Missing x Strong Mechanism). These compare the observed synthetic data and observed real data (note: observed synthetic data is the same as complete synthetic data for the two CC methods because these do not generate synthetic missingness).}
\end{figure}

\subsubsection{Scenarios 3A and 3B (MAR x 10\% Missing x Strong Mechanism)}

\begin{figure}[H]
    \centering
    \includegraphics[width=0.91\linewidth]{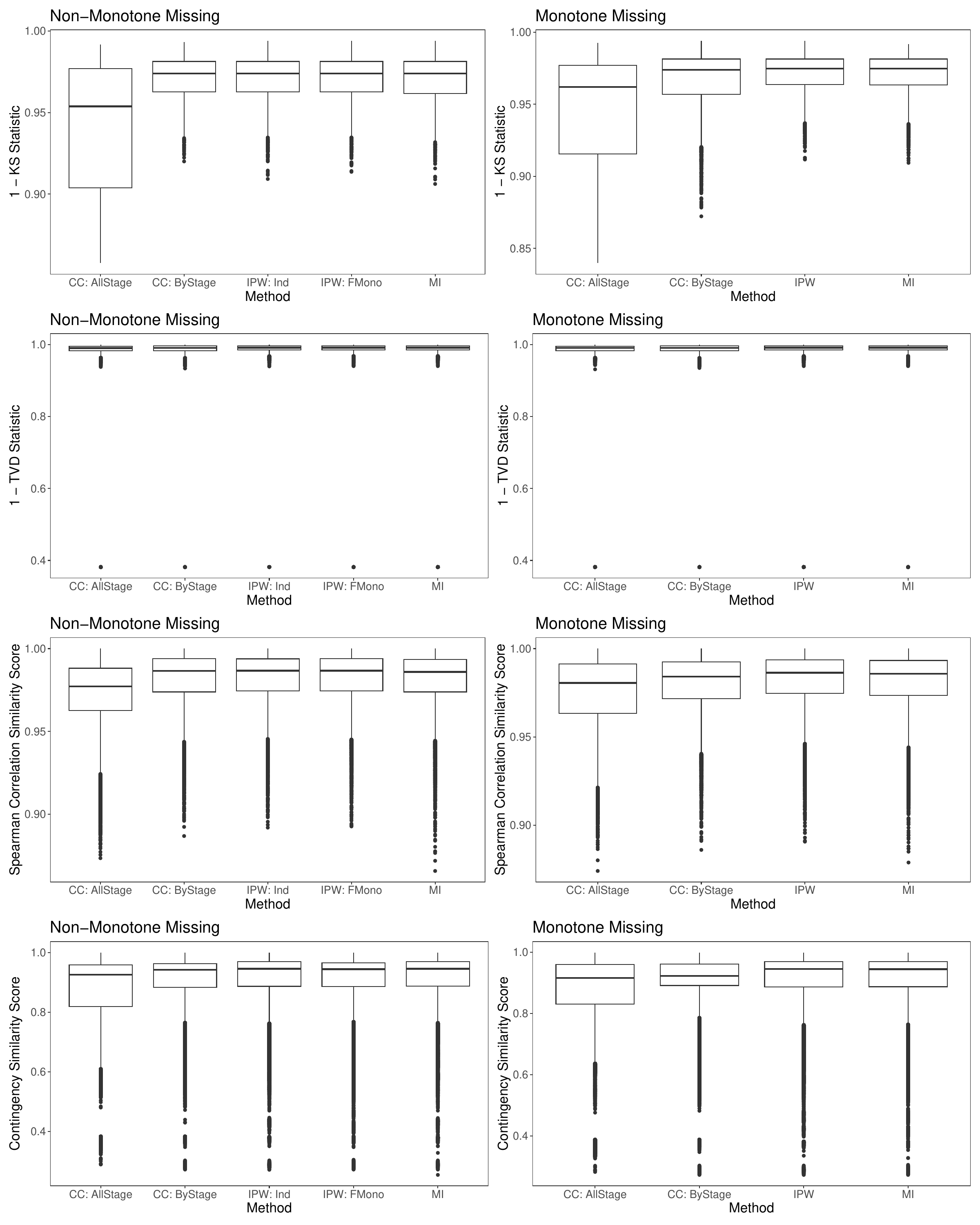}
    \caption{Plot of similarity metrics capturing univariate continuous (first row), univariate discrete (second row), bivariate continuous (third row), and bivariate discrete (fourth row) distributions, across all variables and all 1000 simulation runs for Scenarios 3A and 3B (MAR x 10\% Missing x Strong Mechanism). These compare the observed synthetic data and observed real data (note: observed synthetic data is the same as complete synthetic data for the two CC methods because these do not generate synthetic missingness).}
\end{figure}

\subsubsection{Scenarios 4A and 4B (MAR x 50\% Missing x Strong Mechanism)}

\begin{figure}[H]
    \centering
    \includegraphics[width=0.91\linewidth]{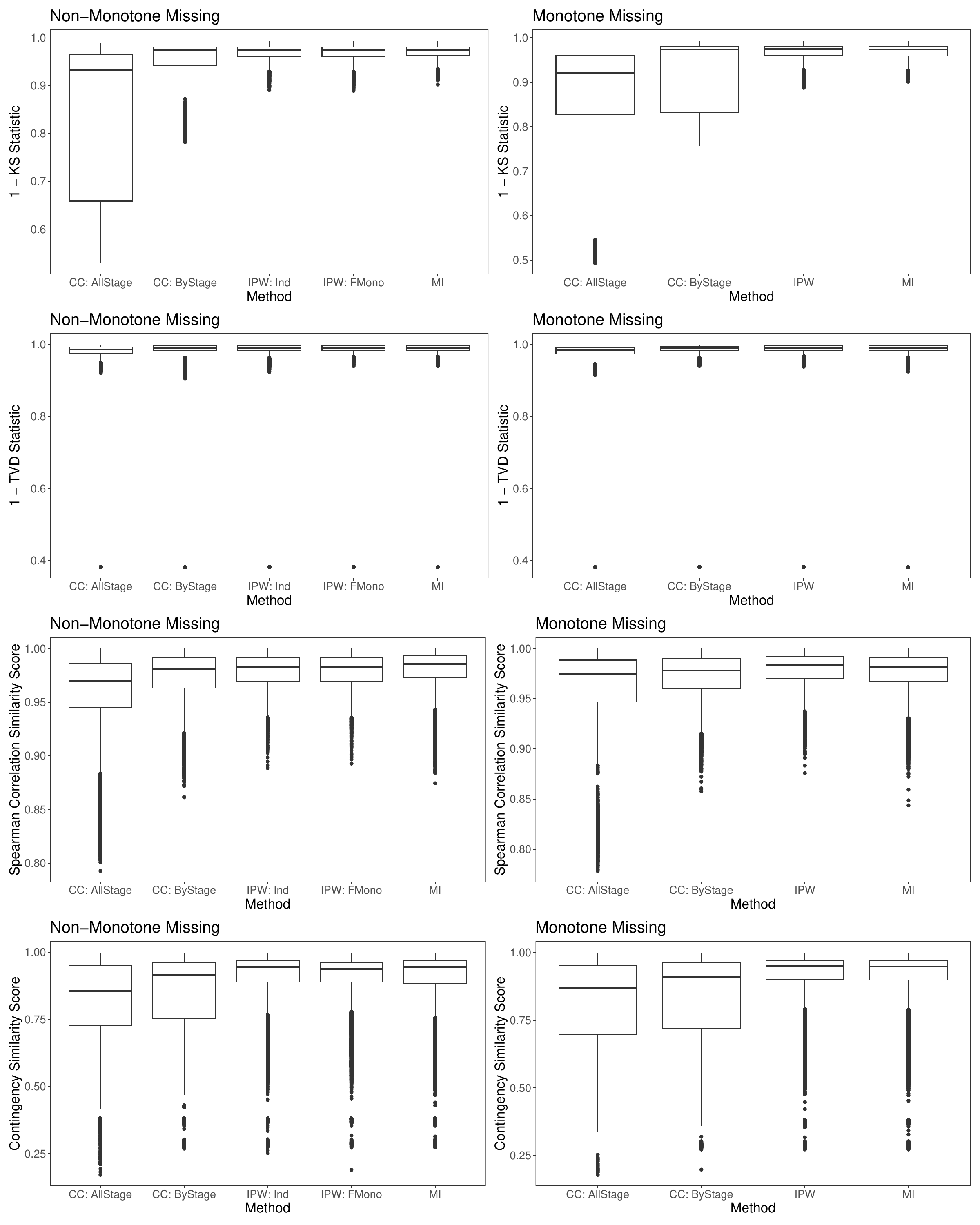}
    \caption{Plot of similarity metrics capturing univariate continuous (first row), univariate discrete (second row), bivariate continuous (third row), and bivariate discrete (fourth row) distributions, across all variables and all 1000 simulation runs for Scenarios 4A and 4B (MAR x 50\% Missing x Strong Mechanism). These compare the observed synthetic data and observed real data (note: observed synthetic data is the same as complete synthetic data for the two CC methods because these do not generate synthetic missingness).}
\end{figure}

\subsubsection{Scenarios 5A and 5B (MAR x 25\% Missing x Weak Mechanism)}

\begin{figure}[H]
    \centering
    \includegraphics[width=0.91\linewidth]{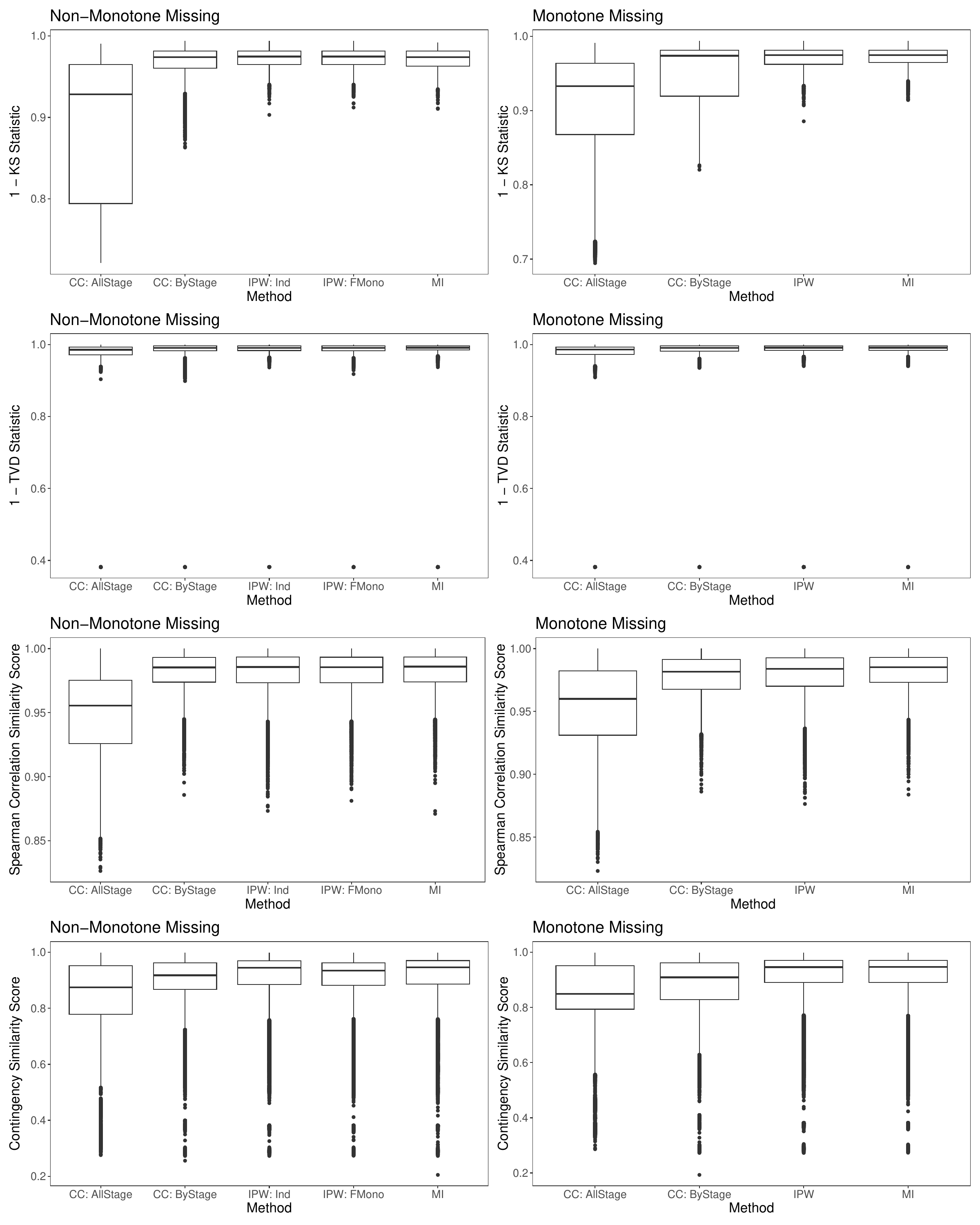}
    \caption{Plot of similarity metrics capturing univariate continuous (first row), univariate discrete (second row), bivariate continuous (third row), and bivariate discrete (fourth row) distributions, across all variables and all 1000 simulation runs for Scenarios 5A and 5B (MAR x 25\% Missing x Weak Mechanism). These compare the observed synthetic data and observed real data (note: observed synthetic data is the same as complete synthetic data for the two CC methods because these do not generate synthetic missingness).}
\end{figure}

\subsubsection{Scenarios 6A and 6B (MAR x 50\% Missing x Weak Mechanism)}

\begin{figure}[H]
    \centering
    \includegraphics[width=0.91\linewidth]{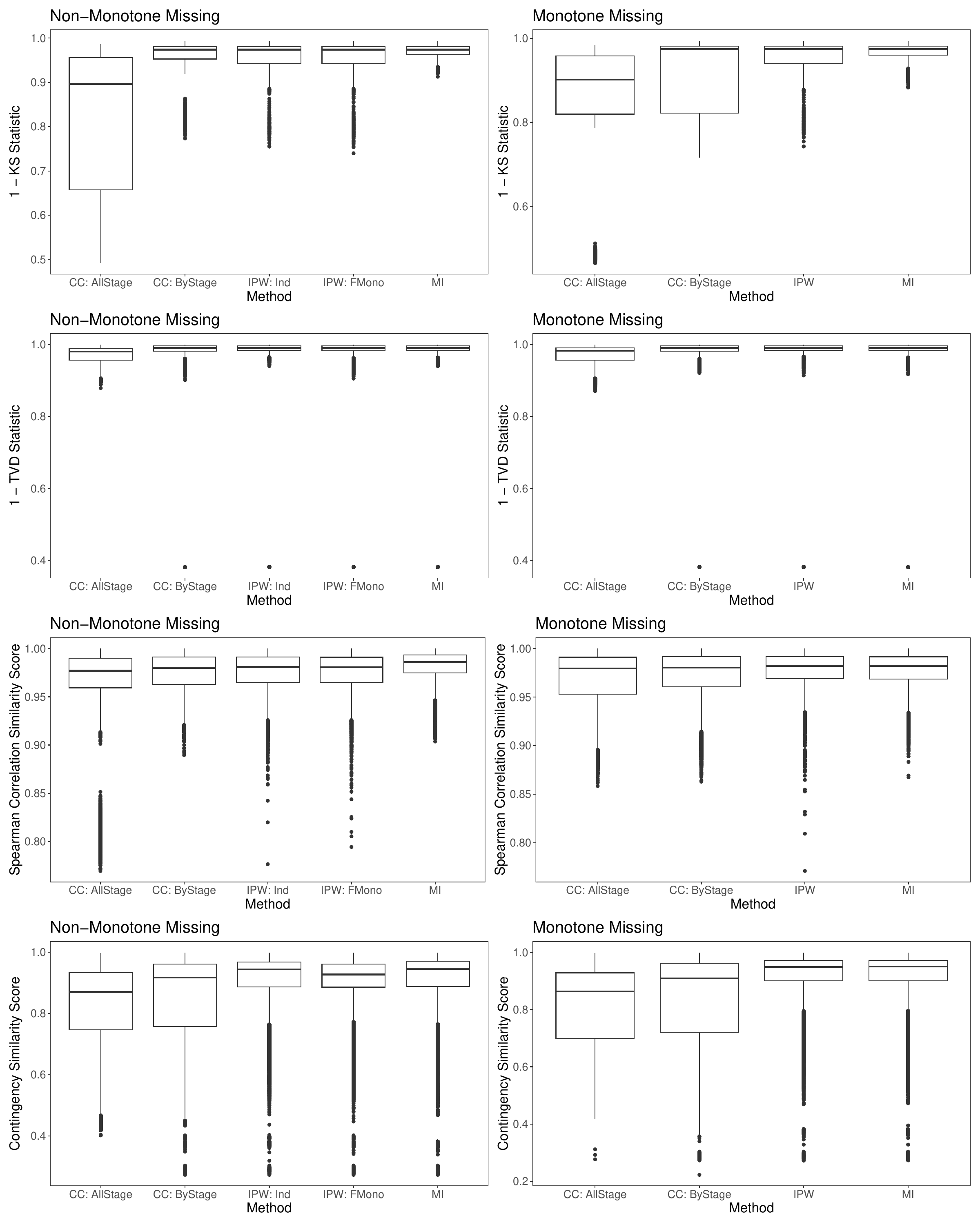}
    \caption{Plot of similarity metrics capturing univariate continuous (first row), univariate discrete (second row), bivariate continuous (third row), and bivariate discrete (fourth row) distributions, across all variables and all 1000 simulation runs for Scenarios 6A and 6B (MAR x 50\% Missing x Weak Mechanism). These compare the observed synthetic data and observed real data (note: observed synthetic data is the same as complete synthetic data for the two CC methods because these do not generate synthetic missingness).}
\end{figure}

\subsection{Multivariate Relationships: PCA Plots}

\subsubsection{Scenarios 2A and 2B (MCAR x 25\% Missing x Strong Mechanism)}

\begin{figure}[h!]
    \centering
    \includegraphics[width=0.91\linewidth]{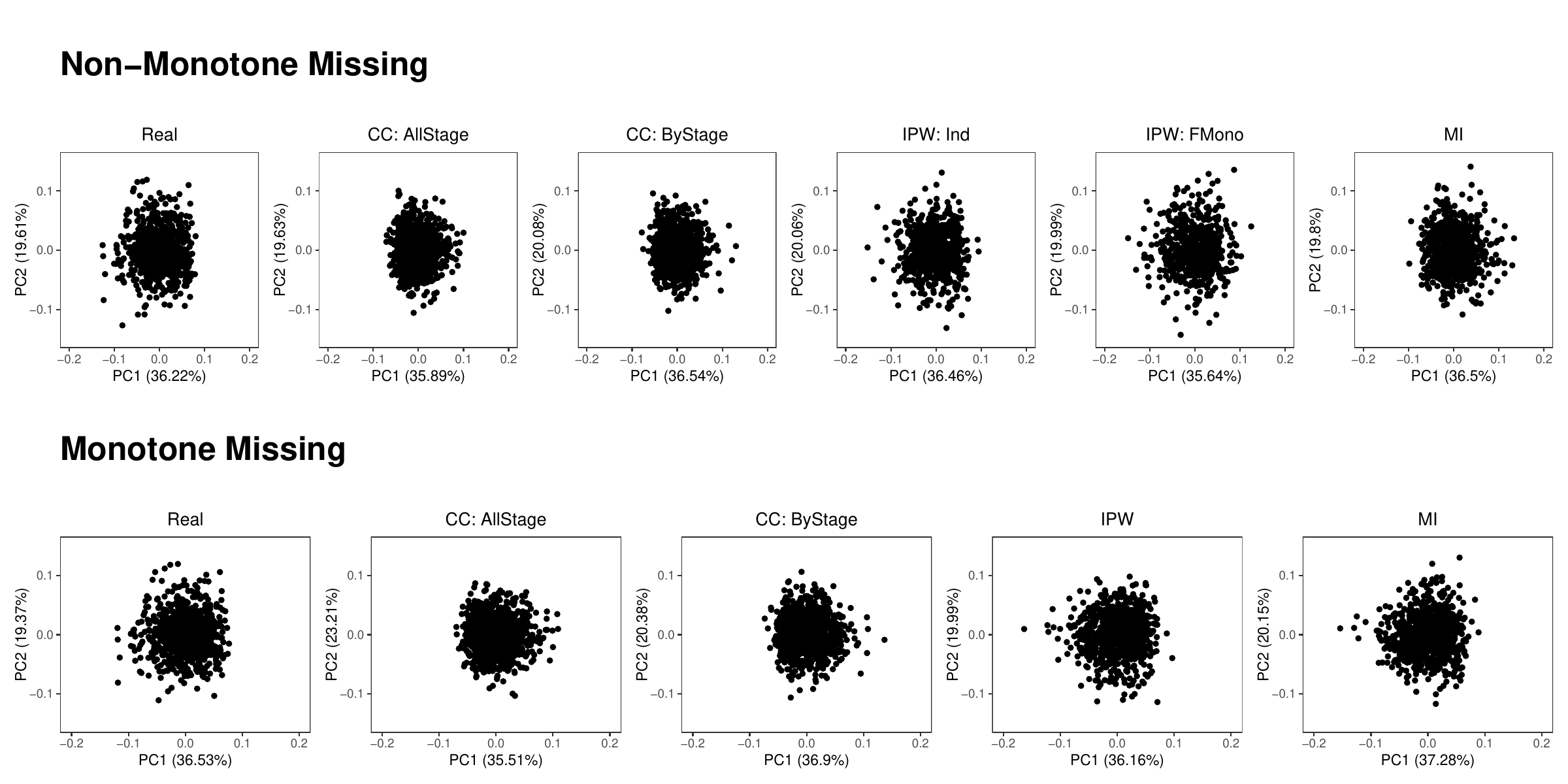}
    \caption{PCA plots of the first two principal components for the real data and for each synthetic data generation framework, taken from one simulation run chosen at random under Scenarios 2A (top row) and 2B (bottom row). The horizontal axis represents the first principal component and the vertical axis represents the second principal component; the proportion of variation from the real data captured by the principal component in indicated along each axis.}
\end{figure}

\subsubsection{Scenarios 3A and 3B (MAR x 10\% Missing x Strong Mechanism)}

\begin{figure}[H]
    \centering
    \includegraphics[width=0.91\linewidth]{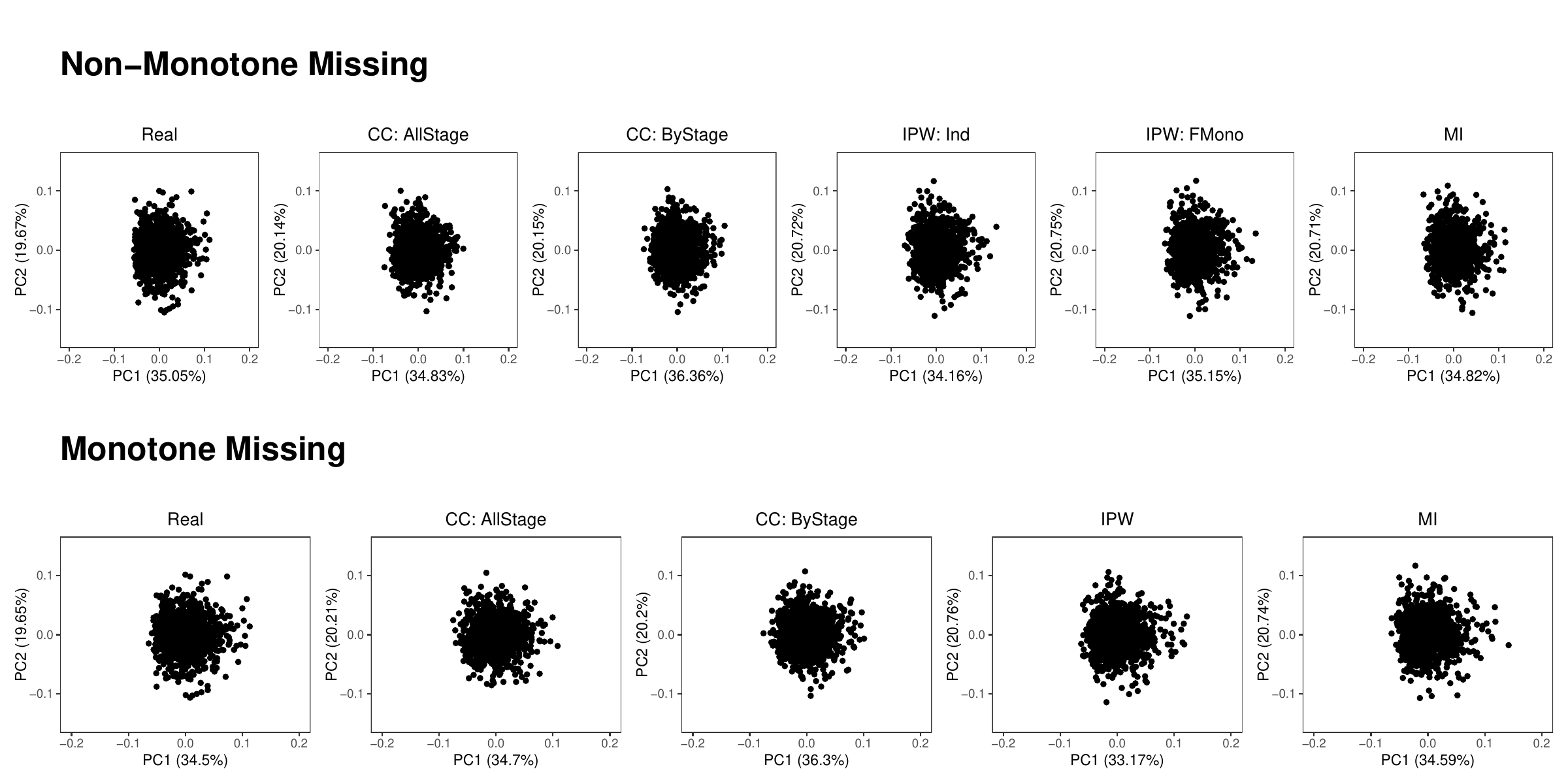}
    \caption{PCA plots of the first two principal components for the real data and for each synthetic data generation framework, taken from one simulation run chosen at random under Scenarios 3A (top row) and 3B (bottom row). The horizontal axis represents the first principal component and the vertical axis represents the second principal component; the proportion of variation from the real data captured by the principal component in indicated along each axis.}
\end{figure}

\subsubsection{Scenarios 4A and 4B (MAR x 50\% Missing x Strong Mechanism)}

\begin{figure}[H]
    \centering
    \includegraphics[width=0.91\linewidth]{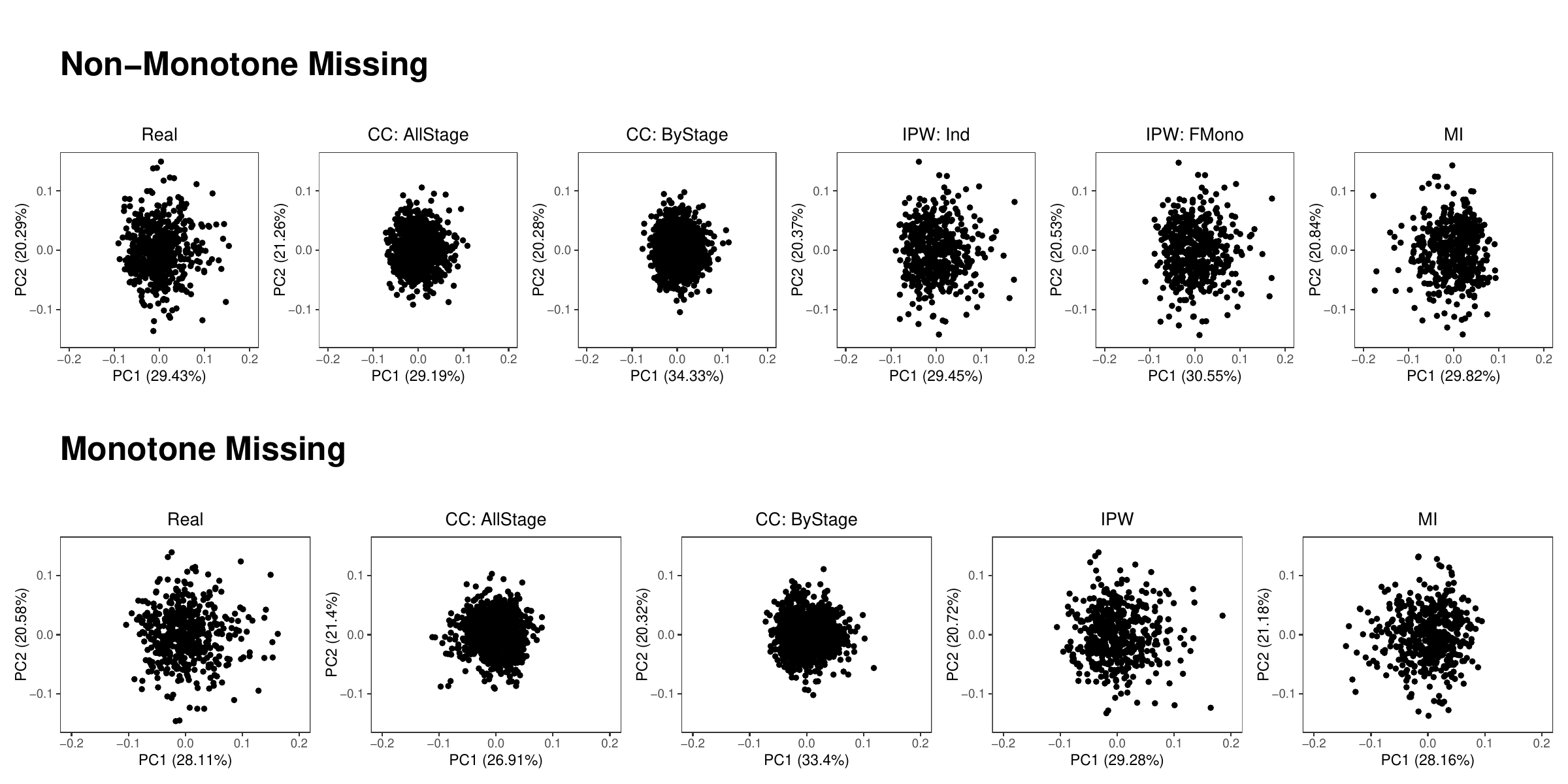}
    \caption{PCA plots of the first two principal components for the real data and for each synthetic data generation framework, taken from one simulation run chosen at random under Scenarios 4A (top row) and 4B (bottom row). The horizontal axis represents the first principal component and the vertical axis represents the second principal component; the proportion of variation from the real data captured by the principal component in indicated along each axis.}
\end{figure}

\subsubsection{Scenarios 5A and 5B (MAR x 25\% Missing x Weak Mechanism)}

\begin{figure}[H]
    \centering
    \includegraphics[width=0.91\linewidth]{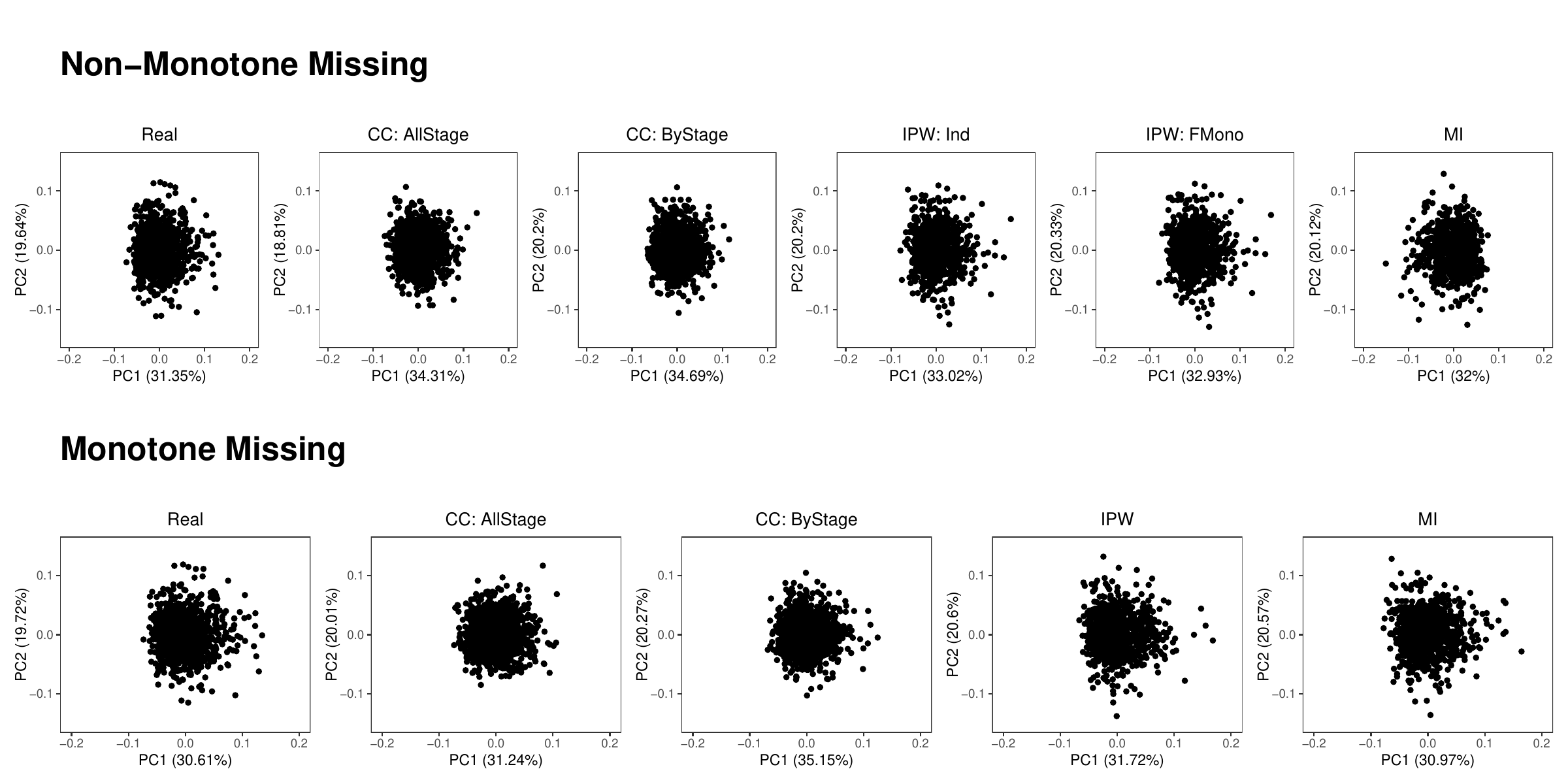}
    \caption{PCA plots of the first two principal components for the real data and for each synthetic data generation framework, taken from one simulation run chosen at random under Scenarios 5A (top row) and 5B (bottom row). The horizontal axis represents the first principal component and the vertical axis represents the second principal component; the proportion of variation from the real data captured by the principal component in indicated along each axis.}
\end{figure}

\subsubsection{Scenarios 6A and 6B (MAR x 50\% Missing x Weak Mechanism)}

\FloatBarrier
\begin{figure}[H]
    \centering
    \includegraphics[width=0.91\linewidth]{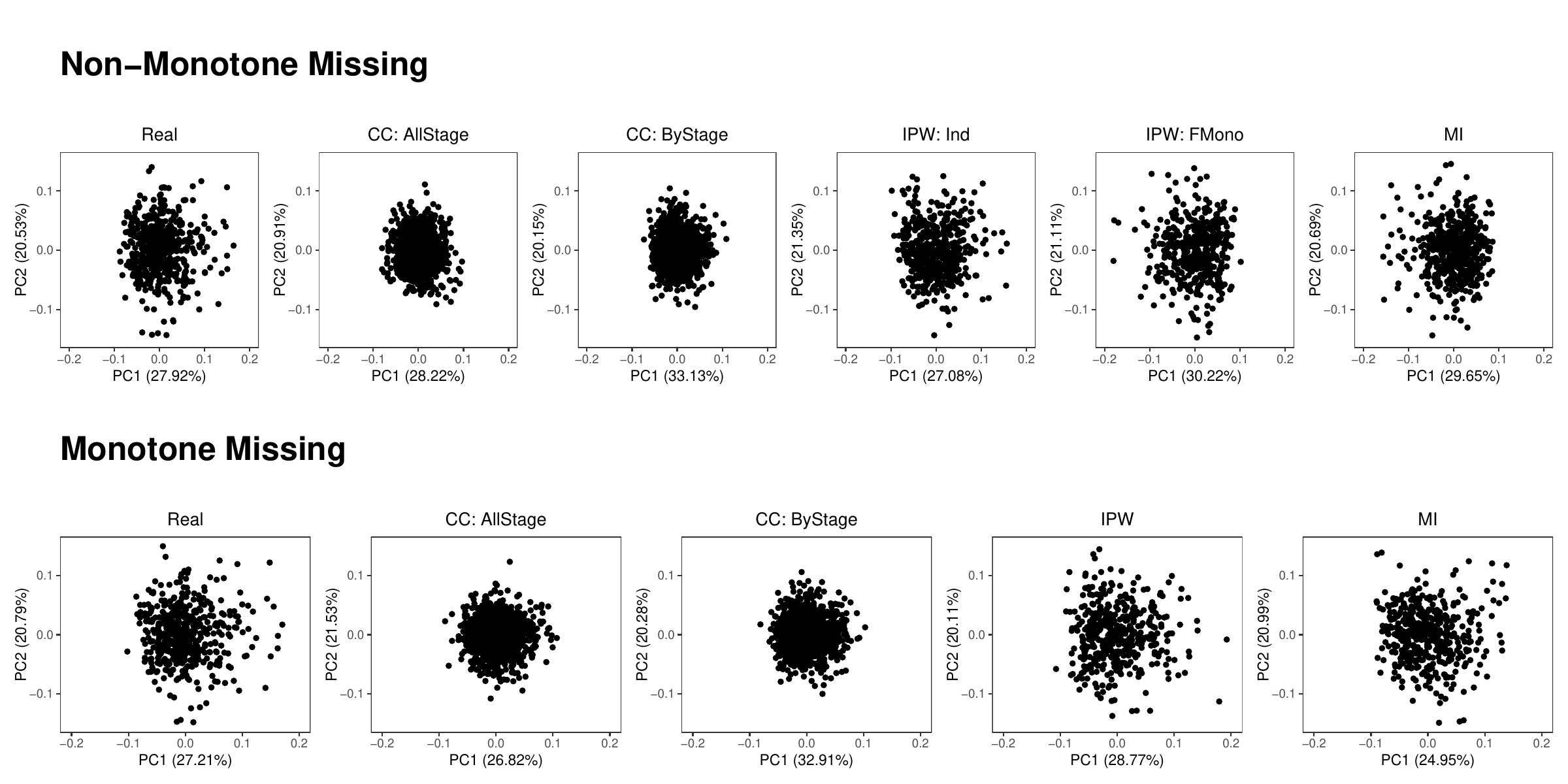}
    \caption{PCA plots of the first two principal components for the real data and for each synthetic data generation framework, taken from one simulation run chosen at random under Scenarios 6A (top row) and 6B (bottom row). The horizontal axis represents the first principal component and the vertical axis represents the second principal component; the proportion of variation from the real data captured by the principal component in indicated along each axis.}
\end{figure}

\subsection{ML Efficacy Metrics}

\subsubsection{Scenarios 2A and 2B (MCAR x 25\% Missing x Strong Mechanism)}

\FloatBarrier
\begin{figure}[H]
    \centering
    \includegraphics[width=0.85\linewidth]{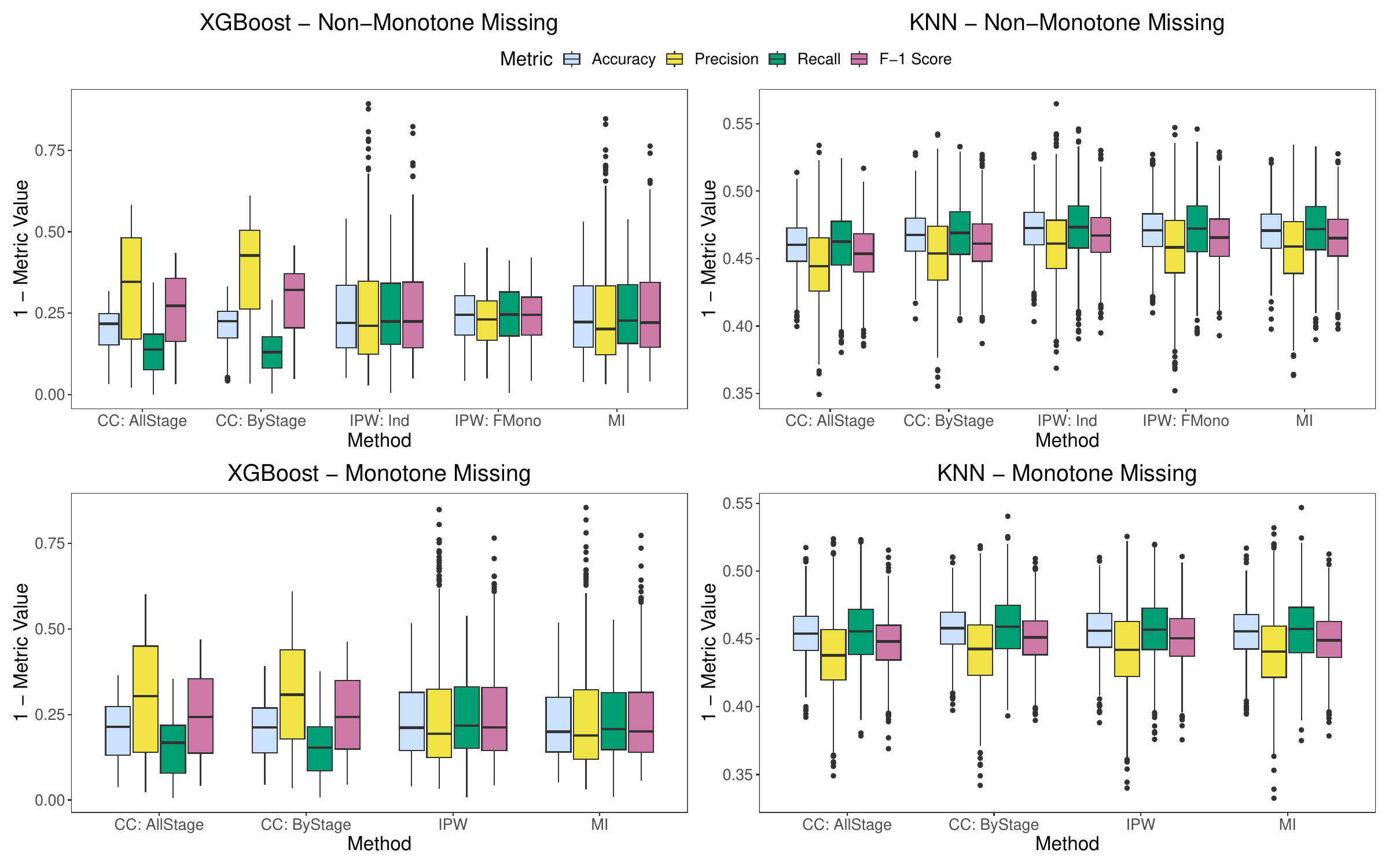}
    \caption{Plot of ML efficacy metrics (accuracy, precision, recall, and F-1 score) for both the XGBoost classifier and KNN classifier across all frameworks and 1000 simulation runs, for Scenarios 2A and 2B (MCAR x 25\% Missing x Strong Mechanism). The classification task was to correctly identify whether a row was from the real data or the synthetic data.}
\end{figure}

\subsubsection{Scenarios 3A and 3B (MAR x 10\% Missing x Strong Mechanism)}

\FloatBarrier
\begin{figure}[H]
    \centering
    \includegraphics[width=0.85\linewidth]{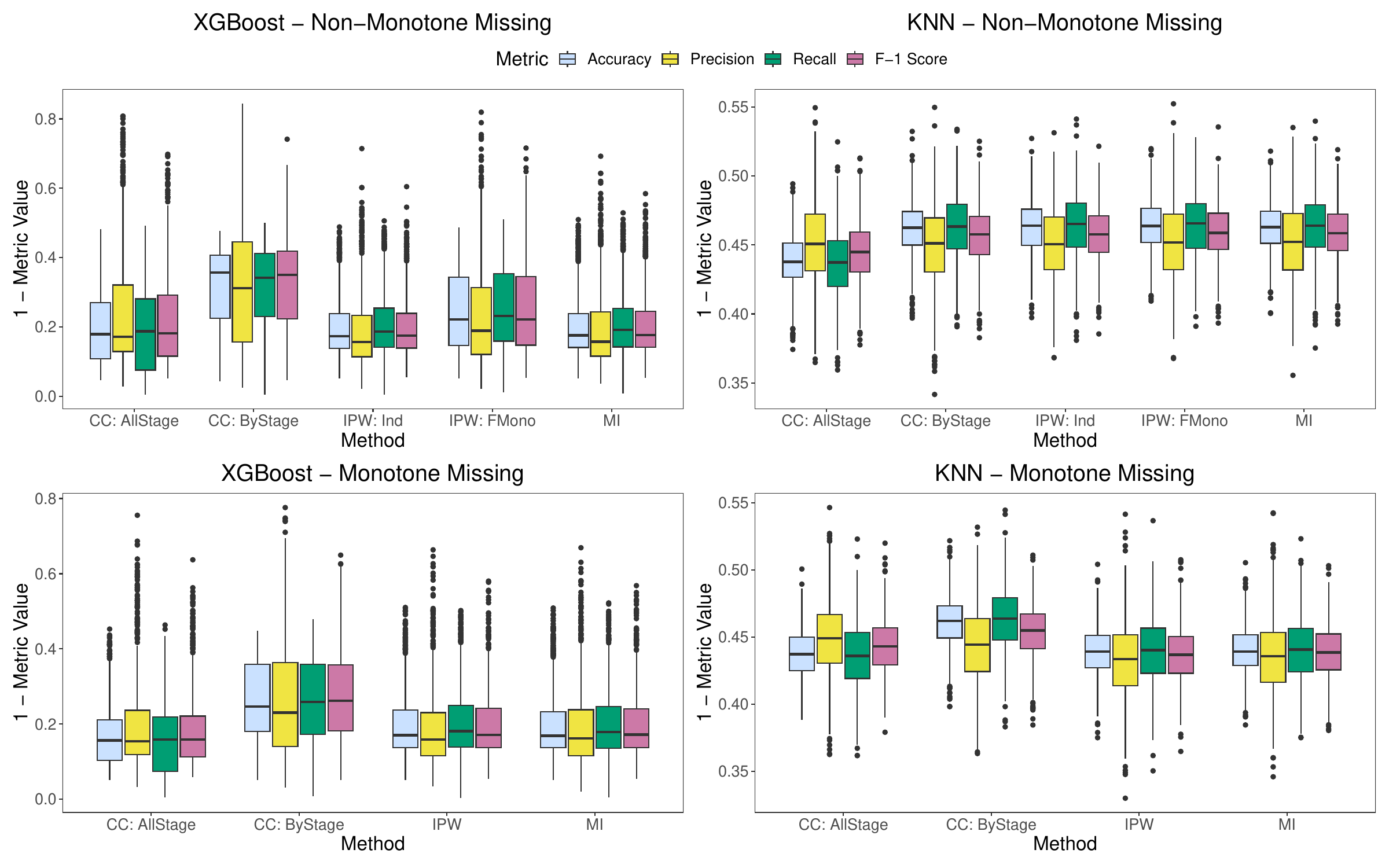}
    \caption{Plot of ML efficacy metrics (accuracy, precision, recall, and F-1 score) for both the XGBoost classifier and KNN classifier across all frameworks and 1000 simulation runs, for Scenarios 3A and 3B (MAR x 10\% Missing x Strong Mechanism). The classification task was to correctly identify whether a row was from the real data or the synthetic data.}
\end{figure}

\subsubsection{Scenarios 4A and 4B (MAR x 50\% Missing x Strong Mechanism)}

\FloatBarrier
\begin{figure}[H]
    \centering
    \includegraphics[width=0.85\linewidth]{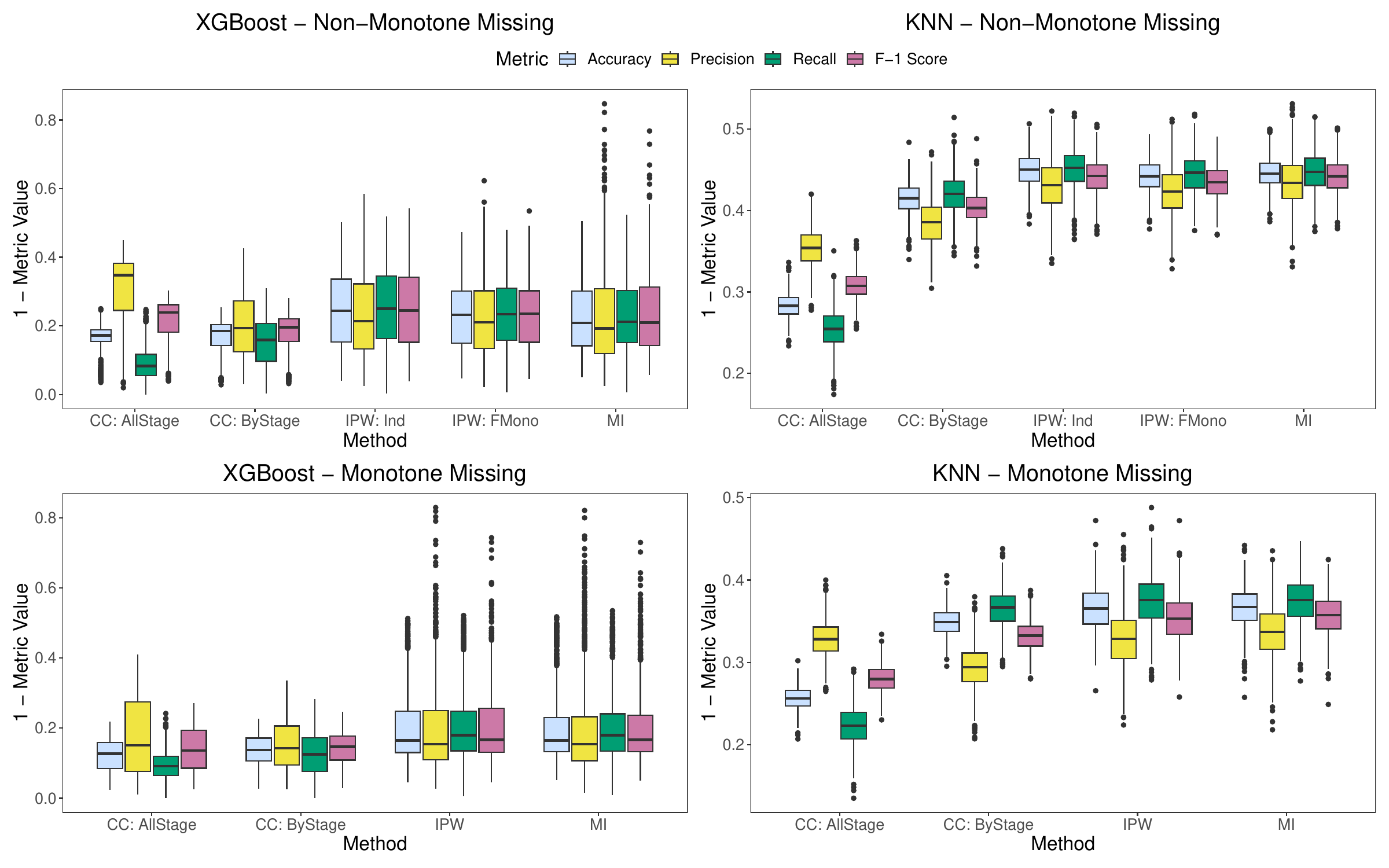}
    \caption{Plot of ML efficacy metrics (accuracy, precision, recall, and F-1 score) for both the XGBoost classifier and KNN classifier across all frameworks and 1000 simulation runs, for Scenarios 4A and 4B (MAR x 50\% Missing x Strong Mechanism). The classification task was to correctly identify whether a row was from the real data or the synthetic data.}
\end{figure}

\subsubsection{Scenarios 5A and 5B (MAR x 25\% Missing x Weak Mechanism)}

\FloatBarrier
\begin{figure}[H]
    \centering
    \includegraphics[width=0.85\linewidth]{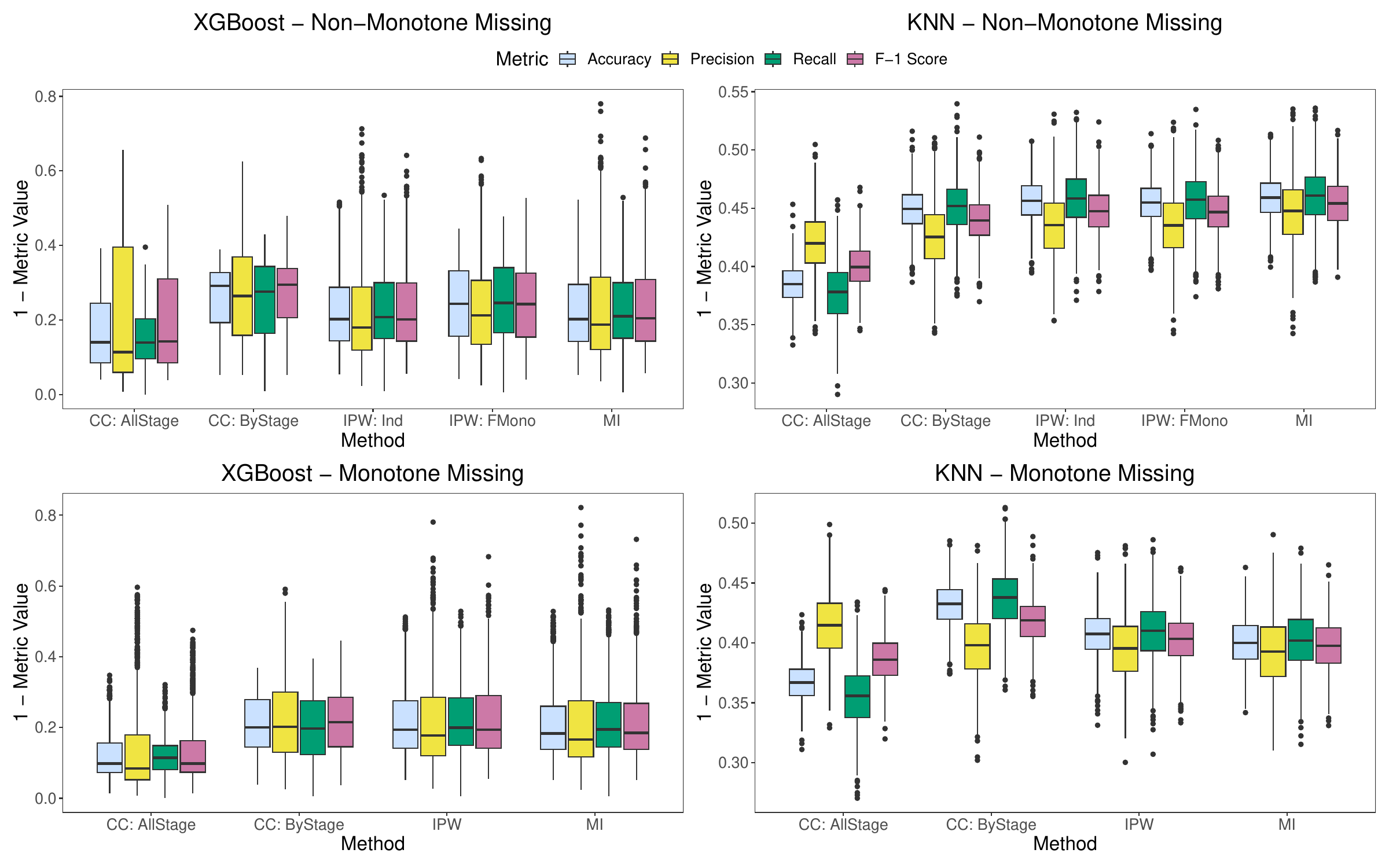}
    \caption{Plot of ML efficacy metrics (accuracy, precision, recall, and F-1 score) for both the XGBoost classifier and KNN classifier across all frameworks and 1000 simulation runs, for Scenarios 5A and 5B (MAR x 25\% Missing x Weak Mechanism). The classification task was to correctly identify whether a row was from the real data or the synthetic data.}
\end{figure}

\subsubsection{Scenarios 6A and 6B (MAR x 50\% Missing x Weak Mechanism)}

\FloatBarrier
\begin{figure}[H]
    \centering
    \includegraphics[width=0.85\linewidth]{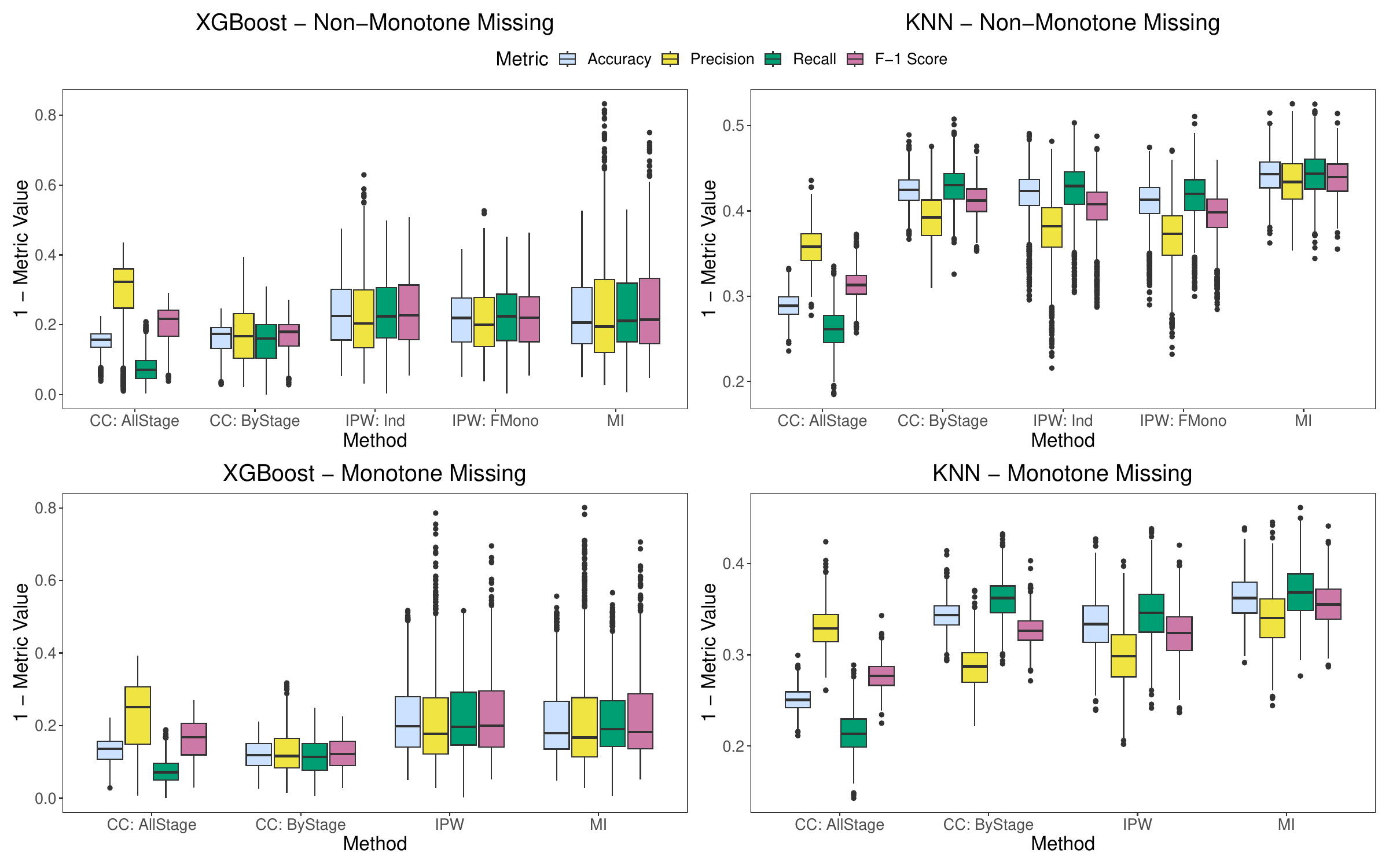}
    \caption{Plot of ML efficacy metrics (accuracy, precision, recall, and F-1 score) for both the XGBoost classifier and KNN classifier across all frameworks and 1000 simulation runs, for Scenarios 6A and 6B (MAR x 50\% Missing x Weak Mechanism). The classification task was to correctly identify whether a row was from the real data or the synthetic data.}
\end{figure}

\subsection{Trial Inference - Complete Data Comparisons}

\subsubsection{Scenarios 2A and 2B (MCAR x 25\% Missing x Strong Mechanism)}

\begin{figure}[H]
    \centering
    \includegraphics[width=0.95\linewidth]{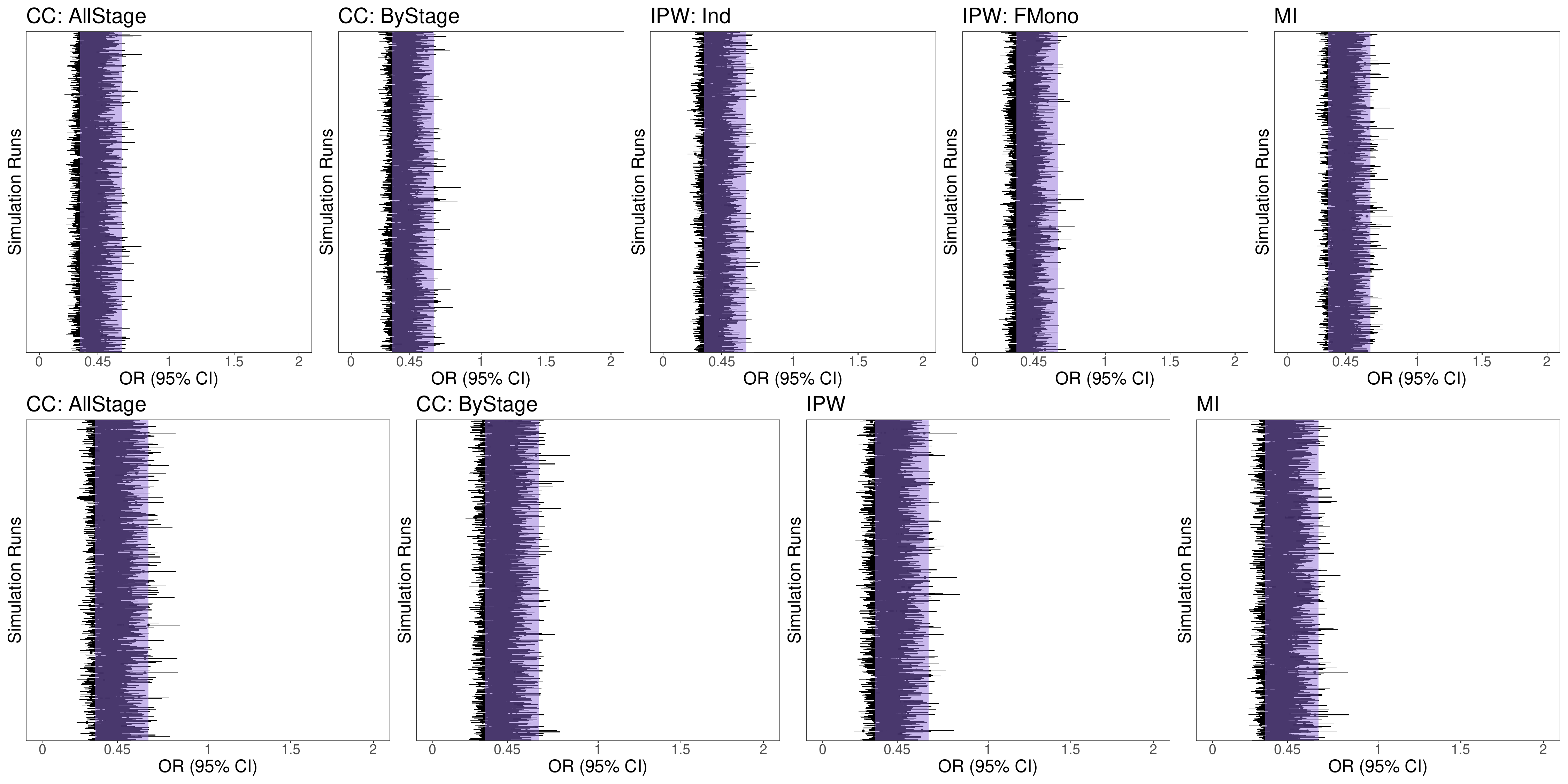}
    \caption{Plot of trial inference results utilizing complete real data and complete synthetic data, for all frameworks under both non-monotone missingness (Scenario 2A, top row) and monotone missingness (Scenario 2B, bottom row). The real (complete) data OR estimate was approximately 0.45, and the real (complete) data 95\% CI is shown by the purple shaded region. Each horizontal bar represents the synthetic data 95\% CI for one simulation run.}
\end{figure}

\subsubsection{Scenarios 3A and 3B (MAR x 10\% Missing x Strong Mechanism)}

\begin{figure}[H]
    \centering
    \includegraphics[width=0.95\linewidth]{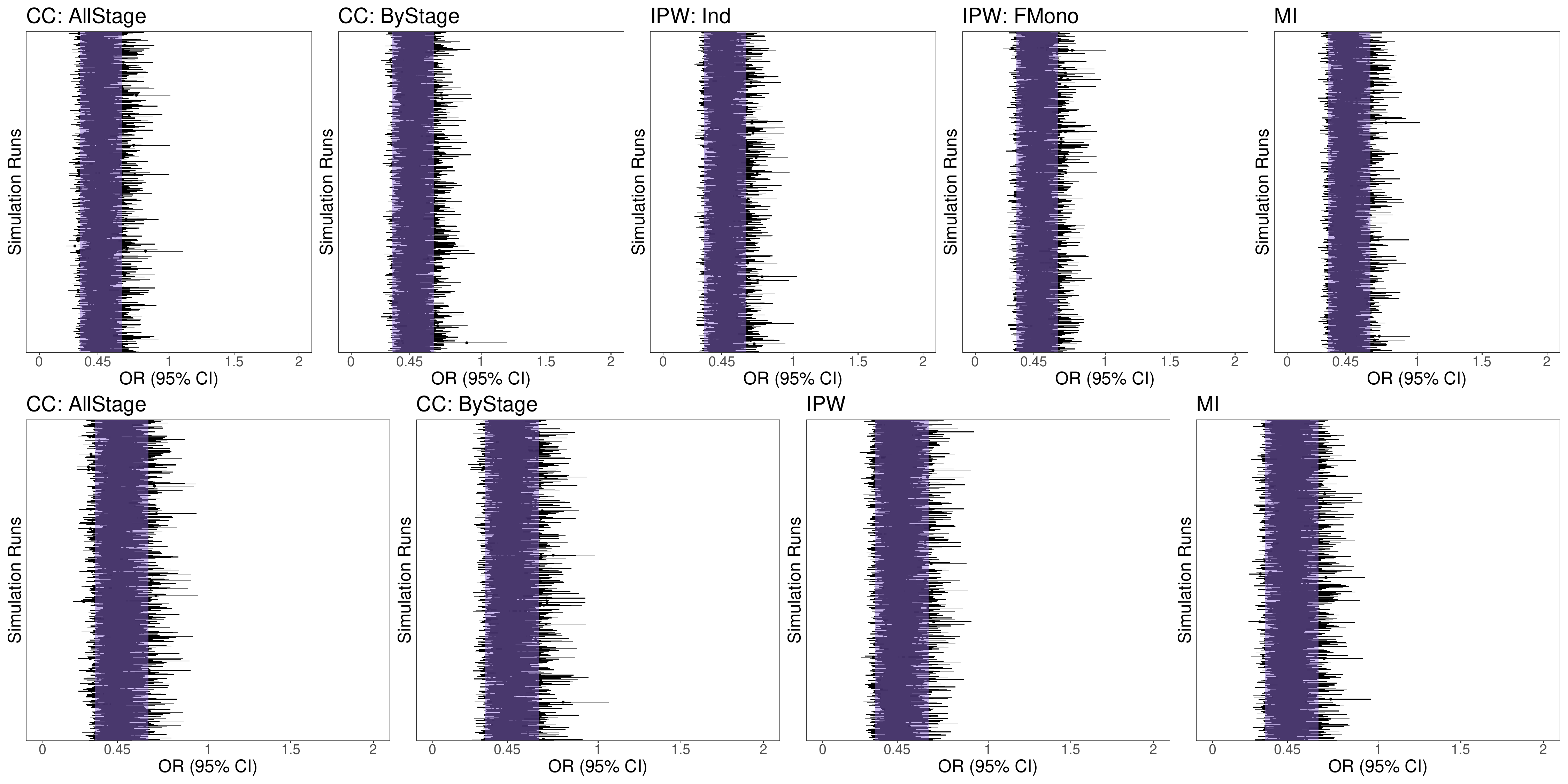}
    \caption{Plot of trial inference results utilizing complete real data and complete synthetic data, for all frameworks under both non-monotone missingness (Scenario 3A, top row) and monotone missingness (Scenario 3B, bottom row). The real (complete) data OR estimate was approximately 0.45, and the real (complete) data 95\% CI is shown by the purple shaded region. Each horizontal bar represents the synthetic data 95\% CI for one simulation run.}
\end{figure}

\subsubsection{Scenarios 4A and 4B (MAR x 50\% Missing x Strong Mechanism)}

\begin{figure}[H]
    \centering
    \includegraphics[width=\linewidth]{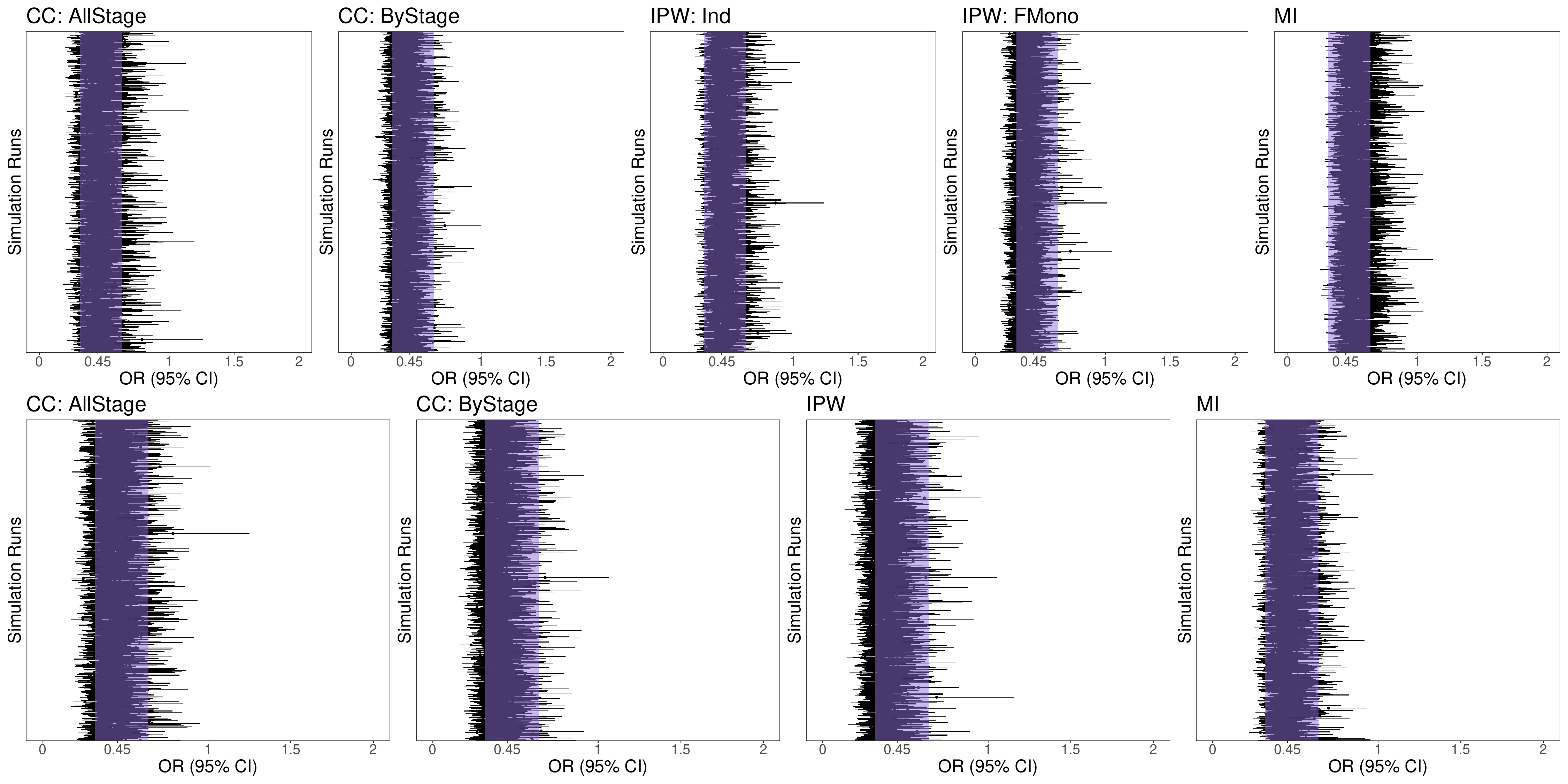}
    \caption{Plot of trial inference results utilizing complete real data and complete synthetic data, for all frameworks under both non-monotone missingness (Scenario 4A, top row) and monotone missingness (Scenario 4B, bottom row). The real (complete) data OR estimate was approximately 0.45, and the real (complete) data 95\% CI is shown by the purple shaded region. Each horizontal bar represents the synthetic data 95\% CI for one simulation run.}
\end{figure}

\subsubsection{Scenarios 5A and 5B (MAR x 25\% Missing x Weak Mechanism)}

\begin{figure}[H]
    \centering
    \includegraphics[width=\linewidth]{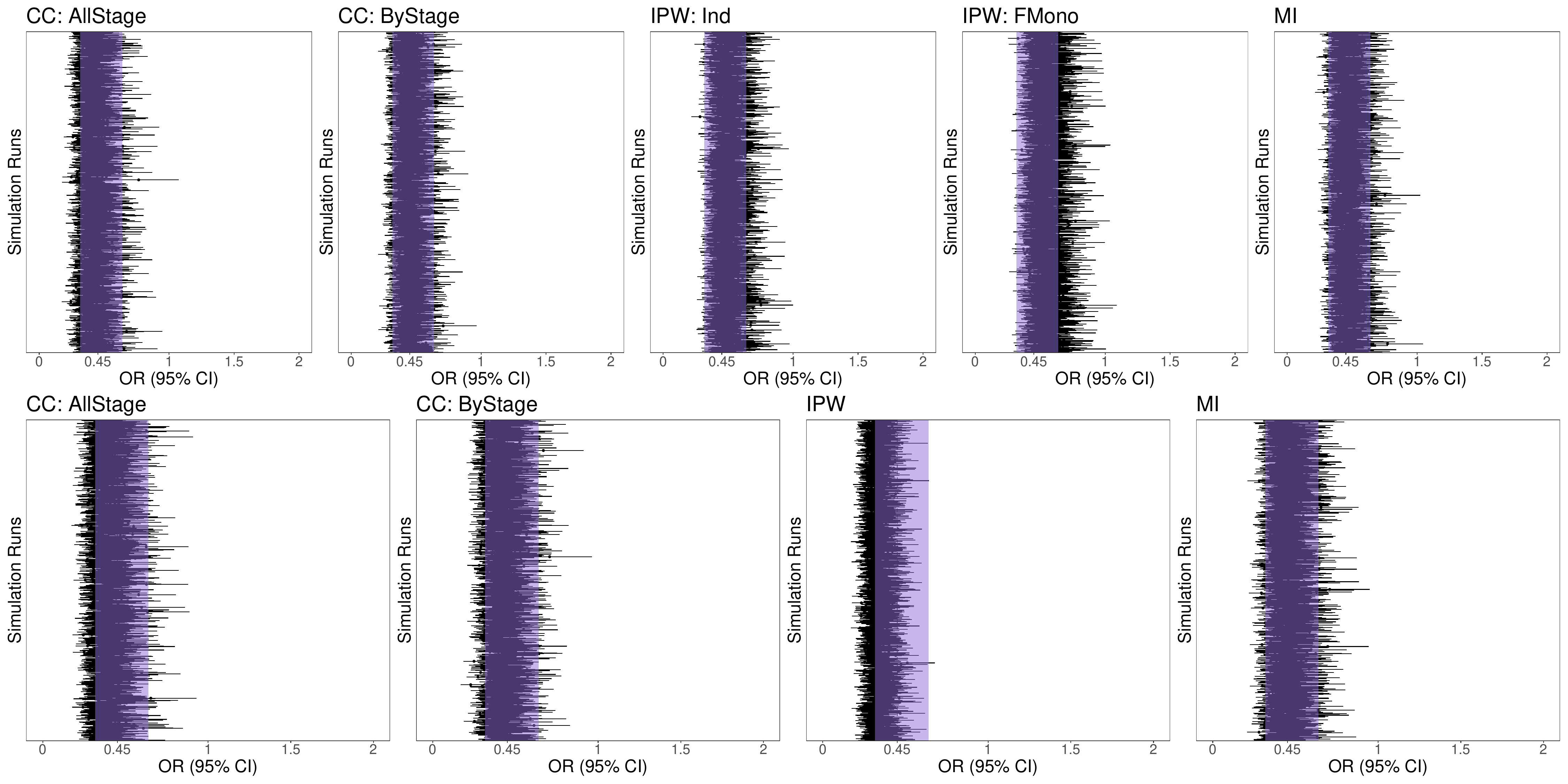}
    \caption{Plot of trial inference results utilizing complete real data and complete synthetic data, for all frameworks under both non-monotone missingness (Scenario 5A, top row) and monotone missingness (Scenario 5B, bottom row). The real (complete) data OR estimate was approximately 0.45, and the real (complete) data 95\% CI is shown by the purple shaded region. Each horizontal bar represents the synthetic data 95\% CI for one simulation run.}
\end{figure}

\subsubsection{Scenarios 6A and 6B (MAR x 50\% Missing x Weak Mechanism)}

\begin{figure}[H]
    \centering
    \includegraphics[width=\linewidth]{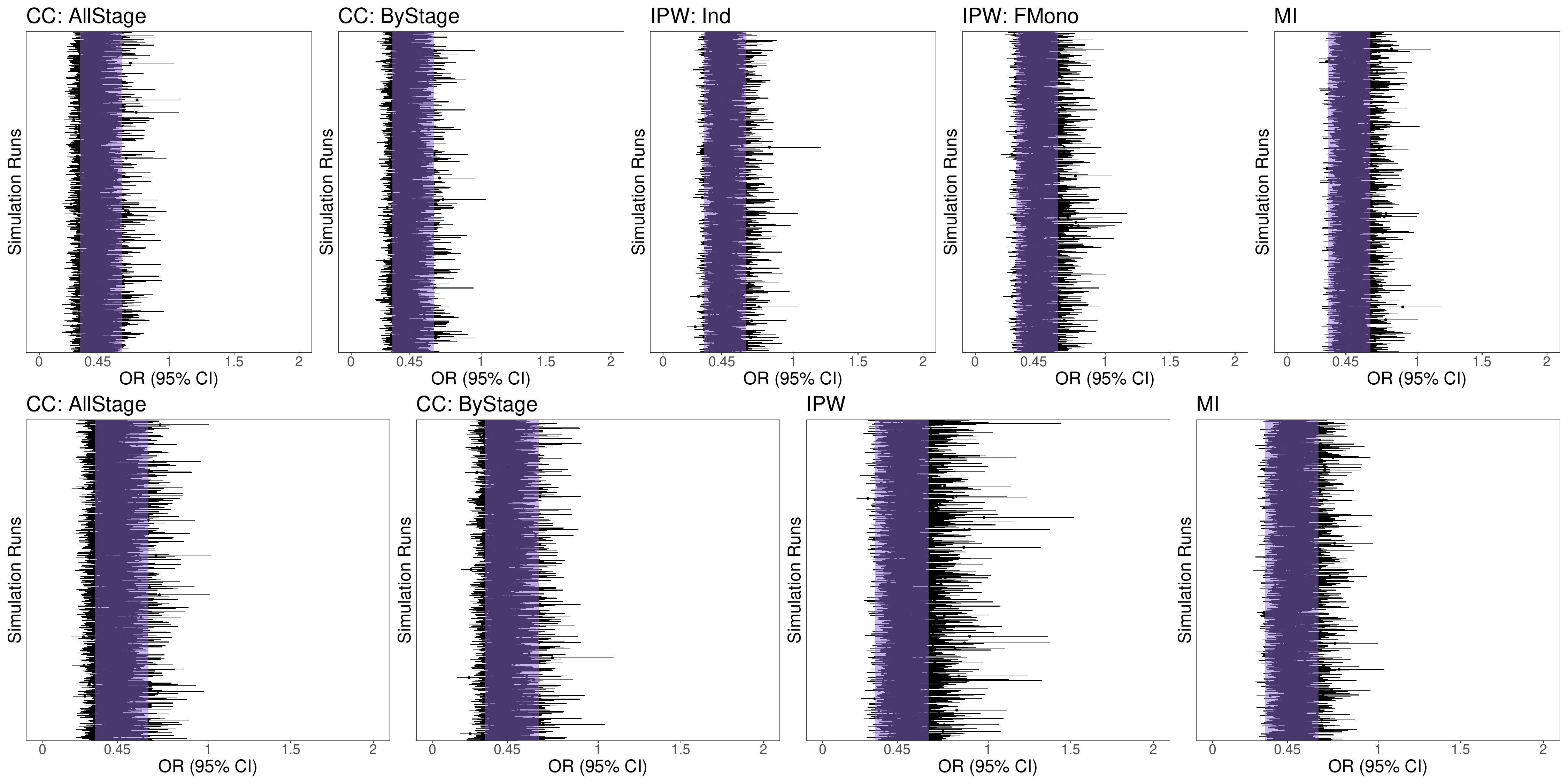}
    \caption{Plot of trial inference results utilizing complete real data and complete synthetic data, for all frameworks under both non-monotone missingness (Scenario 6A, top row) and monotone missingness (Scenario 6B, bottom row). The real (complete) data OR estimate was approximately 0.45, and the real (complete) data 95\% CI is shown by the purple shaded region. Each horizontal bar represents the synthetic data 95\% CI for one simulation run.}
\end{figure}

\newpage

\subsection{Trial Inference - Observed Data Comparisons}

\subsubsection{Scenarios 1A and 1B (MAR x 25\% Missing x Strong Mechanism)}

\begin{figure}[H]
    \centering
    \includegraphics[width=0.95\linewidth]{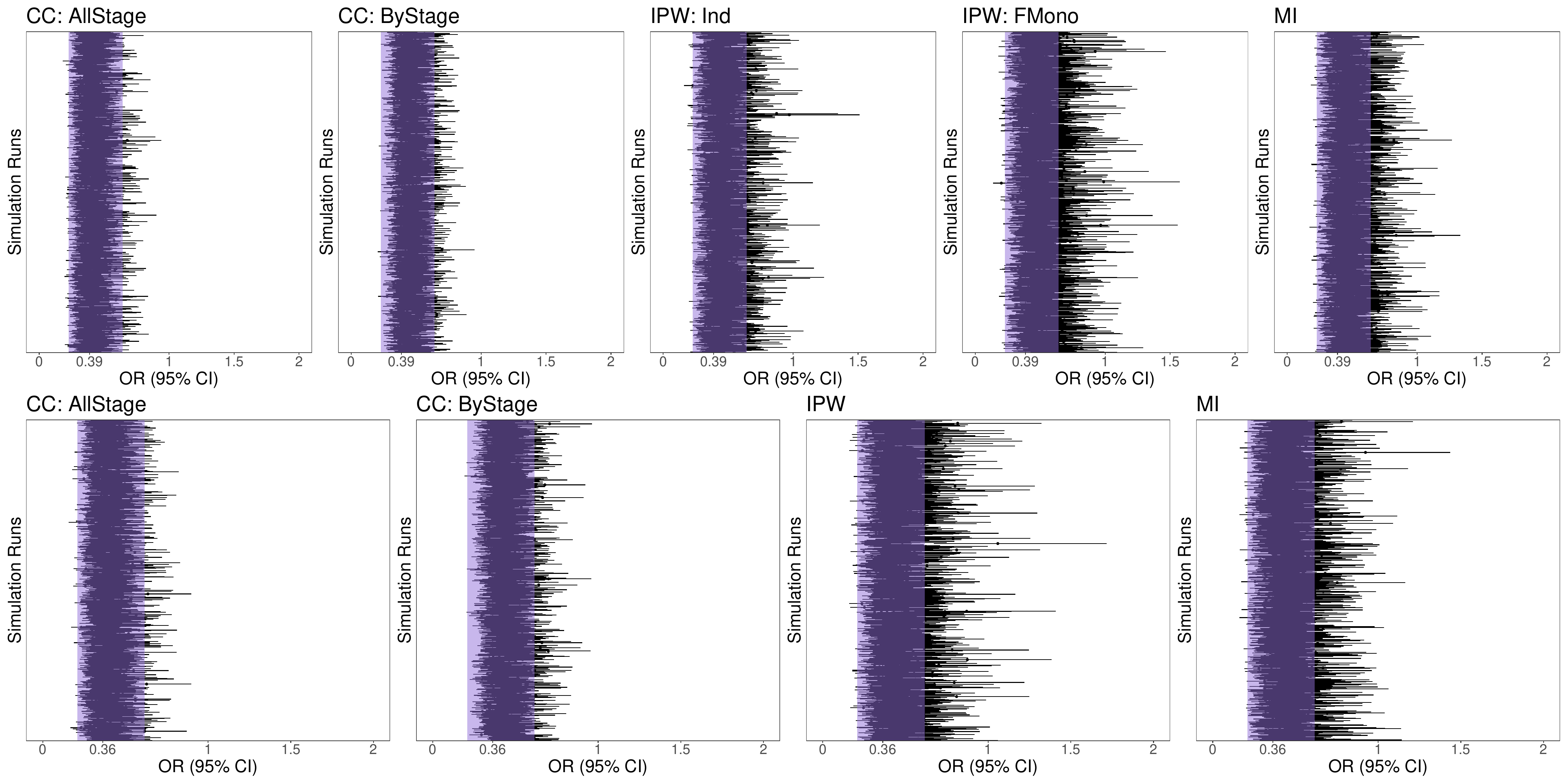}
    \caption{Plot of trial inference results utilizing observed real data and observed synthetic data, for all frameworks under both non-monotone missingness (Scenario 1A, top row) and monotone missingness (Scenario 1B, bottom row). The real (observed) data OR estimate was 0.39 in Scenario 1A and 0.36 in Scenario 1B, and the real (observed) data 95\% CIs are shown by the purple shaded regions. Each horizontal bar represents the synthetic data 95\% CI for one simulation run.}
\end{figure}

\subsubsection{Scenarios 2A and 2B (MCAR x 25\% Missing x Strong Mechanism)}

\begin{figure}[H]
    \centering
    \includegraphics[width=0.95\linewidth]{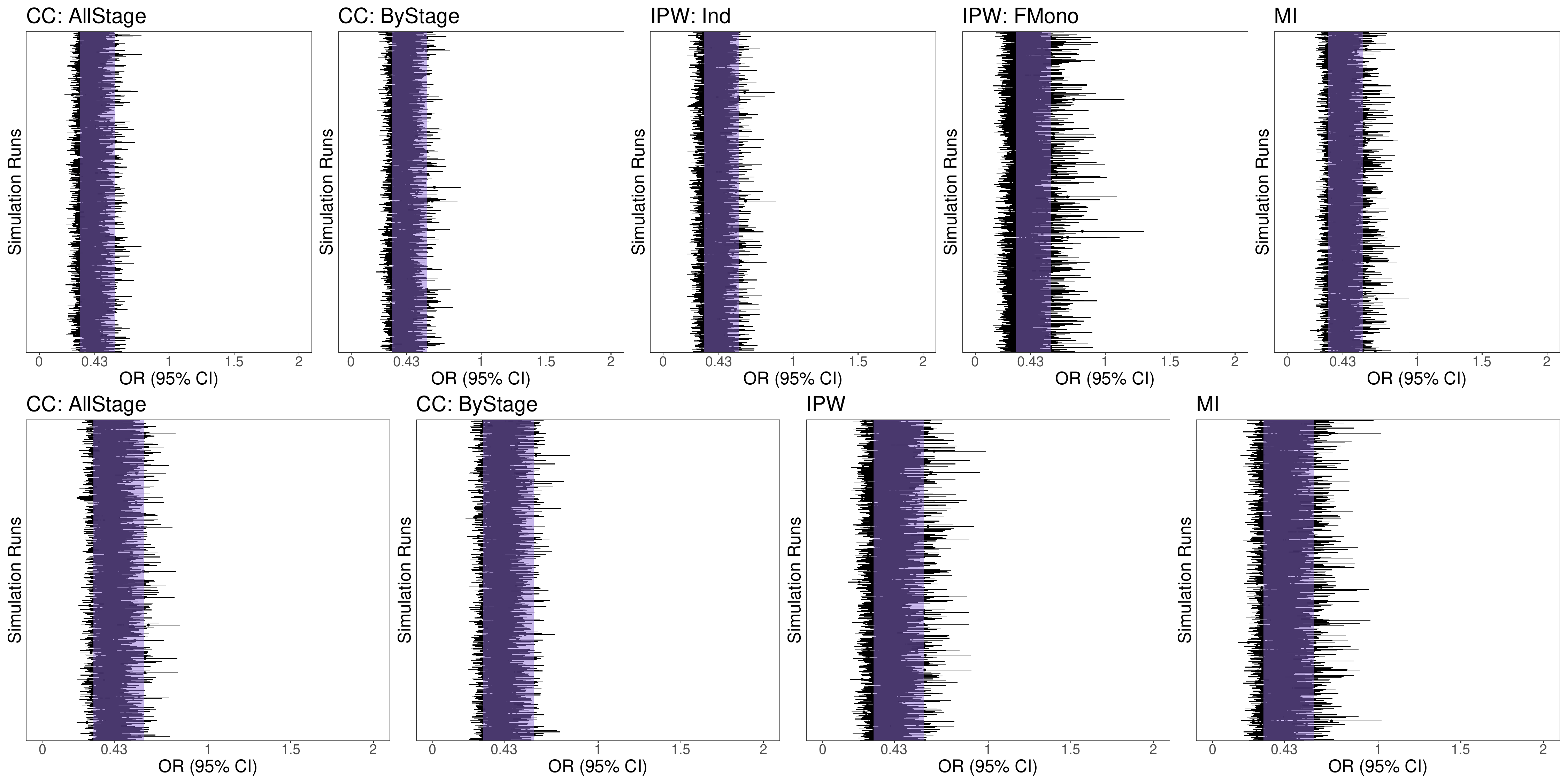}
    \caption{Plot of trial inference results utilizing observed real data and observed synthetic data, for all frameworks under both non-monotone missingness (Scenario 2A, top row) and monotone missingness (Scenario 2B, bottom row). The real (observed) data OR estimate was 0.43 in both Scenarios 2A and 2B, and the real (observed) data 95\% CIs are shown by the purple shaded regions. Each horizontal bar represents the synthetic data 95\% CI for one simulation run.}
\end{figure}

\subsubsection{Scenarios 3A and 3B (MAR x 10\% Missing x Strong Mechanism)}

\begin{figure}[H]
    \centering
    \includegraphics[width=\linewidth]{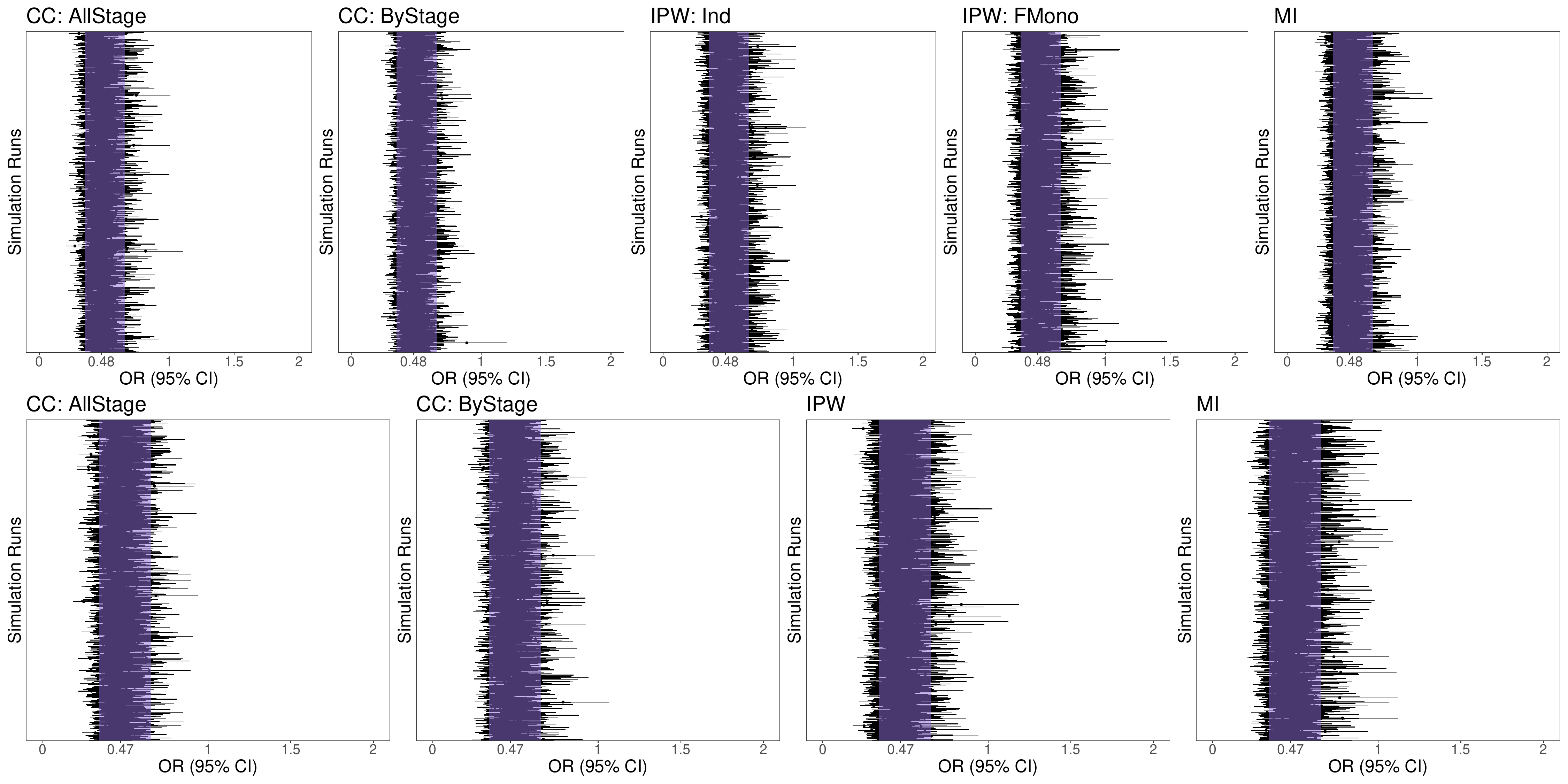}
    \caption{Plot of trial inference results utilizing observed real data and observed synthetic data, for all frameworks under both non-monotone missingness (Scenario 3A, top row) and monotone missingness (Scenario 3B, bottom row). The real (observed) data OR estimate was 0.48 in Scenario 3A and 0.47 in Scenario 3B, and the real (observed) data 95\% CIs are shown by the purple shaded regions. Each horizontal bar represents the synthetic data 95\% CI for one simulation run.}
\end{figure}

\subsubsection{Scenarios 4A and 4B (MAR x 50\% Missing x Strong Mechanism)}

\begin{figure}[H]
    \centering
    \includegraphics[width=\linewidth]{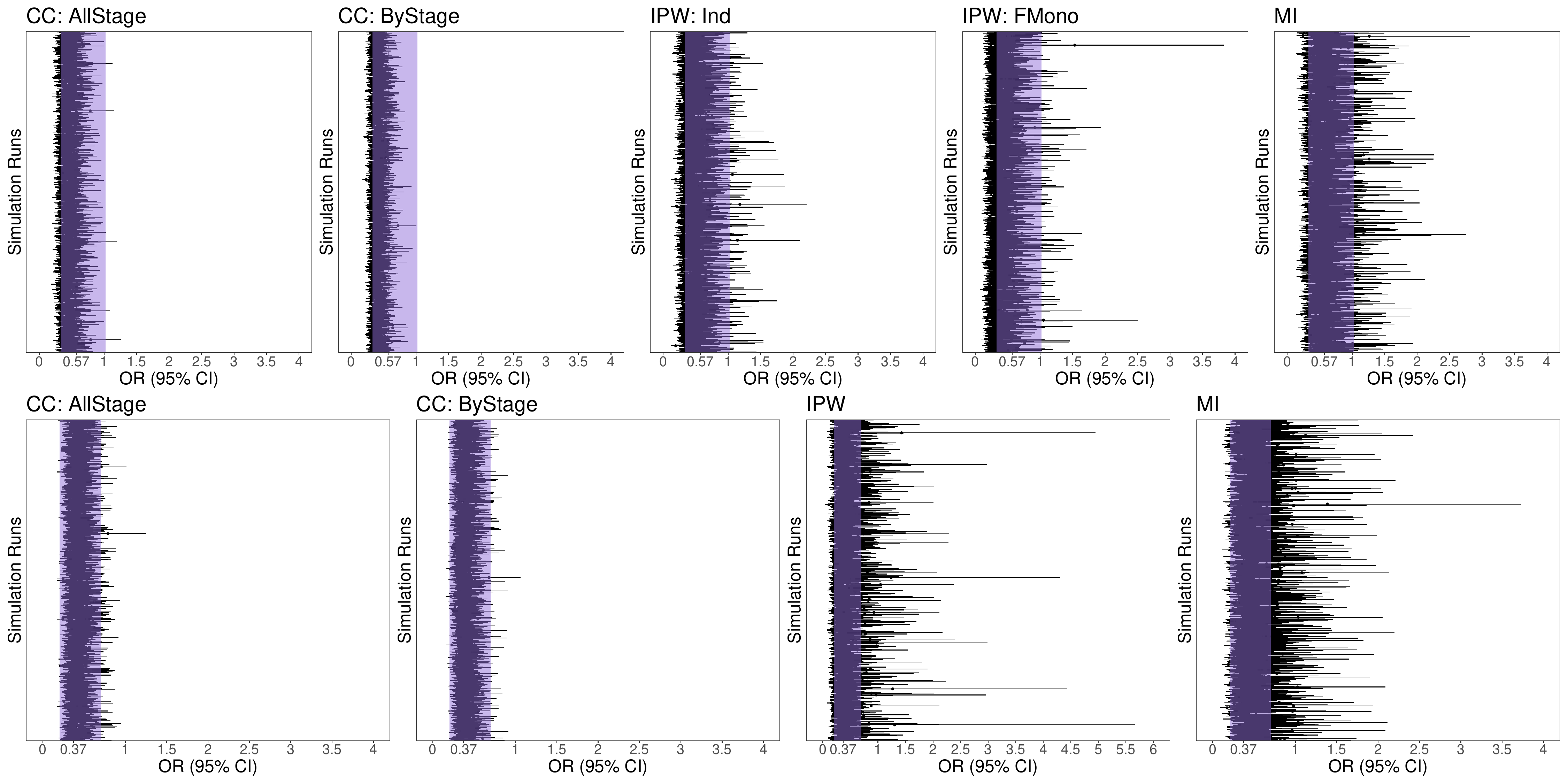}
    \caption{Plot of trial inference results utilizing observed real data and observed synthetic data, for all frameworks under both non-monotone missingness (Scenario 4A, top row) and monotone missingness (Scenario 4B, bottom row). The real (observed) data OR estimate was 0.57 in Scenario 4A and 0.37 in Scenario 4B, and the real (observed) data 95\% CIs are shown by the purple shaded regions. Each horizontal bar represents the synthetic data 95\% CI for one simulation run. Note that the plot for the IPW framework under monotone missingness has a horizontal axis with a larger range than the other plots.}
\end{figure}

\subsubsection{Scenarios 5A and 5B (MAR x 25\% Missing x Weak Mechanism)}

\begin{figure}[H]
    \centering
    \includegraphics[width=\linewidth]{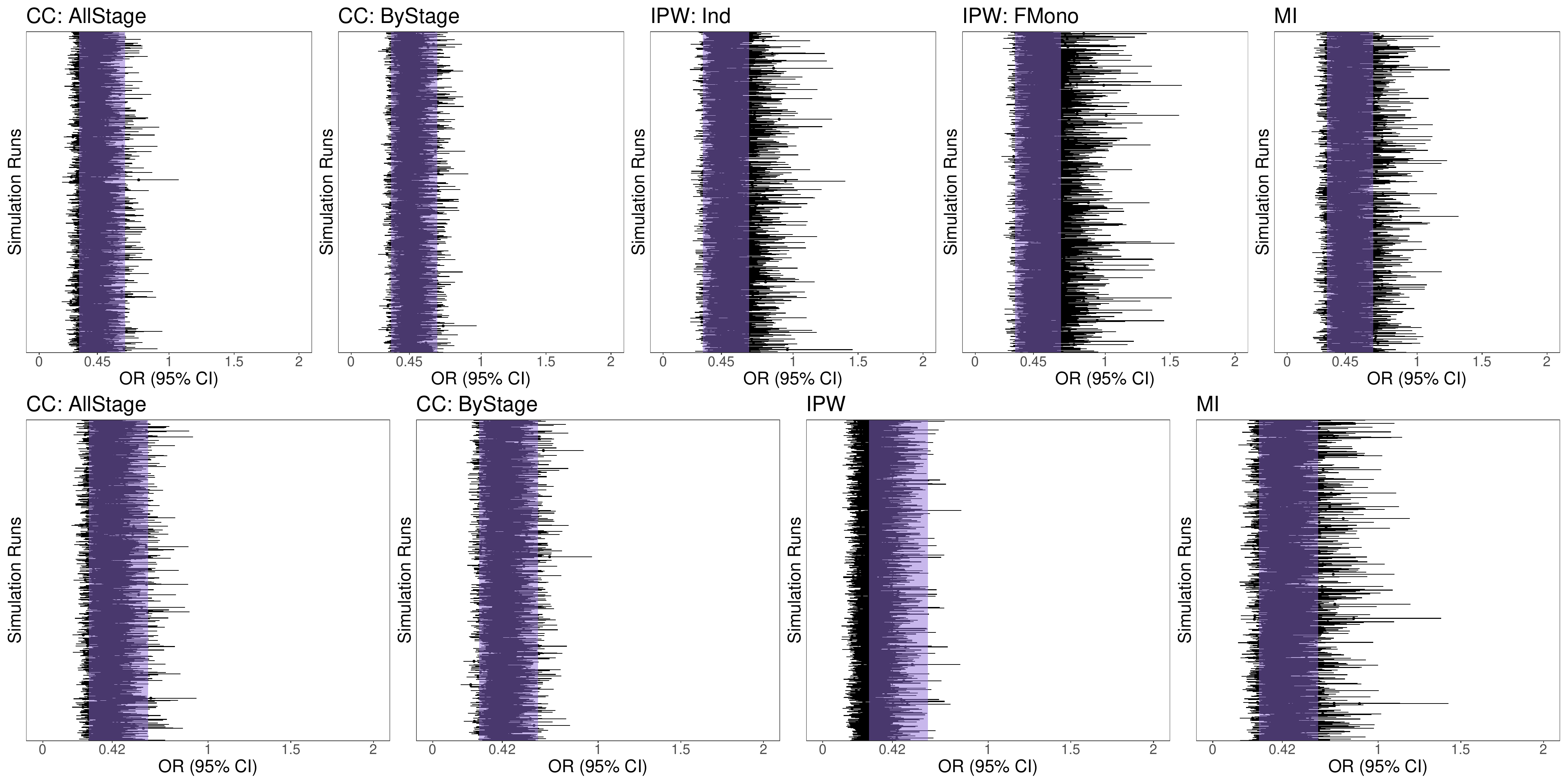}
    \caption{Plot of trial inference results utilizing observed real data and observed synthetic data, for all frameworks under both non-monotone missingness (Scenario 5A, top row) and monotone missingness (Scenario 5B, bottom row). The real (observed) data OR estimate was 0.45 in Scenario 5A and 0.42 in Scenario 5B, and the real (observed) data 95\% CIs are shown by the purple shaded regions. Each horizontal bar represents the synthetic data 95\% CI for one simulation run.}
\end{figure}

\subsubsection{Scenarios 6A and 6B (MAR x 50\% Missing x Weak Mechanism)}

\begin{figure}[H]
    \centering
    \includegraphics[width=\linewidth]{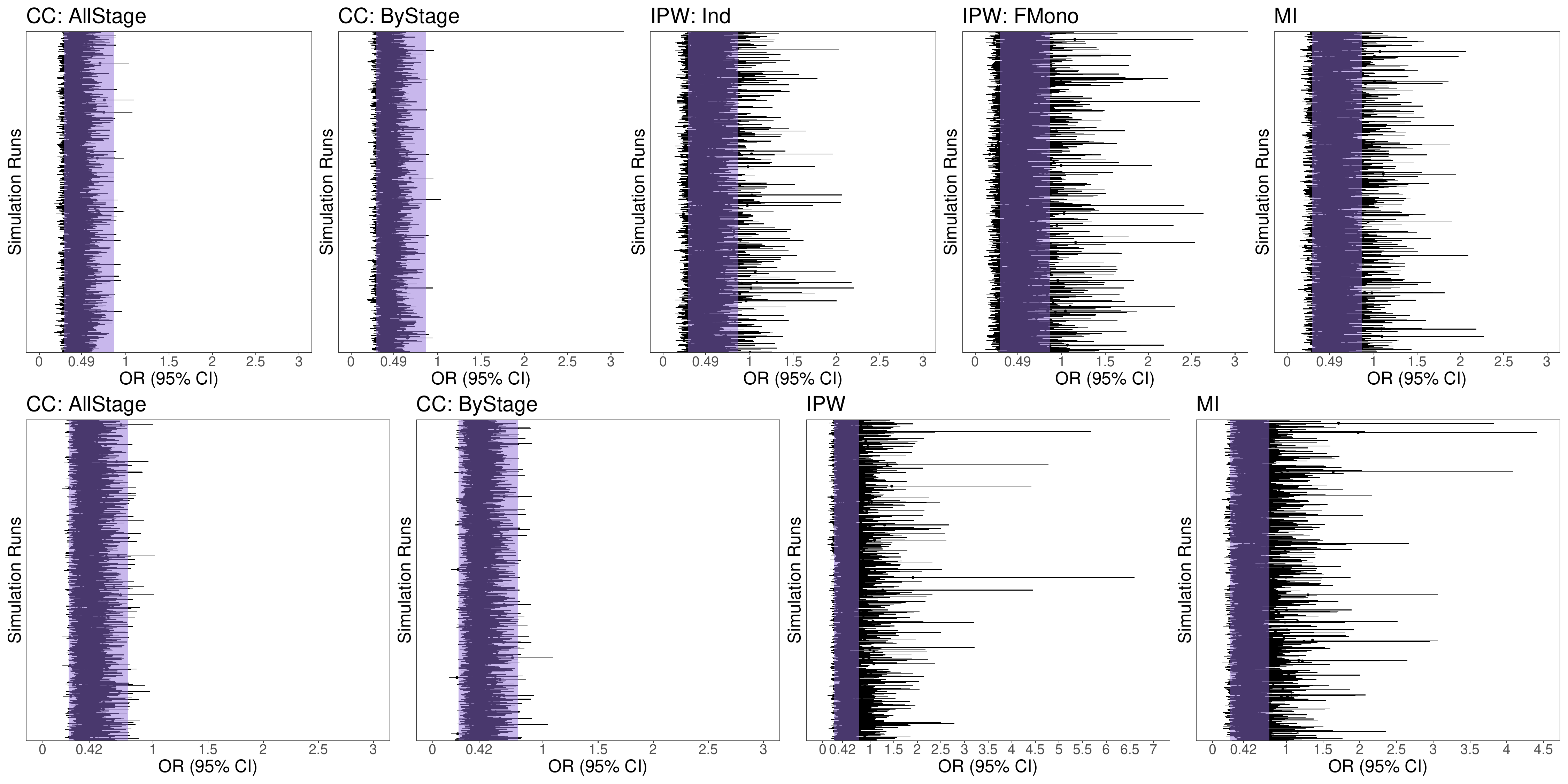}
    \caption{Plot of trial inference results utilizing observed real data and observed synthetic data, for all frameworks under both non-monotone missingness (Scenario 6A, top row) and monotone missingness (Scenario 6B, bottom row). The real (observed) data OR estimate was 0.45 in Scenario 1A and 0.42 in Scenario 1B, and the real (observed) data 95\% CIs are shown by the purple shaded regions. Each horizontal bar represents the synthetic data 95\% CI for one simulation run. Note that the plots for the IPW and MI frameworks under monotone missingness have horizontal axes with larger ranges than the other plots.}
\end{figure}

\subsection{Variables with Missing Data}

\subsubsection{Scenarios 2A and 2B (MCAR x 25\% Missing x Strong Mechanism)}

\begin{figure}[H]
    \centering
    \includegraphics[width=\linewidth]{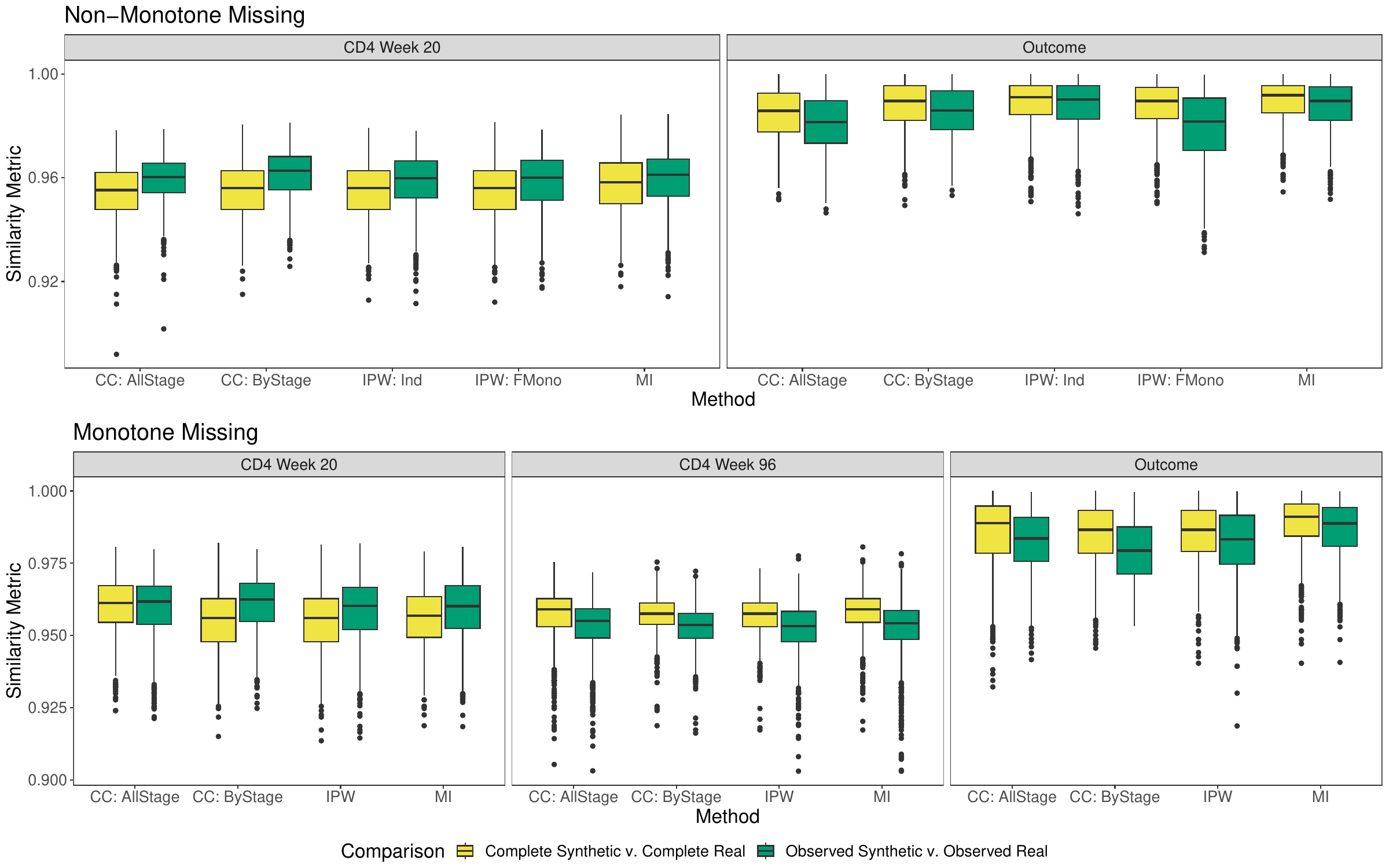}
    \caption{Scenarios 2A and 2B (MCAR x 25\% Missing x Strong Mechanism); plot of similarity metrics for variables with missing values -- 1-KS statistic for continuous variables CD4 count week 20 ($Z_1$) and week 96 ($Z_2$), 1-TVD for the discrete outcome ($Y$). In yellow: comparison of complete synthetic data to complete real data; in green: comparison of observed synthetic data to observed real data. Recall that for CC: All Stage and CC: By Stage, the complete synthetic data and observed synthetic data were the same, as these frameworks do not generate any missing values.}
\end{figure}

\newpage

\subsubsection{Scenarios 3A and 3B (MAR x 10\% Missing x Strong Mechanism)}

\begin{figure}[H]
    \centering
    \includegraphics[width=\linewidth]{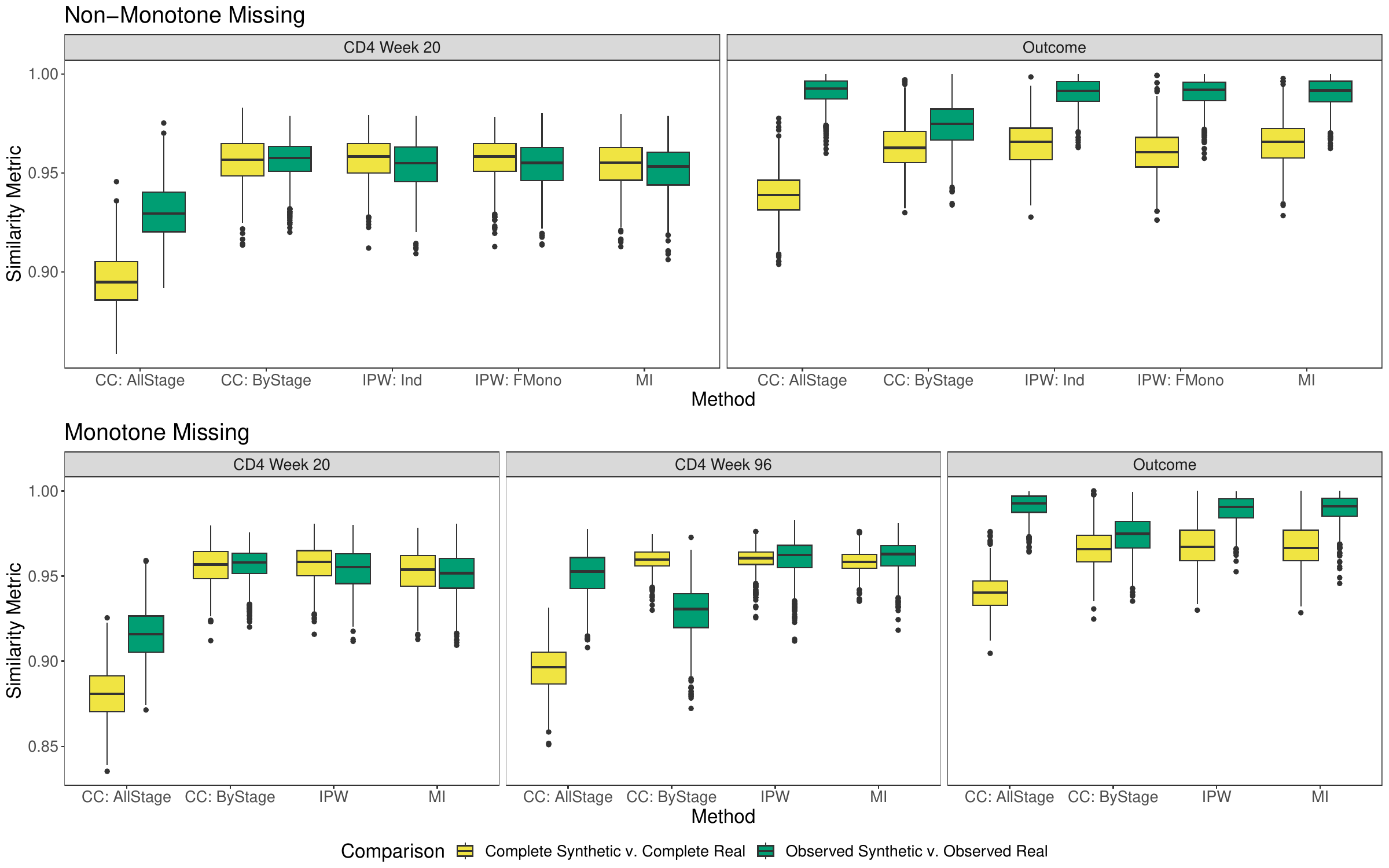}
    \caption{Scenarios 3A and 3B (MAR x 10\% Missing x Strong Mechanism); plot of similarity metrics for variables with missing values -- 1-KS statistic for continuous variables CD4 count week 20 ($Z_1$) and week 96 ($Z_2$), 1-TVD for the discrete outcome ($Y$). In yellow: comparison of complete synthetic data to complete real data; in green: comparison of observed synthetic data to observed real data. Recall that for CC: All Stage and CC: By Stage, the complete synthetic data and observed synthetic data were the same, as these frameworks do not generate any missing values.}
\end{figure}

\newpage

\subsubsection{Scenarios 4A and 4B (MAR x 50\% Missing x Strong Mechanism)}

\begin{figure}[H]
    \centering
    \includegraphics[width=\linewidth]{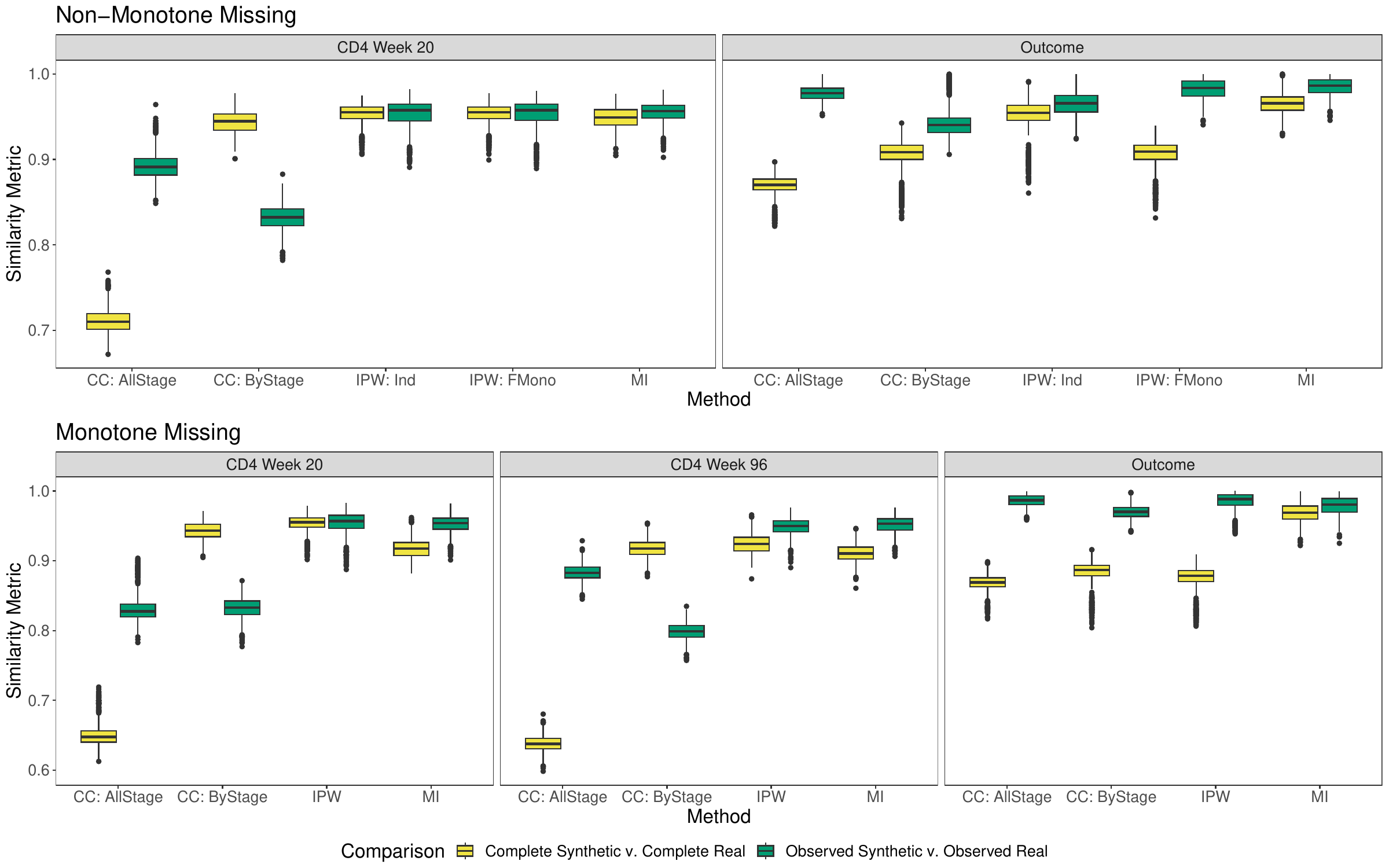}
    \caption{Scenarios 4A and 4B (MAR x 50\% Missing x Strong Mechanism); plot of similarity metrics for variables with missing values -- 1-KS statistic for continuous variables CD4 count week 20 ($Z_1$) and week 96 ($Z_2$), 1-TVD for the discrete outcome ($Y$). In yellow: comparison of complete synthetic data to complete real data; in green: comparison of observed synthetic data to observed real data. Recall that for CC: All Stage and CC: By Stage, the complete synthetic data and observed synthetic data were the same, as these frameworks do not generate any missing values.}
\end{figure}

\newpage

\subsubsection{Scenarios 5A and 5B (MAR x 25\% Missing x Weak Mechanism)}

\begin{figure}[H]
    \centering
    \includegraphics[width=\linewidth]{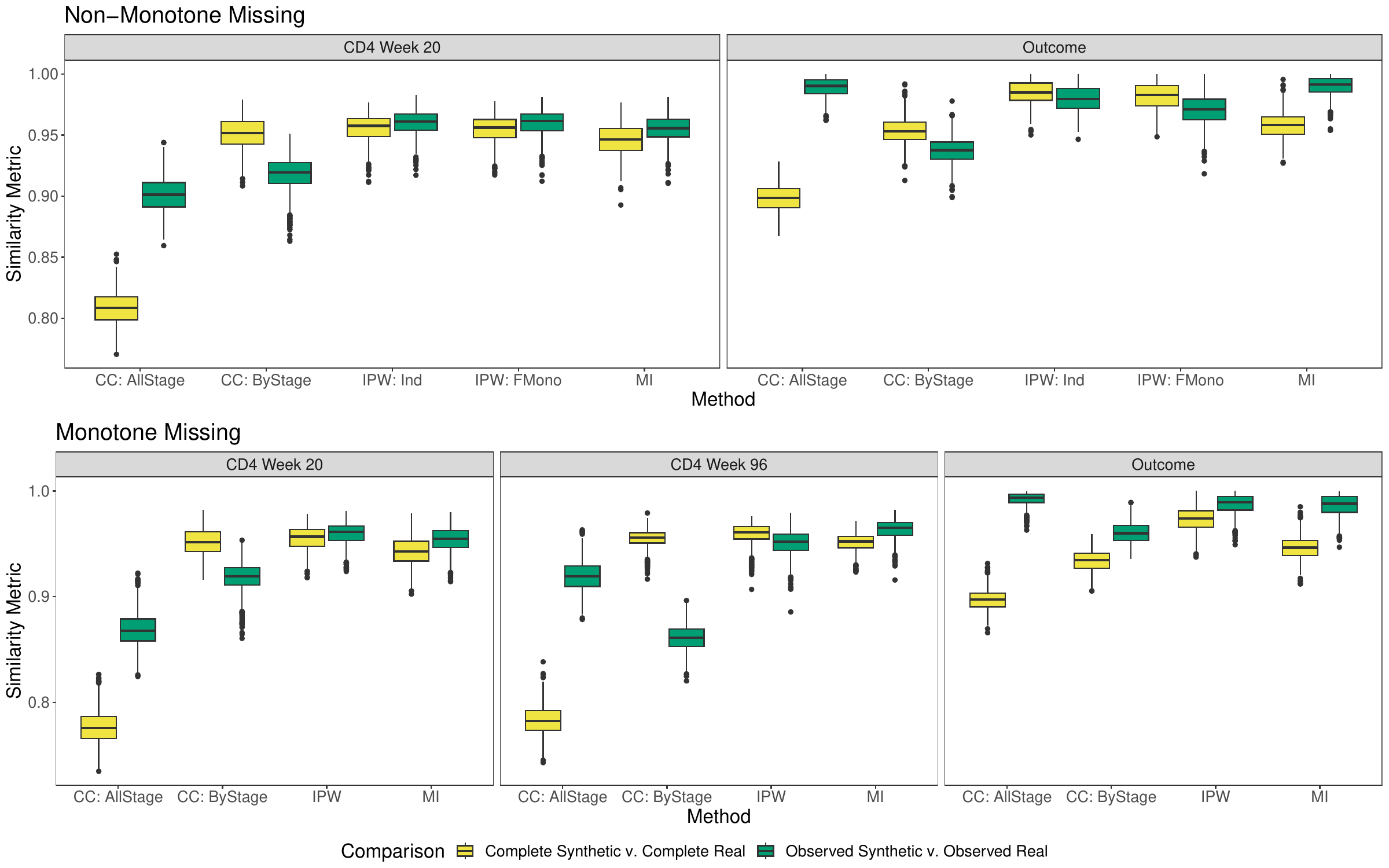}
    \caption{Scenarios 5A and 5B (MAR x 25\% Missing x Weak Mechanism); plot of similarity metrics for variables with missing values -- 1-KS statistic for continuous variables CD4 count week 20 ($Z_1$) and week 96 ($Z_2$), 1-TVD for the discrete outcome ($Y$). In yellow: comparison of complete synthetic data to complete real data; in green: comparison of observed synthetic data to observed real data. Recall that for CC: All Stage and CC: By Stage, the complete synthetic data and observed synthetic data were the same, as these frameworks do not generate any missing values.}
\end{figure}

\newpage

\subsubsection{Scenarios 6A and 6B (MAR x 50\% Missing x Weak Mechanism)}

\begin{figure}[H]
    \centering
    \includegraphics[width=\linewidth]{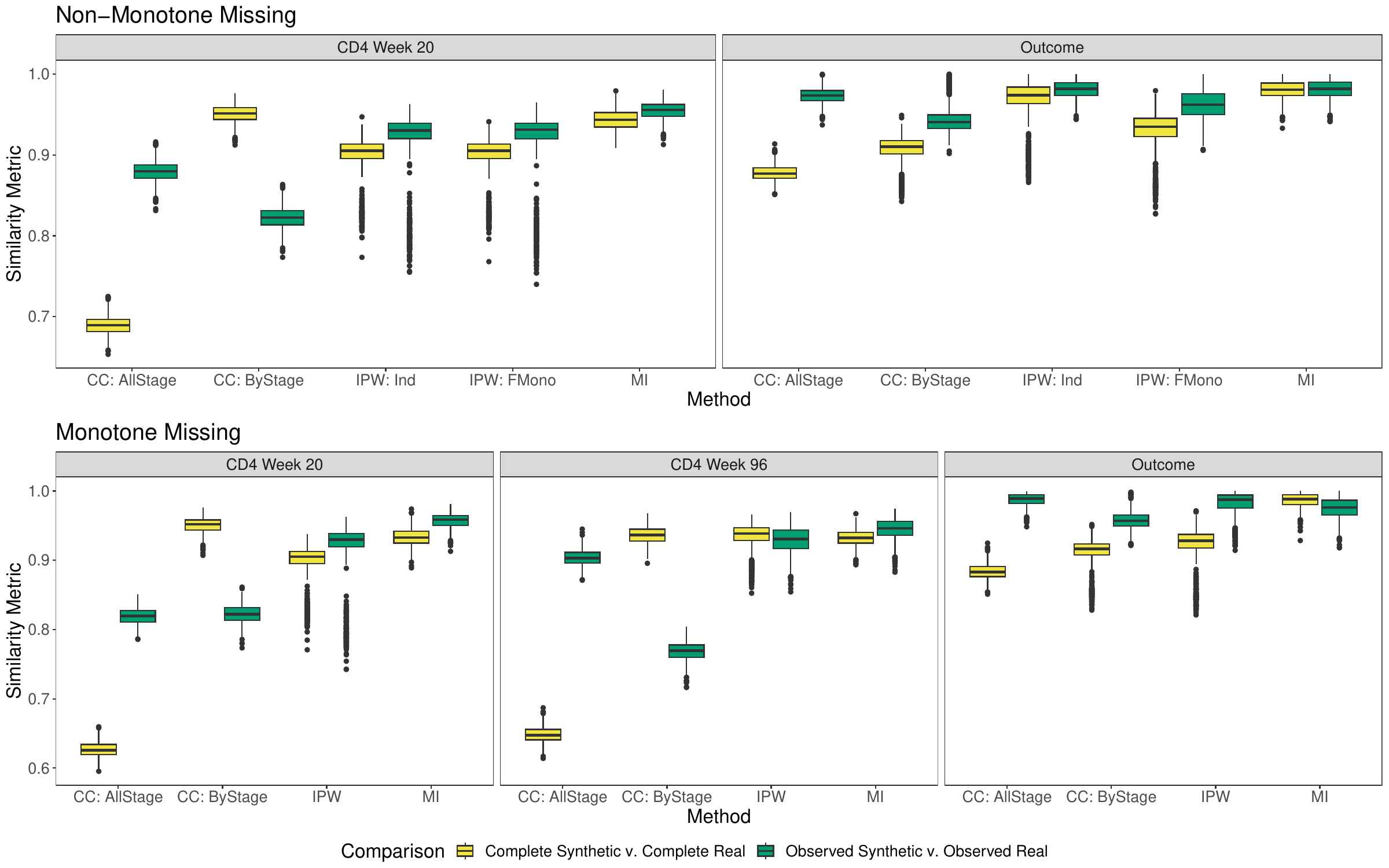}
    \caption{Scenarios 6A and 6B (MAR x 50\% Missing x Weak Mechanism); plot of similarity metrics for variables with missing values -- 1-KS statistic for continuous variables CD4 count week 20 ($Z_1$) and week 96 ($Z_2$), 1-TVD for the discrete outcome ($Y$). In yellow: comparison of complete synthetic data to complete real data; in green: comparison of observed synthetic data to observed real data. Recall that for CC: All Stage and CC: By Stage, the complete synthetic data and observed synthetic data were the same, as these frameworks do not generate any missing values.}
\end{figure}

\newpage

\subsection{Proportion of Synthetic Missingness}

\subsubsection{Scenarios 1A and 1B (MAR x 25\% Missing x Strong Mechanism)}

\FloatBarrier
\begin{figure}[H]
    \centering
    \includegraphics[width=0.95\linewidth]{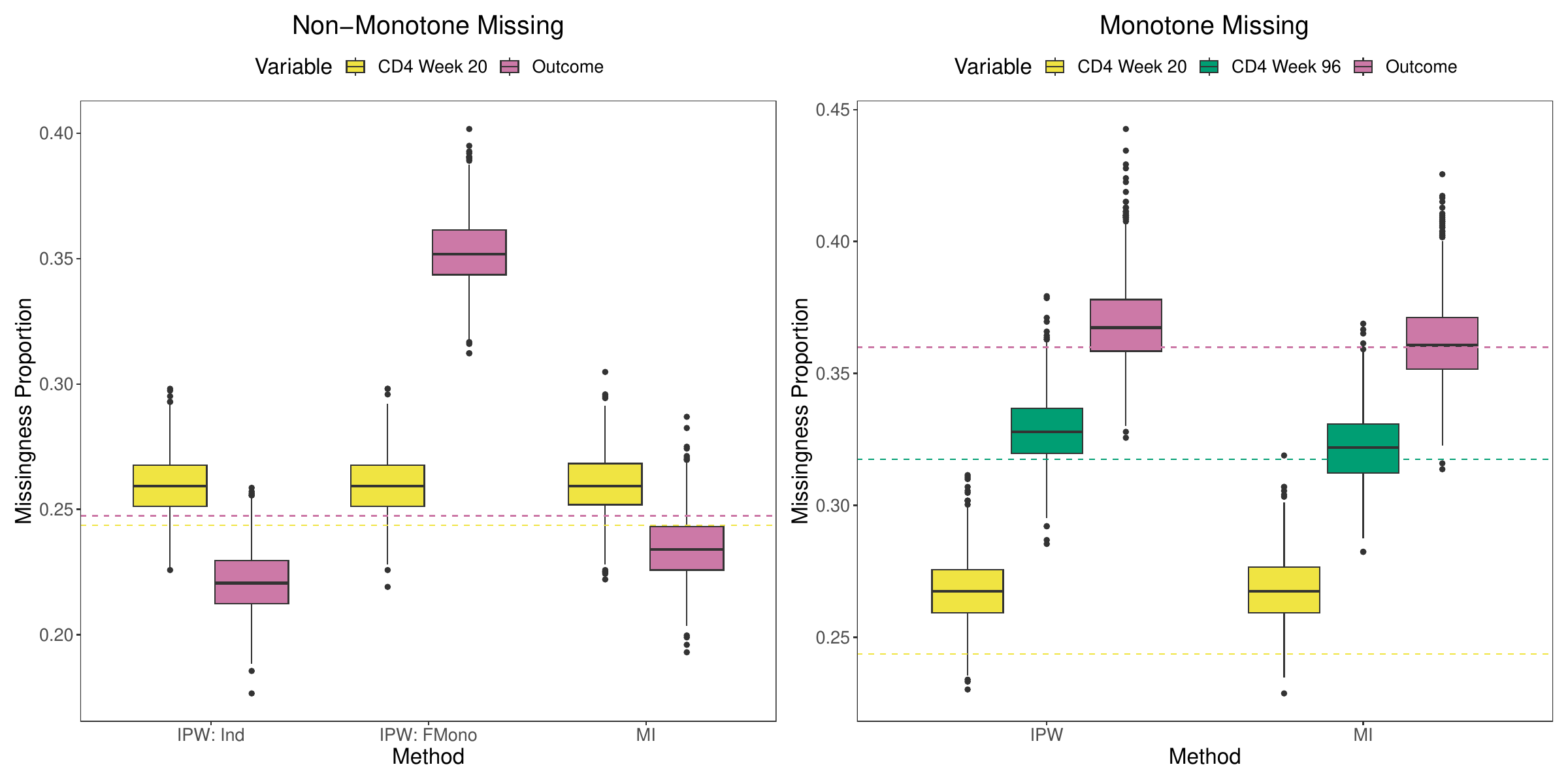}
    \caption{Plot of proportions of missing values generated in the synthetic data for CD4 count at week 20 and the outcome under non-monotone missingness (Scenario 1A) and for CD4 count at week 20, CD4 count at week 96, and the outcome under monotone missingness (Scenario 1B). The horizontal dashed lines represent the proportion of missingness for a given variable (indicated by the colour of the line) in the real data.}
\end{figure}

\subsubsection{Scenarios 2A and 2B (MCAR x 25\% Missing x Strong Mechanism)}

\FloatBarrier
\begin{figure}[H]
    \centering
    \includegraphics[width=0.95\linewidth]{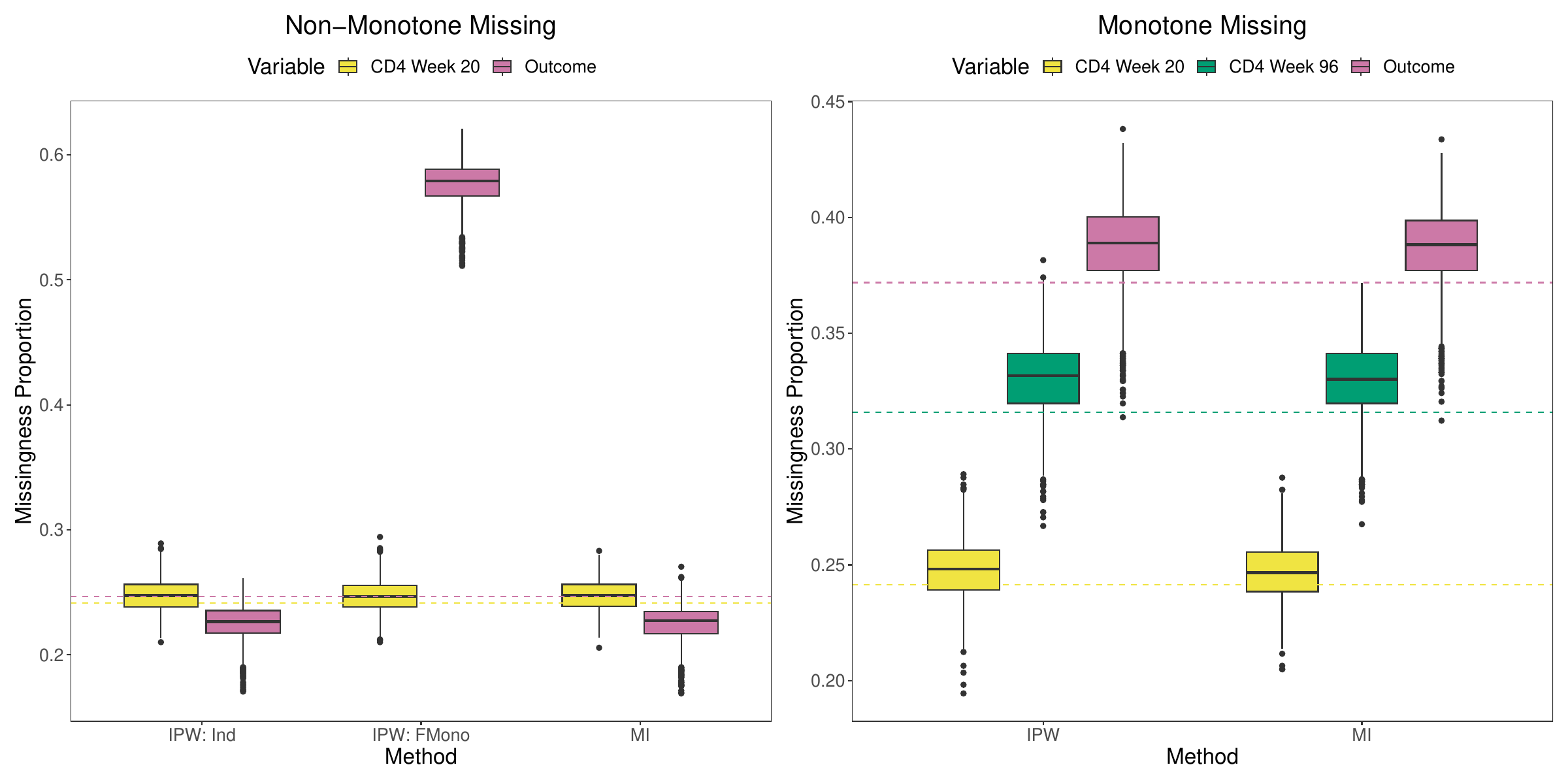}
    \caption{Plot of proportions of missing values generated in the synthetic data for CD4 count at week 20 and the outcome under non-monotone missingness (Scenario 2A) and for CD4 count at week 20, CD4 count at week 96, and the outcome under monotone missingness (Scenario 2B). The horizontal dashed lines represent the proportion of missingness for a given variable (indicated by the colour of the line) in the real data.}
\end{figure}

\subsubsection{Scenarios 3A and 3B (MAR x 10\% Missing x Strong Mechanism)}

\FloatBarrier
\begin{figure}[H]
    \centering
    \includegraphics[width=\linewidth]{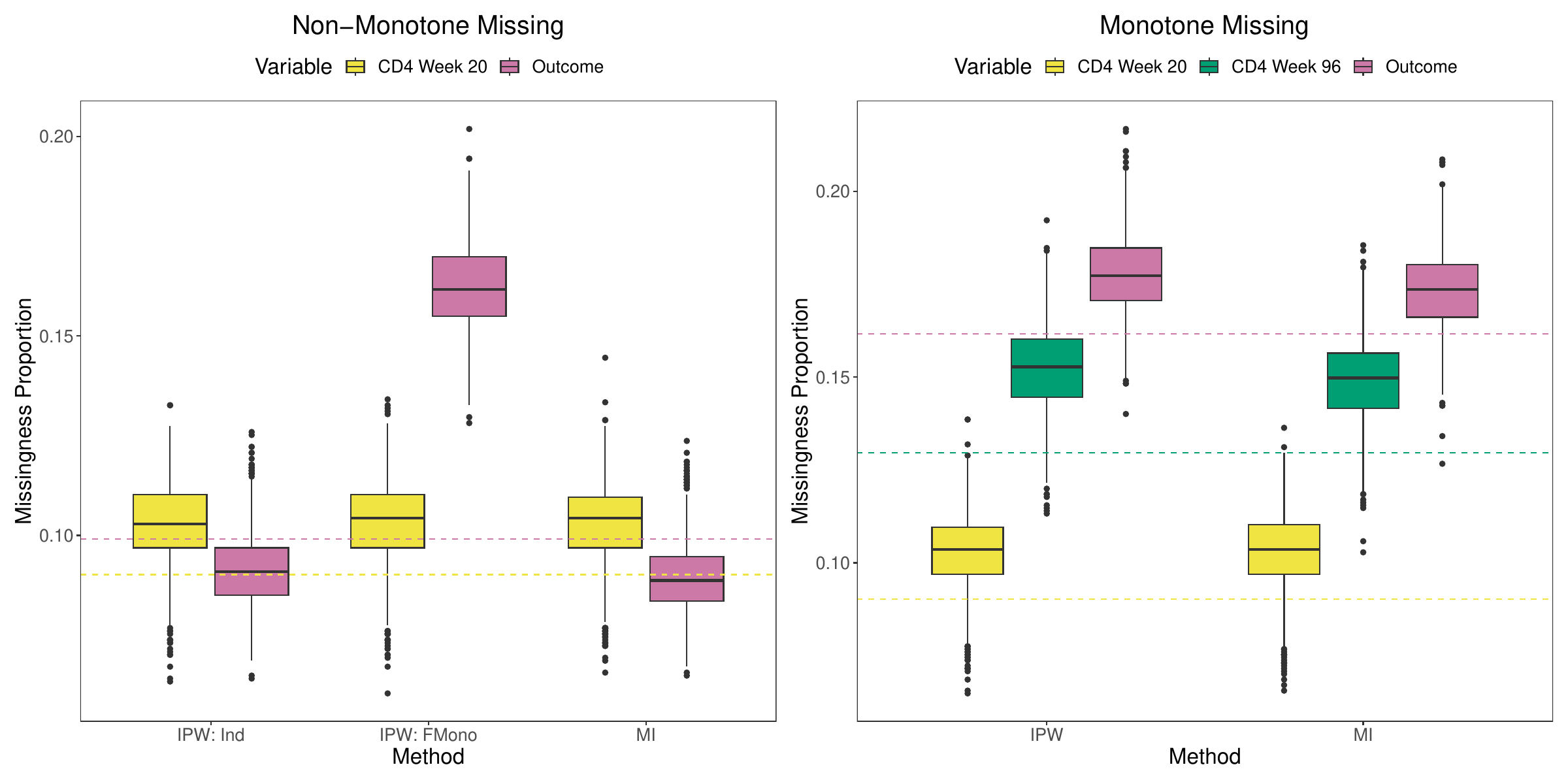}
    \caption{Plot of proportions of missing values generated in the synthetic data for CD4 count at week 20 and the outcome under non-monotone missingness (Scenario 3A) and for CD4 count at week 20, CD4 count at week 96, and the outcome under monotone missingness (Scenario 3B). The horizontal dashed lines represent the proportion of missingness for a given variable (indicated by the colour of the line) in the real data.}
\end{figure}

\subsubsection{Scenarios 4A and 4B (MAR x 50\% Missing x Strong Mechanism)}

\FloatBarrier
\begin{figure}[H]
    \centering
    \includegraphics[width=\linewidth]{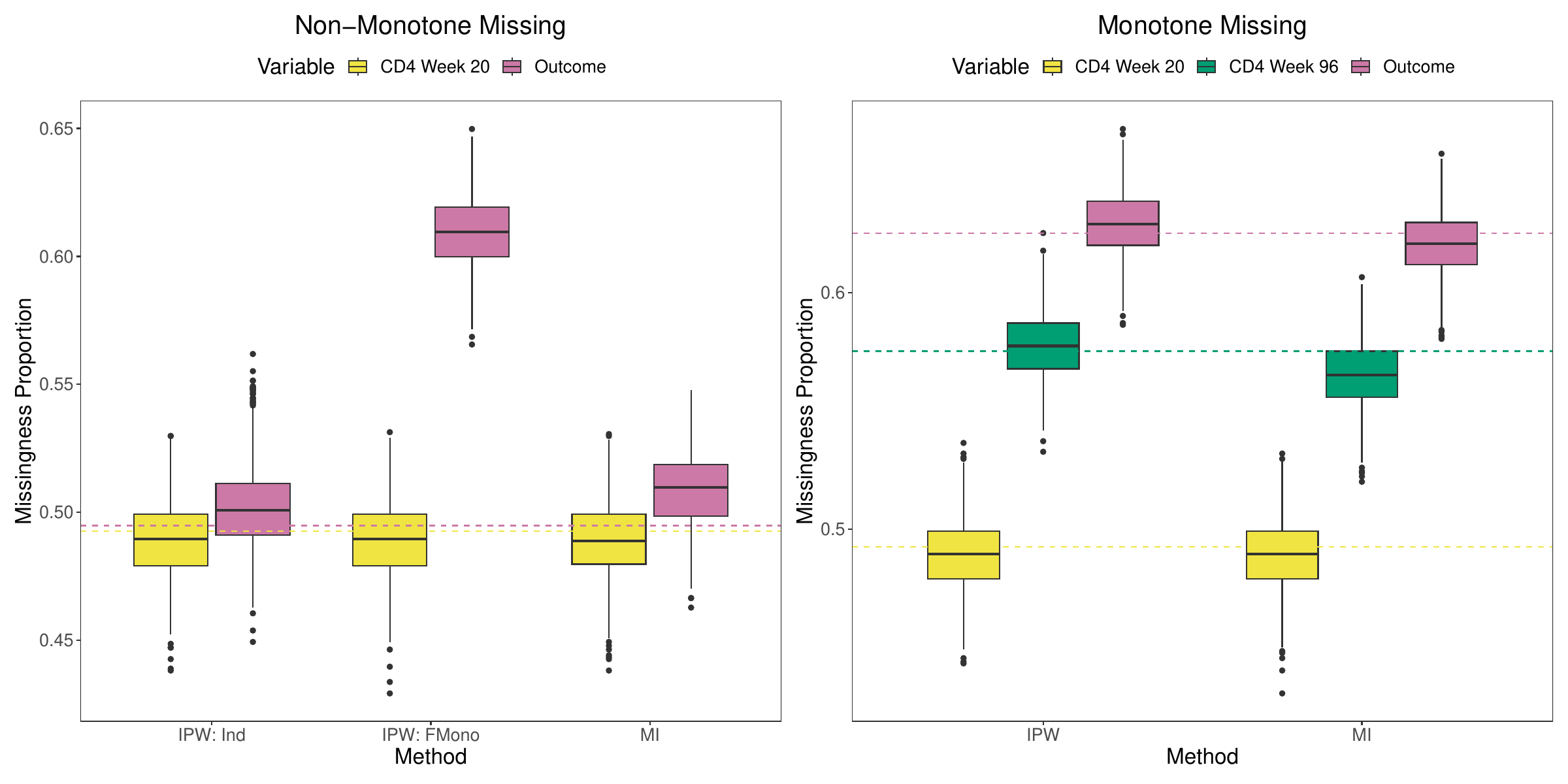}
    \caption{Plot of proportions of missing values generated in the synthetic data for CD4 count at week 20 and the outcome under non-monotone missingness (Scenario 4A) and for CD4 count at week 20, CD4 count at week 96, and the outcome under monotone missingness (Scenario 4B). The horizontal dashed lines represent the proportion of missingness for a given variable (indicated by the colour of the line) in the real data.}
\end{figure}

\subsubsection{Scenarios 5A and 5B (MAR x 25\% Missing x Weak Mechanism)}

\FloatBarrier
\begin{figure}[H]
    \centering
    \includegraphics[width=\linewidth]{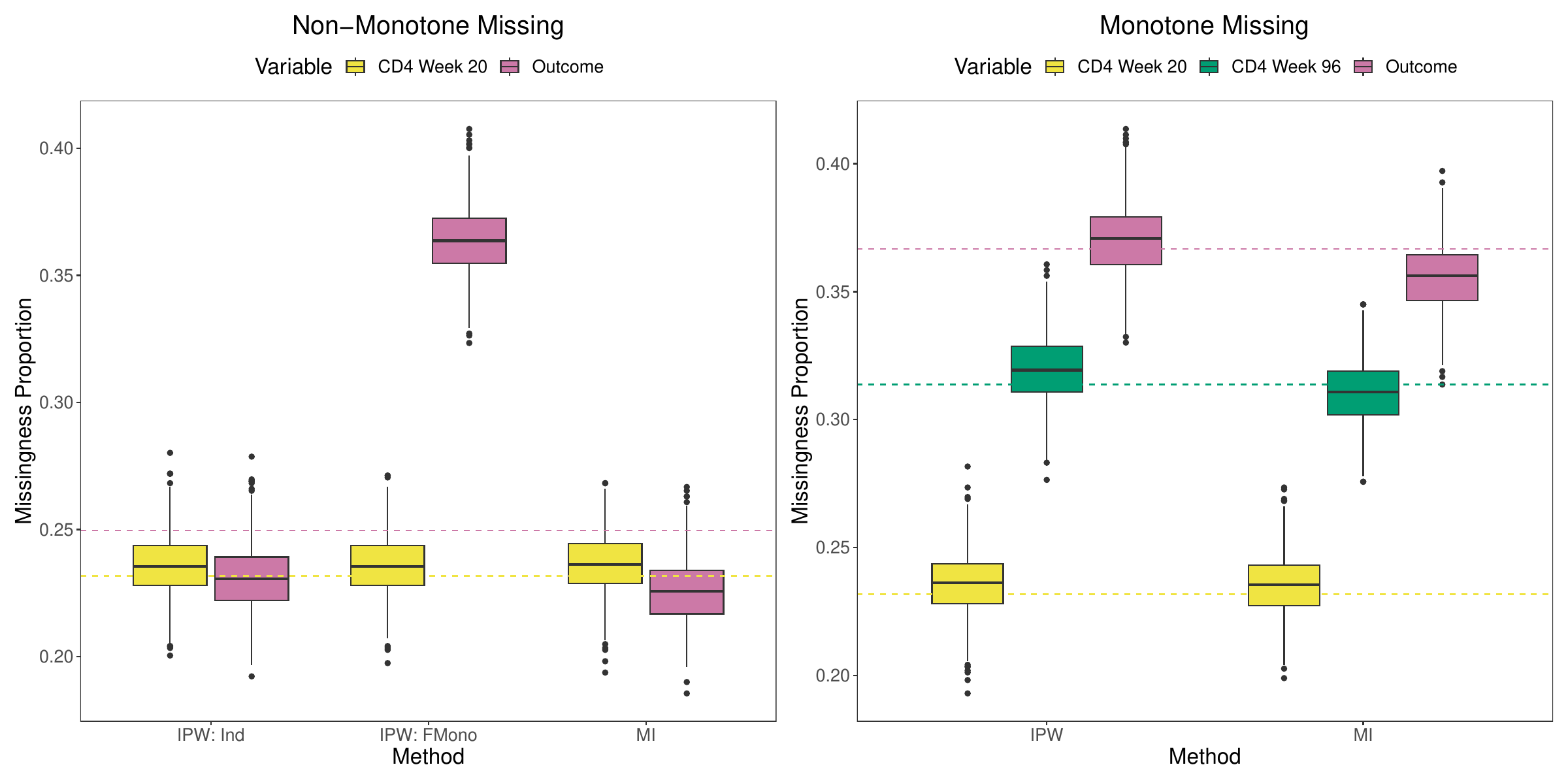}
    \caption{Plot of proportions of missing values generated in the synthetic data for CD4 count at week 20 and the outcome under non-monotone missingness (Scenario 5A) and for CD4 count at week 20, CD4 count at week 96, and the outcome under monotone missingness (Scenario 5B). The horizontal dashed lines represent the proportion of missingness for a given variable (indicated by the colour of the line) in the real data.}
\end{figure}

\subsubsection{Scenarios 6A and 6B (MAR x 50\% Missing x Weak Mechanism)}

\FloatBarrier
\begin{figure}[H]
    \centering
    \includegraphics[width=\linewidth]{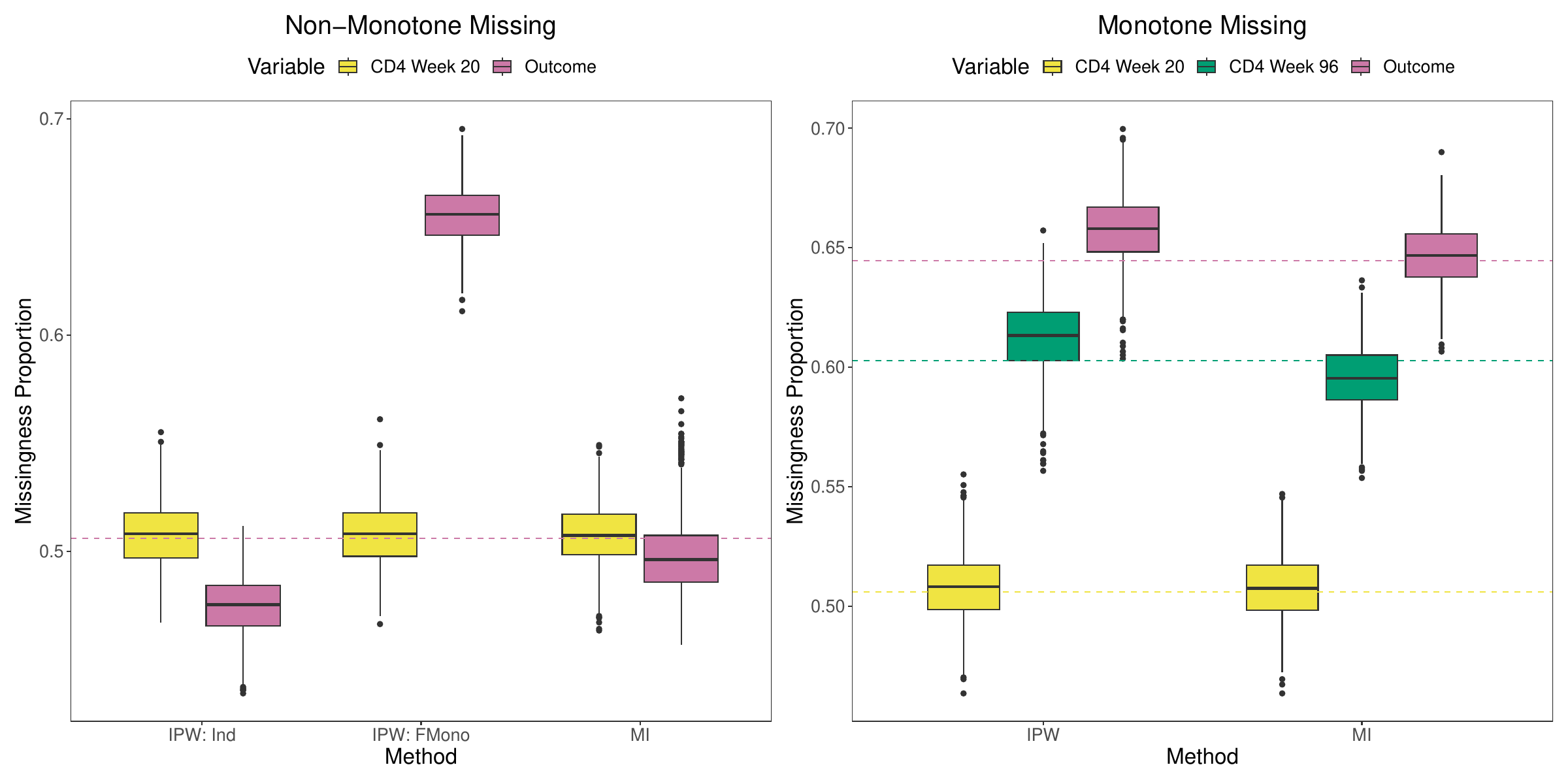}
    \caption{Plot of proportions of missing values generated in the synthetic data for CD4 count at week 20 and the outcome under non-monotone missingness (Scenario 5A) and for CD4 count at week 20, CD4 count at week 96, and the outcome under monotone missingness (Scenario 6B). The horizontal dashed lines represent the proportion of missingness for a given variable (indicated by the colour of the line) in the real data. Note: in the non-monotone missing plot (left hand side), the yellow dashed line does not appear since both variables had approximately the same proportion of missingness.}
\end{figure}

\newpage

\subsection{Computation Time}

Here we present the total computation time for each framework, to generate a synthetic data set and calculate metrics across 1000 parallelized simulation runs (using 10 cores).

\smallskip

\noindent
\textit{A: Non-Monotone Setting}

\vspace{-1cm}
\begin{table}[H]
\caption*{}
\begin{tabular}{|| c c c c c c c ||}
\hline
 & & & \textbf{Scenarios} & & & \\
\hline
\textbf{Framework} & \textbf{1A} & \textbf{2A} & \textbf{3A} & \textbf{4A} & \textbf{5A} & \textbf{6A} \\
\hline
CC: All Stage & 01:13:39 & 01:03:40 & 01:29:35 & 00:53:20 & 01:00:51 & 01:00:29 \\
\hline
CC: By Stage & 01:31:15 & 01:31:05 & 01:29:49 & 01:30:20 & 01:25:11 & 01:32:54 \\ 
\hline
IPW: Indicator Method & 01:32:45 & 01:32:55 & 01:31:23 & 01:31:52 & 01:27:31 & 01:35:19 \\ 
\hline
IPW: Force Monotonicity & 01:33:21 & 01:33:26 & 01:31:00 & 01:32:25 & 01:27:51 & 01:35:42 \\ 
\hline
MI & 01:59:18 & 01:59:52 & 01:42:07 & 02:20:35 & 01:51:56 & 02:28:38 \\ 
\hline
\end{tabular}
\end{table}

\noindent
\textit{B: Monotone Setting}

\vspace{-1cm}
\begin{table}[H]
\caption*{}
\begin{tabular}{|| c c c c c c c ||}
\hline
 & & & \textbf{Scenarios} & & & \\
\hline
\textbf{Framework} & \textbf{1B} & \textbf{2B} & \textbf{3B} & \textbf{4B} & \textbf{5B} & \textbf{6B} \\
\hline
CC: All Stage & 01:17:31 & 01:31:58 & 01:11:29 & 00:43:51 & 00:58:42 & 00:55:39 \\
\hline
CC: By Stage & 01:29:02 & 01:32:58 & 01:25:55 & 01:29:16 & 01:30:01 & 01:33:38 \\ 
\hline
IPW & 01:31:46 & 01:35:36 & 01:28:36 & 01:32:59 & 01:32:46 & 01:38:37 \\ 
\hline
MI & 02:12:39 & 02:21:32 & 01:47:11 & 02:43:29 & 02:14:48 & 02:54:09 \\ 
\hline
\end{tabular}
\end{table}

\end{document}